\documentclass{optica-article}

\journal{opticajournal} 

\articletype{Review}

\usepackage[nolist]{acronym}
\usepackage{physics}
\usepackage{soul}
\usepackage{wrapfig}
\usepackage[utf8]{inputenc}

\newcommand{\bfr}{\mathbf{r}}
\newcommand{\bfR}{\mathbf{R}}
\newcommand{\bfk}{\mathbf{k}}
\newcommand{\bfp}{\mathbf{p}}
\newcommand{\hbfp}{\hat{\bfp}}
\newcommand{\bfP}{\mathbf{P}}
\newcommand{\hbfP}{\hat{\bfP}}
\newcommand{\bfq}{\mathbf{q}}

\newcommand{\bfE}{\mathbf{E}}
\newcommand{\hbfE}{\hat{\bfE}}
\newcommand{\hbfEperp}{\hat{\bfE}_{\perp}}
\newcommand{\bfB}{\mathbf{B}}
\newcommand{\hbfB}{\hat{\bfB}}
\newcommand{\bfJtot}{\mathbf{J}}
\newcommand{\bfj}{\mathbf{j}}
\newcommand{\vacep}{\epsilon_{0}}
\newcommand{\vacmu}{\mu_{0}}
\newcommand{\divg}{\nabla\cdot}
\newcommand{\pdt}{\partial_{t}}
\newcommand{\bfA}{\mathbf{A}}

\newcommand{\bfAperp}{\bfA_{\perp}}
\newcommand{\hbfAperp}{\hat{\bfA}_{\perp}}
\newcommand{\bfJperp}{\bfJtot_{\perp}}
\newcommand{\bfJpara}{\bfJtot_{\parallel}}
\newcommand{\wavepol}{\boldsymbol{\varepsilon}}
\newcommand{\paulimat}{\boldsymbol{\sigma}}
\newcommand{\mebare}{m_{e,\rm{b}}}

\begin{acronym}
\acro{2D}[2D]{two-dimensional}
\acro{QVFs}[QVFs]{quantum vacuum fluctuations}
\acro{FFs}[FFs]{field fluctuations}
\acro{IR}[IR]{infrared}
\acro{cQED}[cQED]{cavity quantum electrodynamics}
\acro{EM}[EM]{electromagnetic}
\acro{QED}[QED]{quantum electrodynamics}
\acro{PF}[PF]{Pauli-Fierz}
\acro{cBOA}[cBOA]{cavity-Born-Oppenheimer}
\acro{PoPES}[PoPES]{polaritonic-potential-energy surfaces}
\acro{LWA}[LWA]{\textit{long-wavelength} approximation}
\acro{DGF}[DGF]{dyadic Green's function}
\acro{DFT}[DFT]{density-functional theory}
\acro{QEDFT}[QEDFT]{quantum-electrodynamical density-functional theory}
\acro{CC-QED}[cc-QED]{coupled-cluster QED}
\acro{QHE}[QHE]{quantum Hall effect}
\acro{IQHE}[IQHE]{integer quantum Hall effect}
\acro{FQHE}[FQHE]{fractional quantum Hall effect}
\acro{2DEG}[2DEG]{two-dimensional electron gas}
\acro{MIT}[MIT]{metal-to-insulator transition}
\acro{CDW}[CDW]{charge density wave}
\acro{SoD}[SoD]{Star-of-David}
\acro{BvK}[BvK]{Born-von Karman}
\acro{MQED}[MQED]{Macroscopic quantum electrodynamics}
\acro{pHEG}[pHEG]{photon-coupled homogeneous electron gas}
\acro{TMD}[TMD]{transition metal dichalcogenide}
\acro{TMDs}[TMDs]{transition metal dichalcogenides}
\acro{hBN}[hBN]{hexagonal boron nitride}
\acro{QSL}[QSL]{quantum spin liquid}
\acro{AFM}[AFM]{antiferromagnetic}
\acro{BZ}[BZ]{Brillouin zone}
\acro{BOA}[BOA]{Born-Oppenheimer approximation}
\acro{KS}[KS]{Kohn-Sham}
\acro{xc}[xc]{exchange-correlation}
\acro{DFPT}[DFPT]{density functional perturbation theory}
\acro{TB}[TB]{tight-binding}
\acro{MLWFs}[MLWFs]{maximally localized Wannier functions}
\acro{ML}[ML]{Machine learning}
\acro{HEG}[HEG]{homogeneous electron gas}
\acro{pHEG}[pHEG]{photon-coupled homogeneous electron gas}
\acro{CNA}[CNA]{clamped-nuclei approximation}
\acro{MP}[MP]{Maxwell-Pauli}
\acro{Mxc}[Mxc]{mean-field exchange-correlation}
\acro{OEP}[OEP]{optimized effective potential}
\acro{DMFT}[DMFT]{dynamical mean-field theory}
\acro{SSH}[SSH]{Su-Schrieffer-Heeger}
\acro{PAW}[PAW]{projector augmented wave}
\acro{PS}[PS]{polystyrene}
\acro{FPC}[FPC]{Fabry-Pérot cavity}
\end{acronym}

\begin{document}

\title{Cavity engineering of solid-state materials without external driving}
\author{I-Te Lu\authormark{1,*}, Dongbin Shin\authormark{1,2}, Mark Kamper Svendsen\authormark{1,3}, Simone Latini\authormark{1,4}, Hannes Hübener\authormark{1}, Michael Ruggenthaler\authormark{1,$\dagger$}, and Angel Rubio\authormark{1,5,$\dagger\dagger$}}

\address{\authormark{1}Max Planck Institute for the Structure and Dynamics of Matter and Center for Free-Electron Laser Science, Luruper Chaussee 149, Hamburg 22761, Germany\\
\authormark{2}Department of Physics and Photon Science, Gwangju Institute of Science and Technology (GIST), Gwangju 61005, Republic of Korea\\
\authormark{3}{Novo Nordisk Foundation Quantum Computing Programme, Niels Bohr Institute,
University of Copenhagen, Universitetsparken 5, 2100 Copenhagen, Denmark}\\
\authormark{4}Department of Physics, Technical University of Denmark, 2800 Kgs. Lyngby, Denmark\\
\authormark{5}Center for Computational Quantum Physics (CCQ), The Flatiron Institute, 162 Fifth avenue, New York, New York 10010, United States of America}
\email{\authormark{*}i-te.lu@mpsd.mpg.de}
\email{\authormark{$\dagger$}michael.ruggenthaler@mpsd.mpg.de}
\email{\authormark{$\dagger\dagger$}angel.rubio@mpsd.mpg.de} 


\begin{abstract*} 
Confining electromagnetic fields inside an optical cavity can enhance the light-matter coupling between quantum materials embedded inside the cavity and the confined photon fields. When the interaction between the matter and the photon fields is strong enough, even the quantum vacuum field fluctuations of the photons confined in the cavity can alter the properties of the cavity-embedded solid-state materials at equilibrium and room temperature. This approach to engineering materials with light avoids fundamental issues of laser-induced transient matter states. To clearly differentiate this field from phenomena in driven systems, we call this emerging field \textit{cavity materials engineering}. In this review, we first present theoretical frameworks, especially, \textit{ab initio} methods, for describing light-matter interactions in solid-state materials embedded inside a realistic optical cavity. Next, we overview a few experimental breakthroughs in this domain, detailing how the ground state properties of materials can be altered within such confined photonic environments. Moreover, we discuss state-of-the-art theoretical proposals for tailoring material properties within cavities. Finally, we outline the key challenges and promising avenues for future research in this exciting field. 
\end{abstract*}

\tableofcontents

\section{Introduction to cavity materials engineering}
\label{sec:intro}
\subsection{Materials engineering and light-matter interaction}
Solid-state materials play crucial roles in modern technology like energy~\cite{amiri.shahbazian-yassar_2021,was.petti.ea_2019}, information technology~\cite{liu.hersam_2019,deleon.itoh.ea_2021,atature.englund.ea_2018}, and space exploration~\cite{pernigoni.grande_2023,levchenko.bazaka.ea_2018}. 
The performance of devices in those technologies is intricately linked to the properties of the materials from which they are constructed. These material properties -- whether electrical, magnetic, optical, thermal, mechanical, or chemical -- stem from a complex hierarchy of structures, ranging from the atomic scale to the macroscopic level~\cite{jr.rethwisch_2020,smith.hashemi_2011}. 
Mastering the engineering of these structures is essential for creating materials with novel functionalities.

At the most fundamental level, the atomic structure -- defined by the type and arrangement of atoms within a material -- is what determines the properties of solid-state materials~\cite{ashcroft.mermin_2011}. The behavior of these atoms and their interactions is governed by the \ac{EM} forces between charged particles (electrons and nuclei/ions), manifesting as both repulsive and attractive interactions. By controlling the number, arrangement, and interaction of these charged particles, a diverse array of material phases and properties can be achieved by applying temperature, pressure, strain, or doping. In \ac{2D} van der Waals materials, additional techniques, such as dielectric screening, electrical gating, and varying the twisting angles between layers, offer further control over \ac{EM} interactions, leading to the discovery of unique electronic behaviors and novel phases~\cite{li.qian.ea_2021,lin.zhang.ea_2022}. Condensed matter physics, or solid-state physics, is fundamentally concerned with understanding these complex interactions and their profound influence on material properties~\cite{altland.simons_2023,coleman_2015,grosso.parravicini_2013,bruus.flensberg_2004,ashcroft.mermin_2011}.

In addition to electrons and nuclei/ions, photons as quantum particles play a critical, yet often overlooked, role in the fundamental atomic structure of materials and the control of materials phenomena. Coulomb interactions between charged particles arise from the longitudinal component of quantized \ac{EM} fields~\cite{cohen-tannoudji.dupont-roc.ea_1989,cohen-tannoudji.dupont-roc.ea_1998,mandl.shaw_2013}, making them an intrinsic aspect of matter. 
Traditionally, the transverse component of the \ac{EM} fields, or photon fields, has only been considered as an external factor used to probe the information of materials or excite charged particles in materials. However, this component is also intrinsic to the atomic structure and can significantly influence atomic or material properties. For instance, the well-known Lamb shift in the hydrogen atom in free space~\cite{lamb.retherford_1947,maclay_2020} and the Casimir effect between two material slabs~\cite{casimir_1948,woods.dalvit.ea_2016} are both consequences of quantum vacuum field fluctuations of virtual photons -- random fluctuations in the \ac{EM} fields that persist even in the absence of any external light source. Additionally, the physical (measured) electron mass in free space includes contributions from the photon clouds surrounding the charged particle~\cite{mandl.shaw_2013,peskin_2018}. These effects illustrate that photon fields are not merely passive but active in shaping the atomic structure. 

Apart from \ac{QED}, which treats both charged particles and \ac{EM} fields as quantum entities and is the primary focus of this review, significant advances in understanding light-matter interactions in solids have historically been made using the semi-classical approach~\cite{basov.averitt.ea_2017,oka.kitamura_2019,rudner.lindner_2020,disa.nova.ea_2021,delatorre.kennes.ea_2021,bao.tang.ea_2022,bloch.cavalleri.ea_2022}, where matter is treated quantum mechanically, but \ac{EM} fields are described by Maxwell's equations~\cite{fox_2001,grosso.parravicini_2013}. High-intensity lasers, for instance, provide a powerful tool for engineering material properties through mechanisms like strong-field physics, ultrafast dynamics, and Floquet engineering~\cite{yang.li.ea_2023,bao.tang.ea_2022,delatorre.kennes.ea_2021,disa.nova.ea_2021,rudner.lindner_2020,oka.kitamura_2019}. When a material is excited into a non-equilibrium state by an intense laser pulse, it can exhibit properties vastly different from those in its ground state. For example, excitation of the \ac{IR} active phonon mode that is strongly coupled with lattice strain can induce a ferroelectric transition by directly coupling light with ionic motion~\cite{nova.disa.ea_2019}. Floquet engineering, another laser-based technique, enables the modification of electronic structures, such as inducing the anomalous Hall effect in graphene under visible light irradiation~\cite{mciver.schulte.ea_2020}, an effect absent in its equilibrium state~\cite{sato.giovannini.ea_2020,aeschlimann.sato.ea_2021,castro.degiovannini.ea_2022}. This approach modifies electronic band structures and Berry curvatures through interactions with the oscillating \ac{EM} field~\cite{rudner.lindner_2020}. Additionally, intense laser excitation can cause structural distortions, transforming the material from a strong to a weak topological insulator~\cite{vaswani.wang.ea_2020} or in affecting the shear mode in layered materials leading to a topological phase transition~\cite{sie.nyby.ea_2019}. These examples illustrate how strong laser fields can manipulate material properties in non-equilibrium states.

While laser-based techniques hold great potential for modifying material properties, they also present several significant challenges. One of the primary challenges in Floquet engineering is the short lifetime of the induced non-equilibrium states, which often decay rapidly due to energy dissipation and interactions with the environment~\cite{aeschlimann.sato.ea_2021}. Precise control over the parameters of the periodic driving field is crucial, yet achieving the desired material properties frequently involves complex optimization processes\cite{castro.degiovannini.ea_2022,castro.giovannini.ea_2023} that may not always be feasible. Additionally, laser-induced modifications in solid-state materials can lead to significant thermal effects, causing unwanted phase transformations and degradation in material properties\cite{sato.giovannini.ea_2020,mori_2023}. These challenges underscore the difficulty of reliably using lasers to engineer and maintain desired matter states or phases. 

\subsection{Cavity quantum electrodynamics (cQED) to enhance light-matter interaction}
To address the limitations of laser-based techniques while maintaining strong light-matter interactions or couplings, concepts and techniques derived from \ac{cQED} offer powerful alternatives by confining \ac{EM} fields within a resonant cavity where the material is strongly coupled to these fields~\cite{kimble_1998,raimond.brune.ea_2001,mabuchi.doherty_2002,cirac.zoller_2004,fleischhauer.imamoglu.ea_2005,walther.varcoe.ea_2006,chang.douglas.ea_2018,mivehvar.piazza.ea_2021}.
\ac{cQED} has its roots in the study of atomic and molecular systems, where early experiments demonstrated the ability to control and manipulate individual atoms and photons within optical cavities~\cite{haroche.raimond_2006}. A key aspect of \ac{cQED} is the transition from weak to strong coupling regimes~\cite{forn-diaz.lamata.ea_2019,friskkockum.miranowicz.ea_2019}, where in the weak coupling regime, the interaction primarily results in phenomena like the Purcell effect~\cite{purcell_1995}, which enhances or inhibits spontaneous emission rates~\cite{goy.raimond.ea_1983,yablonovitch_1987}. As the coupling strength increases, the system enters the strong coupling regime, leading to the formation of hybrid light-matter states (or polaritons) and phenomena such as vacuum Rabi splitting~\cite{brune.schmidt-kaler.ea_1996}. These foundational studies in atomic \ac{cQED} significantly advanced quantum optics, enabling the observation of entangled photon states and laying the groundwork for quantum information science~\cite{haroche.raimond_2006}.

Extending these principles to solid-state platforms has opened up new avenues of research, particularly in the exploration of novel quantum phases. For instance, in semiconductor microcavities, the strong coupling between excitons and photons can give rise to exciton-polaritons~\cite{weisbuch.nishioka.ea_1992}, which under certain conditions can form Bose-Einstein condensates~\cite{kasprzak.richard.ea_2006}, a coherent quantum phase where many polaritons occupy the same quantum state. Similarly, in systems with multiple quantum emitters, such as quantum dots or color centers, strong coupling is thought to lead to superradiant phases~\cite{wang.hioe_1973}, where the collective emission of light is significantly enhanced compared to the sum of individual emissions. This ongoing research primarily focuses on the hybridization of light and material \textit{excitations}~\cite{basov.asenjo-garcia.ea_2021}, i.e. polariton physics, and its potential applications~\cite{rahimi-iman_2020}. However, polariton physics~\cite{deng.haug.ea_2010,byrnes.kim.ea_2014,gao.li.ea_2018,li.bamba.ea_2018a,basov.asenjo-garcia.ea_2021,rahimi-iman_2020,gu.walther.ea_2021,zarerameshti.violakusminskiy.ea_2022,datta.khatoniar.ea_2022,dirnberger.bushati.ea_2022,deshmukh.zhao.ea_2023,kritzell.baydin.ea_2024,galiffi.carini.ea_2024,kritzell.doumani.ea_2024,kim.hou.ea_2024} and chemistry~\cite{ebbesen.rubio.ea_2023} are not within the scope of the current review.

\subsection{Cavity materials engineering without external driving: a new frontier}
 
Cavity engineering uses optical cavities to confine \ac{EM} fields, enabling precise control of material properties by amplifying specific field modes. When a solid-state material is placed inside such a cavity, the interactions between the confined \ac{EM} fields from the cavity and the charged particles inside the material are significantly enhanced, enabling modifications that extend beyond the previously mentioned traditional material engineering techniques like external fields, chemical treatments, or mechanical processing~\cite{ruggenthaler.flick.ea_2014,ebbesen_2016,ruggenthaler.tancogne-dejean.ea_2018}. 
A unique feature of cavity engineering is the ability to control photon vacuum field strengths through optical cavity design.
These field fluctuations interact with the material's electronic or vibrational states, leading to alterations in its \textit{ground state} properties inside~\cite{peraca.baydin.ea_2020,hubener.degiovannini.ea_2021,garcia-vidal.ciuti.ea_2021,schlawin.kennes.ea_2022}, which is the main topic of this review.

A key distinction in the effects of strong light-matter coupling between cavity-engineered and laser-engineered materials lies in the ability to create light-matter hybrid states or phases at equilibrium, where both quantum and thermal fluctuations play a significant role~\cite{ruggenthaler.sidler.ea_2023}. In cavity-engineered systems, this capability allows the formation of new types of quasi-particles and phases that are not achievable with traditional laser techniques. In such systems, the conventional understanding of quasi-particles must be expanded to incorporate photons in the quasi-particle dressing processes~\cite{rivera.kaminer_2020}. 
While photons are constituent particles in these systems, they can be precisely engineered through the design of the cavity environment. 
These fluctuating photonic fields can be tailored to interact with electronic and ionic degrees of freedom, manipulating fundamental symmetries like time-reversal and crystalline symmetries~\cite{hubener.degiovannini.ea_2021}.
This is the underlying principle behind the idea of cavity-engineering solid-state materials.

Cavity engineering of solid-state materials~\cite{hubener.degiovannini.ea_2021,hubener.bostrom.ea_2024}, or cavity materials engineering, is an emerging field that enables the precise manipulation of material properties by controlling the fundamental interactions between electrons, nuclei, and photons. At the atomic level, these three components -- electrons, nuclei/ions, and \textit{photons} -- are the building blocks of matter, with their interactions determining the material properties. Traditionally, most effects in materials engineering (without any external driving force) are assumed to be dominated by the Coulomb forces between charged particles, which are only the longitudinal part of the quantized \ac{EM} field. In cavity materials engineering, however, we also control the transverse electron-photon interactions, introducing a new dimension to influence the entire spectrum of many-body physics within the material. This approach also takes into account that altering one type of interaction, such as the coupling between electrons and photons, inevitably impacts other dynamics, such as electron-electron or nucleus-photon interactions. Consequently, the theory of cavity engineering aims to predict and realize new material phases by controlling these many-body interactions in ways previously unattainable. It is important to note that the focus here is on the internal dynamics of the matter subsystem of the light-coupled solid-state system, rather than on the photon properties typically emphasized in quantum optics~\cite{scully.zubairy_1997}. The correlated light-matter hybrids can not only modify the material properties but also imprint correlations in the photon field, but this is not the scope of the present review.

Cavity materials engineering can be better understood by comparing it to related fields like polaritonic chemistry and quantum optics. Polaritonic chemistry modifies chemical reactions and molecular properties by coupling molecules to confined light fields, forming hybrid light-matter states, i.e., polaritons, to alter reaction pathways. Quantum optics explores light-matter interactions at a fundamental level, including phenomena like spontaneous emission and quantum entanglement. While inspired by these fields, cavity materials engineering uniquely focuses on modifying intrinsic material properties at equilibrium through quantum vacuum fluctuations of the \ac{EM} field within a cavity. These fluctuations, even without external light, can influence electron and ion behavior (interaction and manipulation of symmetries), enabling the engineering of novel matter phases.

The rapid advancement of cavity materials engineering is driven by the combination of theoretical frameworks, experimental achievements, and advanced cavity platforms. Theoretical tools like the \ac{PF} Hamiltonian and \ac{QEDFT} enable accurate first-principles descriptions of systems where electrons and photons are coupled~\cite{ruggenthaler.sidler.ea_2023}. These provide a solid foundation for predicting how material properties change when placed inside a cavity. In addition, \ac{cQED} models tailored for specific scenarios contribute to the understanding of light-matter interactions for solid-state materials. \ac{MQED} plays a crucial role by offering a comprehensive method for quantizing \ac{EM} fields within complex environments~\cite{scheel.buhmann_2009}, successfully connecting theoretical models with experimental setups. Advances in cavity platforms are equally vital. Plasmonic nanocavities and photonic crystals, for instance, have opened doors to explore strong and ultra-strong light-matter coupling regimes\cite{forn-diaz.lamata.ea_2019,friskkockum.miranowicz.ea_2019}. Furthermore, the field is witnessing both theoretical and experimental breakthroughs in manipulating phenomena within cavities. Examples include modifying the quantum Hall effect~\cite{appugliese.enkner.ea_2022,enkner.graziotto.ea_2024}, tuning transitions between metallic and insulating states~\cite{jarc.mathengattil.ea_2023}, and enhancing ferromagnetism~\cite{thomas.devaux.ea_2021}. These discoveries highlight the potential of cavity engineering for controlling material properties and uncovering new quantum phases. The collective effort from different disciplines, including quantum optics, condensed matter physics, and materials science, has been key to broadening the scope of cavity materials engineering.

\begin{figure}[!t]
    \centering
    \includegraphics[width=1.0\textwidth]{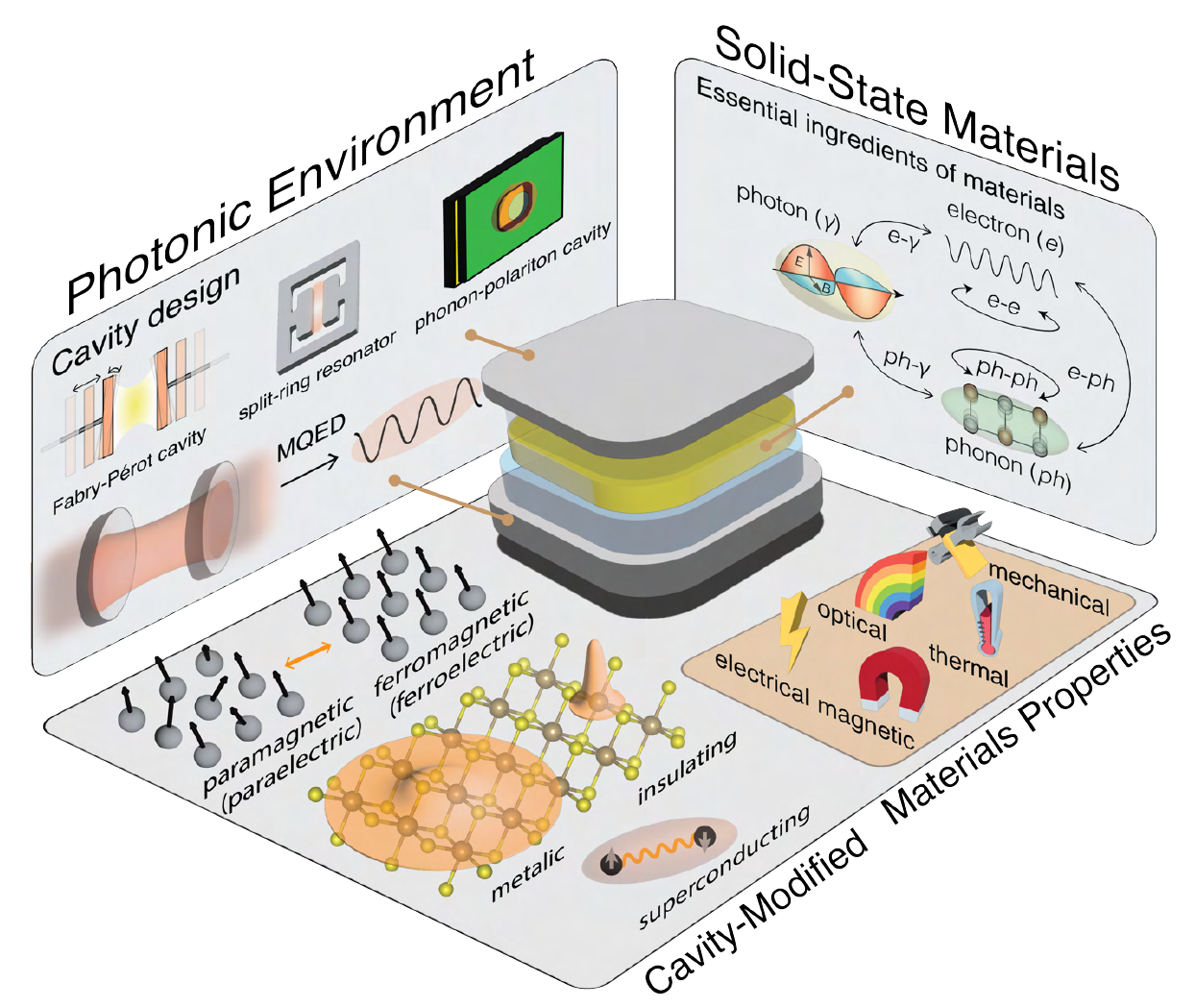}
    \caption{
    \textbf{Cavity materials engineering.} The research field explores the use of quantum fluctuations of photons inside cavities to control the phases and properties of embedded solid-state materials, particularly within dark cavities (i.e., without external driving).
    At the center is a Fabry-Pérot cavity, representing a broad class of cavities with embedded solid-state materials.
    The left panel illustrates how cavity design -- encompassing geometry, configuration, and materials -- can shape the photonic environment, which can described within the framework of macroscopic quantum electrodynamics (MQED).
    The right panel highlights the interplay of key components in solid-state systems, including electrons, nuclei/ions (phonons), and photons, with photons considered as an integral material component under cavity conditions.
    When one component is influenced, for example, controlling photons via the cavity, the others (electrons and phonons) are inevitably impacted, highlighting their interconnected nature. 
    The shaded region at the upper right corner of the bottom panel schematically represents five of six key material property categories -- mechanical, electrical, magnetic, optical, thermal (and chemical) -- that can be modified through cavity engineering with a focus on solid-state physics.
    Specific examples of cavity-modified material properties are also shown at the bottom panel, such as enhanced ferromagnetism, metal-to-insulator phase transitions, and superconductivity.
    }\label{fig:summary-figure}
\end{figure}

\subsection{Research questions and scope of the review}

%
In this review, we focus on the cavity-modified ground state of solid-state materials induced by field fluctuations in a dark cavity -- specifically, without the influence of any external coherent \ac{EM} field pumping; field fluctuations refer to either virtual or thermal fluctuations, or combinations of both. This research field draws upon concepts from condensed matter physics and quantum optics, prompting interest from theorists in both areas. While condensed matter physicists often employ accurate descriptions of electronic structure, their treatment of the \ac{EM} environment may be oversimplified. Conversely, quantum optics researchers often use sophisticated \ac{EM} descriptions but may employ overly simplistic approximations for the electronic structure. These oversimplified descriptions can lead to unphysical predictions, such as misrepresented light-matter interaction strengths and an inability to describe light-induced structural changes in the electronic system. Unlike traditional theoretical models in \ac{cQED}, such as the Jaynes-Cummings model and others primarily developed for quantum optics~\cite{friskkockum.miranowicz.ea_2019}, our focus is on the complex interactions within a general electron-nucleus-photon coupled system. 
To navigate this intricate landscape, several critical research questions must be addresed: Which Hamiltonian should serve as the starting point for such a system?  It needs to be general enough to be able to predict not only the complex nuclear and electronic structure but at the same time also the details of the cavity field. Taking simple \ac{cQED} model Hamiltonians will not suffice since they need input from other theories. How do we rigorously account for the full continuum of photon modes in the Hamiltonian, while simultaneously managing the vast number of degrees of freedom from electrons, nuclei/ions, and photons? Already for the simpler matter-only problem the exponential-wall of the quantum many-body problem prohibits direct solutions in terms of many-body wave functions. Furthermore, how can we design effective model Hamiltonians from the general Hamiltonian for specific scenarios and materials? Most of the readily established models from either solid-state physics or \ac{cQED} will not suffice. For instance, the usual Jaynes-Cummings and Dicke Hamiltonians of \ac{cQED} would predict no changes in ground-state properties. Another key consideration is the description of the photonic environment created by a cavity structure, especially when composed of particular materials. Once we establish the general or effective Hamiltonians, the next challenge lies in exploring how these Hamiltonians govern the cavity-modified ground states. Addressing these questions is essential for advancing our understanding of cavity materials engineering.

A few review papers have explored the use of optical cavities to modify material phases and properties~\cite{peraca.baydin.ea_2020,garcia-vidal.ciuti.ea_2021,bloch.cavalleri.ea_2022,schlawin.kennes.ea_2022,ebbesen.rubio.ea_2023}. For example, Ref.~\cite{peraca.baydin.ea_2020} discusses phenomena and novel ground states in engineered semiconductor systems within the ultrastrong coupling regime. Ref.~\cite{garcia-vidal.ciuti.ea_2021} examines the manipulation of matter, including molecules and solid-state materials, through field fluctuations of photons. Ref.~\cite{bloch.cavalleri.ea_2022} reviews how condensed matter systems can be controlled using classical and quantized \ac{EM} fields. Additionally, Ref.~\cite{schlawin.kennes.ea_2022} focuses on theoretical proposals and early experimental demonstrations of cavity control over collective phenomena in quantum materials. Here we also highlight a non-review paper Ref.~\cite{flick.ruggenthaler.ea_2017}, where some of us introduce the field of cavity modification of the ground and excited states of atoms and molecules including the cavity-controlled potential energy surfaces, bond lengths, spectroscopic quantities, conical intersections, and optical control. In contrast to the above review papers, here we focus on how optical cavities modify the ground state of solid-state materials in a dark cavity, emphasizing \textit{ab initio} methods to study these systems. This review explores both foundational advances and emerging opportunities in cavity materials engineering (Fig.~\ref{fig:summary-figure}), assessing their feasibility and potential impact. By providing a structured framework, we aim to guide readers through the key developments, from theoretical models to experimental realizations, and to highlight the most promising directions for future research.

This review paper is organized as follows: Section~\ref{sec:methodoloy} introduces the \textit{ab initio} theoretical tools for studying many-body electron-nucleus-photon coupled systems and the photonic environment within cavities. Section~\ref{sec:experiments} begins with a discussion on polaritonic chemistry, the field that underpins cavity engineering in molecular systems at room temperature, and reviews recent experimental breakthroughs where field fluctuations have been used to alter solid-state material properties and phases, including the \ac{QHE}, metal-to-insulator transitions, and ferromagnetism. Section~\ref{sec:theory-prediction} explores theoretical predictions for using field fluctuations to modify the ground state of materials, covering phenomena such as para-to-ferroelectric phase transitions, magnetism, superconducting phases, strongly correlated systems, and electronic topological properties. Finally, section~\ref{sec:outlook} offers a conclusion, an outlook on the future directions of the field, and open questions and challenges.

\section{Methodology for solid-state materials inside a photonic environment}\label{sec:methodoloy}

This section presents an overview of the theoretical methodologies used to describe and analyze light-matter interactions in solid-state systems embedded within photonic environments. Beginning with foundational principles of \ac{QED}, we explore the quantization of \ac{EM} fields and the construction of the \ac{PF} Hamiltonian to capture coupled light-matter systems. \ac{MQED} is then introduced to account for realistic cavities and resolve some computational challenges. Additionally, \textit{ab initio} approaches, including \ac{QEDFT}, are discussed to integrate photon-mediated effects into electronic structure calculations. Together, these tools provide the necessary foundation for understanding cavity-modified material properties and advancing research in this rapidly evolving field.

\subsection{Coupling charged particles with photons}\label{sec:coupling-charged-particles-with-photons}
%
\ac{QED} is the fundamental theory describing the interaction of light and matter. It successfully accounts for a wide range of phenomena, including spontaneous emission, light scattering, and the Lamb shift~\cite{cohen-tannoudji.dupont-roc.ea_1989,cohen-tannoudji.dupont-roc.ea_1998,grynberg.aspect.ea_2010,haroche.raimond_2006}. \ac{QED} is derived by combining principles of quantum mechanics and special relativity, employing a quantized description of the \ac{EM} and the Dirac field~\cite{srednicki_2007,greiner.reinhardt_2012,mandl.shaw_2013,peskin_2018}. This quantization introduces creation and annihilation field operators, enabling the representation of photons and their interactions with charged particles. While \ac{QED} provides a comprehensive and extremely accurate framework to treat light-matter interactions, it is plagued by various mathematical inconsistencies~\cite{folland_2021,baez.segal.ea_2014}. These problems are overcome in practice by employing approximations, such as restricting to only perturbation theory for high-energy processes~\cite{srednicki_2007,greiner.reinhardt_2012,mandl.shaw_2013,peskin_2018}. In the low-energy regime, where particle velocities are much smaller than the speed of light, a scenario particularly relevant to the behavior of electrons in solid-sate materials, the non-relativistic \ac{PF} Hamiltonian (see Sec.~\ref{subsubsec:PF-Hamiltonian}) emerges as a suitable approximation that allows for a mathematically well-defined formulation~\cite{spohn_2004}. This Hamiltonian, conserving particle number while allowing for photon creation and annihilation, serves as a starting point for studying many-body systems in the presence of quantized \ac{EM} fields.

In the following, we begin by briefly exploring the quantization of \ac{EM} fields in free space, establishing the basis for describing light-matter coupling (Sec.~\ref{subsubsec:EM-quantization}). Building on this by coupling the quantized \ac{EM} fields with charged particles, the \ac{PF} Hamiltonian provides a low-energy approximation that captures the essential physics of these interactions (Sec.~\ref{subsubsec:PF-Hamiltonian}). However, due to the computational difficulty of solving the full Hamiltonian, effective strategies for approximation are critical to making these systems tractable (Sec.~\ref{subsubsec:Approximation_strategies}). Together, these subsections lay the groundwork for studying and engineering cavity-modified materials.

\subsubsection{Quantization of electromagnetic fields in free space}\label{subsubsec:EM-quantization}

%
The classical theory of electromagnetism is governed by Maxwell's equations with the time-dependent electric field $\bfE(\bfr,t)$ and magnetic field $\bfB(\bfr,t)$, both of which are coupled to the time-dependent charge density $\rho(\bfr,t)$ and the current density $\bfJtot(\bfr,t)$ of a matter system, in the SI unit,
\begin{align}
& \label{eq:MWE-BEJ} \curl\bfB(\bfr,t)-\frac{1}{c^{2}}\pdt\bfE(\bfr,t) = \vacmu\bfJtot(\bfr,t), \\
& \label{eq:MWE-BE} \pdt\bfB(\bfr,t) + \curl\bfE(\bfr,t) = 0,  \\
& \label{eq:MWE-Erho} \divg\bfE(\bfr,t) = \frac{\rho(\bfr,t)}{\vacep},  \\
& \label{eq:MWE-B} \divg\bfB(\bfr,t) = 0,
\end{align}
where $c$ is the speed of light, $\vacep$ is the vacuum permittivity, and $\vacmu$ is the vacuum permeability.
The electric and magnetic fields can be expressed via the scalar and vector potentials, $\phi(\bfr,t)$ and $\bfA(\bfr,t)$, respectively, as follows,
\begin{align}
& \label{eq:E-phi-A} \bfE(\bfr,t) = -\grad\phi(\bfr,t)-\pdt\bfA(\bfr,t),  \\
& \label{eq:B-A} \bfB(\bfr,t) = \curl\bfA(\bfr, t). 
\end{align}
Note that for consistency with relativistic notation, one might also set the vector potential to units of Volts, similar to the scalar potential~\cite{ruggenthaler.sidler.ea_2023}. In this work we use the standard non-relativistic convention of V$\cdot$s/m.
In an arbitrary gauge, the scalar and vector potentials thus obey the following equations, which are obtained using Eqs.~\eqref{eq:E-phi-A} and~\eqref{eq:B-A} substituted into Eqs.~\eqref{eq:MWE-BEJ} and~\eqref{eq:MWE-Erho}, 
\begin{align}
& \label{eq:MWE-Aphi-J} \left(\frac{1}{c^{2}}\pdt^{2}-\nabla^{2}\right)\bfA(\bfr,t) + \grad\left(\divg\bfA(\bfr,t) + \frac{1}{c^{2}}\pdt\phi(\bfr,t)\right) = \vacmu\bfJtot(\bfr,t), \\
& \label{eq:MWE-Aphi-rho} -\nabla^{2} \phi(\bfr,t)-\pdt\left(\divg\bfA(\bfr,t)\right) = \frac{\rho(\bfr,t)}{\vacep}.
\end{align}
To eliminate the intrinsic gauge freedom, we adopt the Coulomb gauge, defined by $\divg\bfA(\bfr,t) = 0$. Under this gauge choice, the vector potential $\bfA(\bfr,t)$ thus becomes a transverse field $\bfAperp(\bfr,t)$, and Maxwell's equations [Eqs.~\eqref{eq:MWE-Aphi-J} and~\eqref{eq:MWE-Aphi-rho}] are decoupled into two equations, 
\begin{align}
& \label{eq:Coulomb-gauge-A} \left(\frac{1}{c^{2}}\pdt^{2}-\nabla^{2}\right)\bfAperp(\bfr,t) = \vacmu\bfJperp(\bfr t), \\
& \label{eq:Coulomb-gauge-phi} -\nabla^{2}\phi(\bfr,t) = \frac{\rho(\bfr,t)}{\vacep},
\end{align}
where we have used the continuity equation,
\begin{equation}
    \pdt\rho(\bfr,t) + \divg\bfJtot(\bfr,t) = 0,
\end{equation}
to remove the longitudinal component of the current density $\bfJpara(\bfr,t)$ from Eq.~\eqref{eq:Coulomb-gauge-A}. Thus only the transverse component of the current density $\bfJperp(\bfr,t)$ couples to the transverse vector potential. The resulting equation for the transverse vector potential [Eq.~\eqref{eq:Coulomb-gauge-A}] governs the dynamics of transverse \ac{EM} fields, driven by the transverse component of the current density. The scalar potential, determined solely by the charge distribution through Poisson's equation [Eq.~\eqref{eq:Coulomb-gauge-phi}], leads to the Coulomb interaction between the charged particles~\cite{greiner.reinhardt_2012}. The Coulomb gauge thus makes the longitudinal Coulomb interaction explicit, which is the common standard in condensed matter physics and quantum chemistry. 

%
To quantize the \ac{EM} field, we begin with the source-free Maxwell's equations, i.e., Eqs.~\eqref{eq:MWE-BEJ}$-$\eqref{eq:MWE-B} with no charge density $\rho(\bfr,t) = 0$ and current density $\bfJtot(\bfr,t)=0$. What caveats apply for cavity situations and how to approximately quantize in that case will be discussed below.
Under the Coulomb gauge, Eqs.~\eqref{eq:Coulomb-gauge-A} and~\eqref{eq:Coulomb-gauge-phi} become
\begin{align}
& \label{eq:Coulomb-gauge-A-nos} \left(\frac{1}{c^{2}}\pdt^{2}-\nabla^{2}\right)\bfAperp(\bfr,t) = 0, \\
& \label{eq:Coulomb-gauge-phi-nos} \nabla^{2}\phi(\bfr,t) = 0.
\end{align}
The vector potential satisfies a wave equation, while the equation for the scalar potential allows only the trivial solution, $\phi(\bfr, t) =0$. The uniqueness of the solution for the transverse vector potential needs appropriate boundary conditions. To ensure consistency between the \ac{EM} field and matter systems, given the shared Laplacian operator $\Delta = \nabla^{2}$ in their respective equations (e.g., the wave equation for the vector potential and Schrödinger equation for the matter wave function), \textit{compatible boundary conditions must be imposed on both}. Since both sectors of the theory, i.e., light and matter, are determined by the same generators of the underlying Poincar\'e or Galilean groups~\cite{srednicki_2007, ruggenthaler.sidler.ea_2023}, the standard choice is to have both sectors with the same boundary conditions. To represent these basic symmetries by self-adjoint generators not all choices are possible, and the standard choice is a quantization box with a volume of $V=L^3$ (the side length $L$) and periodic boundary conditions. In solid-state physics those boundary conditions are called \ac{BvK} boundary conditions. This periodicity allows a plane-wave eigenexpansion, i.e., $e^{i\bfk_{n}\cdot\bfr}$ with momenta $\bfk_{n} = \frac{2\pi}{L}(n_{x}\hat{\mathbf{x}} + n_{y}\hat{\mathbf{y}} + n_{z}\hat{\mathbf{z}}) $, where $n_{x}$, $n_{y}$, and $n_{z}$ are integers, of both \ac{EM} fields and electronic wave functions. When $L\rightarrow \infty$, the discrete wave vectors $\bfk_{n}$ become continuous $\bfk$. With the \ac{BvK} boundary condition, the transverse vector potential $\bfAperp(\bfr,t)$ in Eq.~\eqref{eq:Coulomb-gauge-A-nos}, i.e., freely propagating \ac{EM} fields, can be expressed in terms of the plane-wave eigenbasis $\left\{e^{i\bfk_{n}\cdot\bfr}\right\}$ with the associated frequency $\omega_{\bfk_{n}} = c|\bfk_{n}|$. 
Due to the Coulomb gauge $\divg\bfAperp(\bfr,t)$, for each plane-wave (distributional for the case of $V=\mathbb{R}^3$) eigenfunction with the momentum $\bfk$, there are two transverse polarization vectors $\wavepol(\bfk,\lambda)$ where $\lambda = 1, 2$, satisfying 
\begin{equation}
\bfk\cdot\wavepol(\bfk,\lambda)=\wavepol(\bfk,1)\cdot\wavepol(\bfk,2)=0.
\end{equation}
%

The quantization of \ac{EM} fields in free space is a well-established procedure. 
Under the Coulomb gauge and for the quantization volume being $V=\mathbb{R}^3$, the vector potential is promoted to a quantum operator $\hbfAperp(\bfr)$ defined by 
\begin{equation}
\hbfAperp(\bfr) = \sqrt{\frac{\hbar}{\vacep(2\pi)^{3}}}\sum_{\lambda=1}^{2}\int\frac{1}{\sqrt{2\omega_{\bfk}}}
\left(\hat{a}(\bfk,\lambda)e^{i\bfk\cdot\bfr}\wavepol(\bfk,\lambda)+\hat{a}^{\dagger}(\bfk,\lambda)e^{-i\bfk\cdot\bfr}\wavepol^{*}(\bfk,\lambda)\right) d\bfk,
\end{equation}
where $\hat{a}^{\dagger}(\bfk,\lambda)$ and $\hat{a}(\bfk,\lambda)$ are the creation and annihilation field operators for photons in the plane-wave (distributional) eigenmode with a momentum $\bfk$ and a polarization index of $\lambda$, respectively. 
We note that for $V=\mathbb{R}^3$ the \ac{BvK} boundary conditions are replaced by normalization conditions as is the case in quantum mechanics~\cite{thirring_2013}.
The creation and annihilation field operator follows the commutation relations,
\begin{equation}
    [\hat{a}(\bfk',\lambda'),\hat{a}^{\dagger}(\bfk,\lambda)] = \delta_{\lambda,\lambda'}\delta^{3}(\bfk-\bfk').
\end{equation}
Therefore, the electric field operator $\hbfEperp(\bfr)$ and the magnetic field operator $\hbfB(\bfr) = \frac{1}{c}\curl\hbfAperp(\bfr)$ are given by, respectively,
\begin{align}
\hbfEperp(\bfr) & = \sqrt{\frac{\hbar}{\vacep(2\pi)^{3}}}\sum_{\lambda=1}^{2}\int\frac{i\omega_{\bfk}}{\sqrt{2\omega_{\bfk}}}\left(\hat{a}(\bfk,\lambda)e^{i\bfk\cdot\bfr}\wavepol(\bfk,\lambda)-\hat{a}^{\dagger}(\bfk,\lambda)e^{-i\bfk\cdot\bfr}\wavepol^{*}(\bfk,\lambda)\right) d\bfk, \\
\hbfB(\bfr) & = \sqrt{\frac{\hbar}{\vacep(2\pi)^{3}}}\sum_{\lambda=1}^{2}\int\frac{i}{\sqrt{2\omega_{\bfk}}}\left(\hat{a}(\bfk,\lambda)e^{i\bfk\cdot\bfr}\bfk\times\wavepol(\bfk,\lambda)-\hat{a}^{\dagger}(\bfk,\lambda)e^{-i\bfk\cdot\bfr}\bfk\times\wavepol^{*}(\bfk,\lambda)\right) d\bfk.
\end{align}
%

%
The quantization of \ac{EM} fields in free space, as outlined above, establishes the foundational framework for describing light-matter interactions. This formulation, centered on the transverse vector potential $\hbfAperp(\bfr)$, is essential for understanding how quantized photon modes couple to charged particles. Importantly, this quantization procedure underpins the subsequent derivation of Hamiltonians that govern these coupled systems, such as the \ac{PF} Hamiltonian discussed in the next section. Before introducing these frameworks, it is crucial to recognize how the boundary conditions and gauge choices, such as the Coulomb gauge employed here, simplify the resulting equations while preserving physical accuracy.

Upon coupling to matter, the dynamics of the expectation value of the photon field, $\bfAperp(\bfr,t)$, are governed by Maxwell's equation in the Coulomb gauge [Eq.~\eqref{eq:Coulomb-gauge-A}], with the time-dependent matter's current density $\bfJtot(\bfr,t)$. The time-dependent scalar potential $\phi(\bfr,t)$, determined by the time-dependent charge density $\rho(\bfr,t)$ through Poisson's equation [Eq.~\eqref{eq:Coulomb-gauge-phi}], can be expressed in terms of Green's function $G(\bfr,\bfr')$.
\begin{equation}
\phi(\bfr,t) = \int G(\bfr,\bfr')\rho(\bfr',t) d\bfr'.
\end{equation}
The specific form of Green's function depends on the boundary conditions imposed on both the photon field and the matter system. For $V=\mathbb{R}^3$ the Green function of Poisson's equation [Eq.~\eqref{eq:Coulomb-gauge-phi}] is the Coulomb kernel, 
\begin{equation}
G(\bfr,\bfr') = \frac{1}{4\pi\vacep|\bfr-\bfr'|},
\end{equation}
which is the standard choice in solid-state physics to describe the longitudinal interaction among the charged particles. Thus, for a given charge density $\rho(\bfr,t)$, the scalar potential $\phi(\bfr,t)$ becomes the Hartree potential,
\begin{equation}
\phi(\bfr,t) = \int G(\bfr,\bfr')\rho(\bfr') d\bfr' \to \int \frac{\rho(\bfr',t)}{4\pi\vacep|\bfr-\bfr'|} d\bfr'.
\end{equation}
We note that the kernel of Poisson's equation changes once we have a different quantization volume and/or different boundary conditions. Therefore, the choice of boundary conditions for the photon field directly impacts the form of the scalar potential and, consequently, the interaction between light and matter. Consistency in the boundary conditions applied to both the photon field and the matter system is crucial for accurately modeling their coupled dynamics.

\subsubsection{The Pauli-Fierz Hamiltonian}\label{subsubsec:PF-Hamiltonian}

%
The low-energy \ac{QED} Hamiltonian for a light-matter coupled system, which contains $N_{e}$ electrons, $N_{n}$ nuclei/ions, and the full continuum of photon modes, is described by the non-relativistic \ac{PF} Hamiltonian, which is \textit{gauge invariant}, within the Coulomb gauge for $V=\mathbb{R}^3$~\cite{spohn_2004,jestadt.ruggenthaler.ea_2019,ruggenthaler.sidler.ea_2023},  
\begin{equation} \label{eq:HPF-full-continuum}
\begin{aligned}
\hat{H}_{\rm{PF}} & = \sum_{i=1}^{N_{e}}\left[\frac{\left(\hbfp_{i}+|e|\hbfAperp(\bfr_{i})\right)^{2}}{2\mebare} + \frac{|e|\hbar}{2\mebare}\paulimat_{i}\cdot\hbfB(\bfr_{i})\right] + \frac{1}{2}\sum_{i\neq j}^{N_{e}}\frac{e^{2}}{4\pi\vacep|\bfr_{i}-\bfr_{j}|} \\
& + \sum_{I=1}^{N_{n}}\left[\frac{\left(\hbfP_{I}-Z_{I}|e|\hbfAperp(\bfR_{I})\right)^{2}}{2M_{I,\rm{b}}} - \frac{Z_{I}|e|\hbar}{2M_{I,\rm{b}}}\mathbf{S}_{I}\cdot\hbfB(\bfR_{I})\right] + \frac{1}{2}\sum_{I\neq J}^{N_{n}}\frac{Z_{I}Z_{J}e^{2}}{4\pi\vacep|\bfR_{I}-\bfR_{J}|} \\
& -\sum_{i=1}^{N_{e}}\sum_{I=1}^{N_{n}}\frac{Z_{I} e^{2}}{4\pi\vacep|\bfr_{i}-\bfR_{I}|} + \sum_{\lambda=1}^{2}\int \hbar \omega_{\bfk} \hat{a}^{\dagger}(\bfk,\lambda)\hat{a}(\bfk,\lambda)d\bfk.
\end{aligned}
\end{equation}
Here in the first line, $\hbfp_{i}=-i\hbar\nabla_{i}$ represents the momentum operator for the $i$th electron, $|e|$ is the magnitude of electron charge, and $\mebare$ is the electron's \textit{bare} mass, which differs from the electron's \textit{physical} mass $m_{e}$ used in the standard quantum mechanics. An electron at position $\bfr_{i}$ is coupled to the transverse Coulomb-gauged photon field $\hbfAperp(\bfr_{i})$ and its spin $\paulimat_{i}$ (spin-$1/2$ Pauli matrices) is coupled to the magnetic field $\hbfB(\bfr_{i})$. The last term describes the longitudinal Coulomb repulsion among electrons with $\sum_{i \neq j}$ indicating a double sum excluding $i=j$. In the second line, nuclei/ions are modeled with an effective \textit{bare} mass $M_{I,\rm{b}}$, an effective charge $Z_{I}|e|$, and an effective spin $\mathbf{S}_{I}$. $\hbfP_{I}=-i\hbar\nabla_{I}$ is the momentum operator for the $I$th nucleus/ion. A nucleus/ion at position $\bfR_{I}$ is coupled to the photon field $\hbfAperp(\bfR_{I})$ and its spin $\mathbf{S}_{I}$ is also coupled to the magnetic field $\hbfB(\bfR_{I})$. The last term describes the Coulomb repulsion between nuclei/ions. In the last line, the first term describes the Coulomb attraction between electrons and nuclei/ions, while the second term refers to the energy of the free photon field. Due to normal-ordering the inconsequential energy contribution due to $\hbar \omega_{\bfk}/2$ in the free photon field has been discarded.
%

%
In the \ac{PF} Hamiltonian, the electron and nuclear masses are bare masses [Eq.~\eqref{eq:HPF-full-continuum}], distinct from their experimentally observed values. In standard non-relativistic quantum mechanics, the Schrödinger equation without explicitly coupled photons uses the physical electron mass $m_{e}$ instead of $\mebare$. This distinction arises from the explicit inclusion of the transverse photon-particle interactions in the Hamiltonian. 
The physical masses include self-energy contributions from the photon cloud surrounding these particles, encapsulating the effects of transverse field coupling~\cite{mandl.shaw_2013}.
For example, the physical electron mass can be expressed as $m_{e} = \mebare + m_{e,\rm{ph}}$, where $m_{e,\rm{ph}}$ denotes the mass correction due to photon interaction contribution~\cite{hainzl.seiringer_2003,rokaj.ruggenthaler.ea_2022}.
This distinction is essential for accurately modeling low-energy phenomena, as it ensures self-consistent treatment of photon-induced effects without double-counting.

The \ac{PF} Hamiltonian, while essential for describing light-matter interactions, requires regularization to ensure mathematical consistency. Without limiting the coupling between light and matter at arbitrarily small and large wavelengths, the Hamiltonian becomes ill-defined. Regularization typically involves applying hard cutoffs to these wavelengths, and renormalization is the process of systematically removing these cutoffs while maintaining a well-defined theory. Although a full non-perturbative renormalization procedure remains unresolved for the \ac{PF} Hamiltonian~\cite{spohn_2004}, ultraviolet regularization is sufficient for low-energy phenomena, which are the focus here. This approach ensures that the results are qualitatively reliable, as evidenced by studies demonstrating that infra-red regularization can be removed for \ac{PF} ground states without affecting key predictions~\cite{spohn_2004}.

To simplify the \ac{PF} Hamiltonian and make it compatible with standard quantum mechanical frameworks, the Coulomb gauge is employed. This choice separates longitudinal interactions, described by matter-only operators, from transverse photon fields. This separation avoids more complex quantization methods, such as the Gupta-Bleuler approach~\cite{greiner.reinhardt_2012}, while preserving physical photon characteristics. Furthermore, the Coulomb gauge aligns well with practical calculations in quantum optics and polaritonics~\cite{cohen-tannoudji.dupont-roc.ea_1998}. Systems that exhibit equilibrium states in matter-only quantum mechanics can generally admit corresponding equilibrium states in low-energy \ac{QED} within this gauge~\cite{spohn_2004,jestadt.ruggenthaler.ea_2019}. While other gauges might offer alternative perspectives, they often lack rigorous proofs of equivalence or consistent descriptions of Coulomb interactions, which remain open issues~\cite{woolley_2024}. Despite these advantages, solving the full \ac{PF} theory in the Coulomb gauge for realistic systems is computationally demanding, requiring approximations or alternative formulations, such as \ac{QEDFT}~\cite{ruggenthaler.flick.ea_2014}, discussed in Sec.~\ref{subsubsec:QEDFT}.

In practical scenarios, treating the cavity structure fully atomistically in the \ac{PF} framework is often unfeasible due to its macroscopic nature, as seen in a Fabry-P\'erot. Instead, the cavity is typically represented as a modification to the boundary conditions of the photon field. However, altering photon mode structures introduces several challenges.
The simplest way to highlight potential issues is to remember that the Lagrangian (a scalar) of the coupled light-matter system has \textit{one} set of joined (light-matter) boundary conditions~\cite{greiner.reinhardt_2012}. Changing the boundary conditions on a subsystem can break the connection to the corresponding Lagrangian and hence fundamental conservation laws. Moreover, altering the mode structure can impact the relationship between bare and physical masses, potentially affecting renormalization procedures and the accuracy of mass calculations~\cite{ruggenthaler.sidler.ea_2023}. Furthermore, changes to the mode structure can modify the Coulomb interaction (as shown in Sec.~\ref{subsubsec:EM-quantization}), particularly in systems with significant longitudinal photon contributions, such as nanoplasmonic cavities.

To mitigate these issues, the \ac{LWA} is often used, as detailed in Sec.~\ref{subsec:LWA}. The \ac{LWA} ensures that mode consistency between light and matter becomes irrelevant, simplifying theoretical treatments. While implicit treatments of the photonic environment through approximations like the \ac{LWA} are possible, still careful parameter adjustment is required to maintain consistency~\cite{ruggenthaler.sidler.ea_2023} and avoid double-counting free-space photon modes when using physical mass (for details see discussion in Sec.~\ref{subsec:LWA}). In summary, modifying photon-mode structures demands a thorough understanding of the coupled system to ensure accurate and reliable theoretical predictions. This foundational discussion of regularization, gauge choices, and boundary condition considerations sets the stage for the next section, where strategies for approximating the \ac{PF} Hamiltonian are explored.

\subsubsection{Strategies to approximate the Pauli-Fierz Hamiltonian}
\label{subsubsec:Approximation_strategies}

Solving the full \ac{PF} Hamiltonian [Eq.~\eqref{eq:HPF-full-continuum}], which encompasses both matter (that also includes a microscopic description of the cavity itself) and quantized \ac{EM} fields, poses significant challenges due to the increased complexity compared to the already complex electronic structure calculations [see Eq.~\eqref{eq:Matter-H-from-PF}]. To address this, a common strategy is to approximate the \ac{PF} Hamiltonian by treating the cavity in an effective manner and disentangling the subsystems. However, it is not always possible to do so and still capture the relevant physics. For example, in the context of electron-nucleus interactions non-adiabatic quantum effects might be needed to capture the correct behavior of coupled systems. Similarly, processes like radiative dissipation (e.g., spontaneous emission) cannot be captured if only a few effective cavity modes are retained. Yet keeping such restrictions in mind, we will discuss this strategy briefly in this section.

Approximating the \ac{PF} Hamiltonian begins with recognizing that cavities primarily enhance specific free-space photon modes~\cite{svendsen.ruggenthaler.ea_2023}. By implicitly accounting for interactions with the full photon continuum, an effective \ac{PF} Hamiltonian can be constructed. This Hamiltonian incorporates only the cavity-enhanced modes and the material embedded within the cavity~\cite{rokaj.ruggenthaler.ea_2022,svendsen.ruggenthaler.ea_2023}.
The effective \ac{PF} Hamiltonian is given as 
\begin{equation}\label{eq:HPF-eff-modes}
\begin{aligned}
    \hat{H}_{\rm{PF}}^{\rm{eff}} & = \sum_{i=1}^{N_{e}}\left[\frac{\left(\hbfp_{i}+|e|\hbfAperp^{\rm{eff}}(\bfr_{i})\right)^{2}}{2m_{e}} + \frac{|e|\hbar}{2m_{e}}\paulimat_{i}\cdot\hbfB^{\rm{eff}}(\bfr_{i})-|e|\phi^{\rm{eff}}(\bfr_{i})\right] + \frac{1}{2}\sum_{i\neq j}^{N_{e}}\frac{e^{2}}{4\pi\vacep|\bfr_{i}-\bfr_{j}|} \\
    & + \sum_{I=1}^{N_{n}}\left[\frac{\left(\hbfP_{I}-Z_{I}|e|\hbfAperp^{\rm{eff}}(\bfR_{I})\right)^{2}}{2M_{I}} - \frac{Z_{I}|e|\hbar}{2M_{I}}\mathbf{S}_{I}\cdot\hbfB^{\rm{eff}}(\bfR_{I})+Z_{I}|e|\phi^{\rm{eff}}(\bfR_{I})\right] + \frac{1}{2}\sum_{I\neq J}^{N_{n}}\frac{Z_{I}Z_{J}e^{2}}{4\pi\vacep|\bfR_{I}-\bfR_{J}|} \\
    & -\sum_{i=1}^{N_{e}}\sum_{I=1}^{N_{n}}\frac{Z_{I} e^{2}}{4\pi\vacep|\bfr_{i}-\bfR_{I}|} + \sum_{\alpha=1}^{M_{p}} \hbar \omega_{\alpha} \hat{a}_{\alpha}^{\dagger}\hat{a}_{\alpha}.
\end{aligned}
\end{equation}
Here the effective photon field $\hbfAperp^{\rm{eff}}(\bfr)$ composed of $M_{p}$ effective photon modes is 
\begin{equation}\label{eq:photon-field-effective-modes}
\hbfAperp^{\rm{eff}}(\bfr) = \sum_{\alpha=1}^{M_{p}} \frac{\lambda_{\alpha}}{\sqrt{2\omega_{\alpha}}}\left(\hat{a}_{\alpha}\wavepol_{\alpha}e^{i\bfk_{\alpha}\cdot\bfr} + \hat{a}_{\alpha}^{\dagger}\wavepol_{\alpha}^{*}e^{-i\bfk_{\alpha}\cdot\bfr} \right).
\end{equation}
The effective photon modes are indexed by $\alpha$, with corresponding creation and annihilation operators $\hat{a}_{\alpha}^{\dagger}$ and $\hat{a}_{\alpha}$. Each mode is characterized by its frequency $\omega_{\alpha}$, polarization $\wavepol_{\alpha}$, wave vector $\bfk_{\alpha}$, and vacuum field amplitude $\lambda_{\alpha}$. The latter, defined as $\lambda_{\alpha} = \sqrt{\hbar/(\vacep V_{\alpha})}$, incorporates information about the cavity through its effective mode volume $V_{\alpha}$. In cases where the interaction between the embedded material and the cavity material is negligible, such as when the material is located far from the cavity walls, the electric scalar potential from the cavity $\phi^{\rm{eff}}(\bfr)$ can often be neglected. Suppose the \ac{LWA} is assumed (discussed in Sec.~\ref{subsec:LWA}) such that the momentum transfer from the photon field to the matter system is negligible (i.e., the wavelength of the photon field much larger than the extent of the matter system). In that case, we can discard the modes' spatial dependence. This approach enables coupling to any local photonic density of states to emulate a specific cavity structure. This local density of states can be determined, for instance, by using \ac{MQED} (see Sec.~\ref{subsec:describe-photonic-env} for details). 
Notably, without explicit cavity coupling, the effective \ac{PF} Hamiltonian reduces to the standard matter Hamiltonian widely used in solid-state physics,
\begin{equation}\label{eq:Matter-H-from-PF}
\begin{aligned}
    \hat{H}_{\rm{M}} & = \sum_{i=1}^{N_{e}}\frac{\hbfp_{i}^{2}}{2m_{e}} + \frac{1}{2}\sum_{i\neq j}^{N_{e}}\frac{e^{2}}{4\pi\vacep|\bfr_{i}-\bfr_{j}|} -\sum_{i=1}^{N_{e}}\sum_{I=1}^{N_{n}}\frac{Z_{I} e^{2}}{4\pi\vacep|\bfr_{i}-\bfR_{I}|} \\
    & + \sum_{I=1}^{N_{n}}\frac{\hbfP_{I}^{2}}{2M_{I}} + \frac{1}{2}\sum_{I\neq J}^{N_{n}}\frac{Z_{I}Z_{J}e^{2}}{4\pi\vacep|\bfR_{I}-\bfR_{J}|}.
\end{aligned}
\end{equation}
This Hamiltonian includes electron-electron, electron-nucleus, and nucleus-nucleus Coulomb interactions, which arise from the longitudinal components of the \ac{EM} fields. Furthermore, the physical masses of electrons and nuclei/ions incorporate self-energy corrections due to interactions with the free-space quantum vacuum of the photon field.

To simulate complex light-matter systems more efficiently, various approximations techniques are applied. For instance, the generalized Born-Huang expansion expresses the full light-matter wave function as a sum of conditional wave functions~\cite{galego.garcia-vidal.ea_2015,flick.appel.ea_2017,schafer.ruggenthaler.ea_2018,ruggenthaler.sidler.ea_2023}, allowing subsystems -- such as photons, electrons, and nuclei/ions -- to be treated differently based on the effect and system of interest.
Keeping only the lowest-order terms in this expansion results in approximations such as the \ac{PoPES} approximation~\cite{galego.garcia-vidal.ea_2015}, the \ac{cBOA} approximation~\cite{flick.appel.ea_2017}, and the explicit-polariton approximation~\cite{schafer.ruggenthaler.ea_2018}. 
These techniques, detailed in Secs.~\ref{subsec:abinitio-matter} and \ref{subsec:abinitio-lightmatter}, provide a practical means to reduce computational complexity without sacrificing critical dynamics.

Even with approximations, solving the resulting Hamiltonian for the light-matter coupled system remains formidable. Analytical solutions are feasible in rare cases, such as non-interacting free electrons in the \ac{pHEG} model~\cite{rokaj.ruggenthaler.ea_2022,schafer.buchholz.ea_2021,lu.ruggenthaler.ea_2024}. 
However, most scenarios require advanced computational methods, including (see Ref.~\cite{ruggenthaler.sidler.ea_2023} for a more detailed exposition) quantum Monte Carlo~\cite{weber.vinasbostrom.ea_2023}, coupled-cluster techniques~\cite{mordovina.bungey.ea_2020,haugland.ronca.ea_2020,riso.grazioli.ea_2023}, and \ac{QEDFT}~\cite{tokatly_2013,ruggenthaler.flick.ea_2014,pellegrini.flick.ea_2015}. To further reduce complexity, model Hamiltonians are often constructed. These models capture the essential physics needed to explain observed phenomena, but often rely on coefficients derived from \textit{ab initio} calculations or experimental data as input.

The strategies discussed here highlight the importance of balancing computational feasibility with physical accuracy in approximating the \ac{PF} Hamiltonian. The use of effective Hamiltonians, coupled with advanced computational techniques, allows for the practical exploration of cavity-modified systems. The next section describes the \ac{EM} environment inside a cavity, focusing on the role of \ac{MQED} in capturing the complexities of real-world systems.

\subsection{Describing the electromagnetic environment inside a cavity}\label{subsec:describe-photonic-env}
\subsubsection{Possible cavity platforms for cavity materials engineering}\label{subsubsec:cavity-platforms}
The choice of cavity platform is critical in shaping light-matter interactions and the resulting material properties in cavity-modified systems. Each cavity type offers unique features, operational regimes, and energy scales, making them suitable for specific phenomena and advancing the field of cavity materials engineering. Below, we explore several key cavity platforms, highlighting their strengths, limitations, and applications in the context of cavity-induced material modifications. For a more detailed discussion on cavity setups for achieving strong coupling regimes, readers are encouraged to refer to more comprehensive review papers and articles such as Refs.~\cite{forn-diaz.lamata.ea_2019,friskkockum.miranowicz.ea_2019,hubener.degiovannini.ea_2021,zhong.he.ea_2023,hirai.andellhutchison.ea_2024,galiffi.carini.ea_2024}.

Fabry-Pérot cavities are widely used in cavity materials research due to their simple design and broad applicability across a wide frequency range, from infrared to visible light. Consisting of two parallel mirrors that confine \ac{EM} waves via multiple reflections, they support standing wave modes whose resonance frequency and quality factor are tunable through the mirror spacing, reflectivity, and dielectric environment. These cavities are used to studying phenomena such as metal-to-insulator transitions in materials like 1T-TaS$_2$ (see Sec.~\ref{subsec:MIT-TaS2}). Their larger mode volume can limit light-matter coupling strength compared to more confined platforms such as plasmonic cavities, but their simplicity and tunability make them suitable in cavity materials engineering~\cite{jarc.mathengattil.ea_2022,jarc.mathengattil.ea_2023}. Specifically, if we have collective strong coupling situations in mind, where the material itself can substantially enhance the bare-cavity coupling strength and a large effective mode volume is desirable.

Split-ring resonators are designed to manipulate \ac{EM} fields at microwave and terahertz frequencies~\cite{maissen.scalari.ea_2014}. These cavities consist of metallic rings with small gaps that exhibit strong localization of electric and magnetic fields, enabling significant light-matter interaction even in compact geometries. These resonators are particularly effective for coupling to \ac{2D} electron systems like quantum Hall bars, where the enhanced vacuum electric field fluctuations near the split gap can induce observable modifications in electronic properties, such as the breakdown of integer quantum Hall quantization and the stabilization of fractional states~\cite{appugliese.enkner.ea_2022,enkner.graziotto.ea_2024} (see Sec.~\ref{subsec:quantm-Hall-effect}). Split-ring resonators' small footprint and tunable frequency range make them an ideal platform for studying cavity-modified phenomena, particularly in systems requiring localized \ac{EM} fields~\cite{enkner.graziotto.ea_2024}.

Plasmonic cavities exploit the collective oscillations of free electrons, known as surface plasmons, in metallic nanostructures to confine and enhance \ac{EM} fields at subwavelength scales. These cavities offer extremely strong light-matter coupling due to their ability to focus light into very small mode volumes, often just a few nanometers. Plasmonic cavities are well-suited for investigating ultrastrong coupling regimes, where the interaction between light and matter is significantly enhanced. For example, plasmonic cavities have been used to couple surface plasmons to vibrational modes of polymers, tuning ferromagnetism and superconductivity in unconventional superconductors~\cite{thomas.devaux.ea_2021} (see Sec.~\ref{subsec:ferromagnetism}). Operating across a broad range of wavelengths, from visible to near-infrared, plasmonic cavities can be integrated with various solid-state material platforms such as \ac{2D} materials~\cite{kipp.bretscher.ea_2024}. However, a major challenge is the high optical losses due to metal dissipation, which can hinder strong coupling at room temperature. Despite this limitation, plasmonic cavities hold significant promise for exploring cavity quantum materials and quantum technologies.

Phonon-polariton cavities are based on the coupling between optical phonons in polar dielectric materials and confined \ac{EM} fields, particularly in the mid-infrared to terahertz frequency ranges~\cite{galiffi.carini.ea_2024}. These cavities, utilizing materials such as hexagonal boron nitride, provide exceptional field confinement and low optical losses, allowing for strong light-matter coupling~\cite{herzigsheinfux.orsini.ea_2024}. The anisotropic properties of materials like hexagonal boron nitride enable directional control over polariton propagation, making it possible to manipulate material properties spatially within the cavity. Phonon-polariton cavities are potentially useful for modifying ferroelectric behavior (see Sec.~\ref{subsec:theory-prediction-paraferro}) and enhancing superconductivity (see Sec.~\ref{subsec:cavity-modified-superconductivity}). These capabilities make phonon-polariton cavities a powerful tool for exploring the coupling between light and phonons in solid-state systems.

Photonic crystal cavities use periodic dielectric structures to create photonic band gaps, confining light in designated regions. These cavities are highly tunable, provide high-quality factors with minimal losses, and are excellent for studying material responses in the optical and near-infrared regimes. Their ability to localize light in subwavelength regions makes them particularly advantageous for experiments that require long photon lifetimes and precise mode control. Photonic crystal cavities have been applied in studies of light-matter coupled high-mobility \ac{2D} electron gases~\cite{zhang.lou.ea_2016}, where they enable significant modifications to electronic properties through enhanced light-matter interactions.

Each of these cavity platforms offers distinct advantages and challenges, shaped by their geometry, material composition, and operational frequency regimes. Fabry-Pérot cavities, split ring resonators, and plasmonic cavities typically operate in the infrared, terahertz, and visible ranges, making them ideal for studying a broad array of solid-state materials. Microcavities, phonon-polariton cavities, and photonic crystal cavities provide excellent confinement and control at subwavelength scales, suitable for probing low-energy excitations and phonon interactions in solid-state systems. As experimental techniques improve and new cavity designs are developed, these platforms will continue to drive progress in cavity materials engineering, enabling the exploration of new quantum phases and material properties in controlled photonic environments.

\subsubsection{Overview of macroscopic quantum electrodynamics (MQED)}
\label{subsubsec:MQED}
In our earlier discussions (Sec.~\ref{subsubsec:Approximation_strategies}), the \ac{EM} field was simplified as a few discrete cavity modes with freely adjustable coupling strengths [Eq.~\eqref{eq:HPF-eff-modes}]. However, real-world cavities introduce complexities such as continuous mode distribution, $\lambda_\alpha\rightarrow\boldsymbol{\lambda}(\omega)$, and interactions with the polaritonic field of the lossy cavity environment~\cite{huttner.barnett_1992}. Accurately describing these systems requires a comprehensive framework for quantizing the \ac{EM} field in lossy, complex environments. This is where \ac{MQED} becomes indispensable. \ac{MQED} quantized \ac{EM} fields in terms of experimentally relevant cavity configurations and explicitly accounts for material losses. By introducing \ac{DGF}, \ac{MQED} enables a direct connection between the photon fields and experimentally realizable cavity configurations.

In coupled quantum light-matter systems, both the cavity setup and the matter consist of charged particles. Although a full treatment would consider both on the same theoretical footing, practical approaches typically separate the cavity (which shapes the \ac{EM} field) from the material (the primary focus); note that we have mentioned the caveat of such separation in Sec.~\ref{subsubsec:PF-Hamiltonian}. This separation allows the cavity to be treated as an empty structure, whose quantized fields are then coupled to the matter system.

Real cavity setups generally present material losses and quantizing the \ac{EM} field in the presence of such losses is not a trivial generalization of the lossless case. 
Classically, the effect of a material background is captured and incorporated into Maxwell’s equations via material-dependent optical parameters~\cite{saleh.teich_2019}, such as the refractive index $n(\boldsymbol{r},\omega)$.
In the simplest case of a homogeneous medium, the refractive index modifies the Helmholtz equation such that $k=n(\omega)k_0$ where $k_0$ is the free space wave number. As discussed in section \ref{sec:coupling-charged-particles-with-photons}, in the lossless case, the \ac{EM} field is quantized by expanding it in terms of the eigenmodes of the Helmholtz equation with associated creation and annihilation operators. Therefore, one might expect that the eigenmodes of the modified Helmholtz equation can be used as a basis for quantization in lossy media. 
However, in lossy media, the refractive index becomes dispersive, with a non-zero imaginary component, as a result of the causality dictated Kramers-Kronig relations~\cite{saleh.teich_2019}. This results in a non-Hermitian Helmholtz equation, leading to non-orthogonal eigenfunctions. Consequently, traditional lossless quantization schemes fail~\cite{scheel.buhmann_2009}.
The failure of the straightforward quantization approach is due to an inadequate treatment of the underlying matter theory encoded within the frequency-dependent refractive index~\cite{huttner.barnett_1992,huttner.barnett_1992a}. A proper description of quantized fields in lossy media requires explicit consideration of the matter system.

The central idea in \ac{MQED} is to quantize the \ac{EM} field and matter of the \textit{empty cavity} together, then projecting back onto the \ac{EM} degrees of freedom for the coupled system. 
This approach introduces generalized polaritonic modes, $\mathbf{\hat{f}}^\dagger(\bfr, \omega)$, which describe the fundamental excitations of the coupled system~\cite{huttner.barnett_1992,huttner.barnett_1992a}. These modes obey standard bosonic commutation relations, 
\begin{align}\label{eq:theory-chap1:polaritons_commutator}
    \left[\hat{\mathbf{f}}(\bfr, \omega), \hat{\mathbf{f}}^\dagger(\bfr', \omega')\right] = \delta(\omega - \omega')\delta(\bfr - \bfr').
\end{align}
The Hamiltonian for the coupled system can be expressed as~\cite{huttner.barnett_1992,huttner.barnett_1992a}, 
\begin{align}\label{eq:final_MQED_ham}
    \hat{H} = \int_0^\infty d\omega\ \hbar\omega\int d\bfr\ \mathbf{\hat{f}}^\dagger(\bfr, \omega)\cdot\mathbf{\hat{f}}(\bfr, \omega).
\end{align}

In general \ac{EM} environments, the quantized \ac{EM} field operators can be represented in terms of the \ac{DGF}, $\mathbf{G}(\mathbf{r},\mathbf{r'},\omega)$, which relates the electric field at $\bfr$ to a point source at $\bfr'$~\cite{scheel.buhmann_2009}. The \ac{DGF} satisfies a modified Helmoholtz equation~\cite{chew_1999,novotny.hecht_2012},
\begin{align}\label{eq:theory-nano-optics:def_DGF}
    \left[\curl\curl -\frac{\omega^2}{c^2}\right]\mathbf{G}(\mathbf{r},\mathbf{r'}, \omega) -i\mu_0\omega \int d\mathbf{s}\, \mathbf{Q}(\mathbf{r}, \mathbf{s}, \omega)\mathbf{G}(\mathbf{s}, \mathbf{r}, \omega) = \delta(\mathbf{r}-\mathbf{r'}),
\end{align}
where $\mathbf{Q}(\mathbf{r}, \mathbf{r'}, \omega)$ is the complex conductivity tensor of the medium. 
Given the \ac{DGF}, the electric field from an arbitrary source current, $\mathbf{j}_\mathrm{ext}(\mathbf{r'})$, can be computed as
\begin{align}\label{eq:theory-nano-optics:classical_Helmholtz_dyadic_GF_particular}
    \bfE(\bfr) = i\omega\vacmu\int \mathbf{G}(\bfr, \bfr', \omega)\cdot \bfj_{\rm{ext}}(\bfr')d\bfr'.
\end{align}
%
%
%
Physically, the \ac{DGF} encodes information such the cavity’s geometry, \ac{EM} boundary conditions, and material properties, enabling a direct connection between the quantized field operators and the cavity setup. It is a projector from the coupled system onto the \ac{EM} degrees of freedom~\cite{scheel.buhmann_2009}.
The quantized \ac{EM} field operators of an arbitrary cavity setup can be expressed in terms of the \ac{DGF} and the fundamental polaritonic operators in the following way,
\begin{align}\label{eq:theory-chap1:mqed-field-expansion-in-terms-of-f}
    & \hbfE(\bfr, \omega) = i\vacmu\omega\sqrt{\frac{\hbar \omega}{\pi}} \int d\bfr'\,\int d\mathbf{s}\, \mathbf{G}(\bfr, \bfr', \omega)\cdot\mathbf{K}(\bfr', \mathbf{s}, \omega)\cdot  \hat{\mathbf{f}}(\mathbf{s}, \omega), \\
    & \hbfB(\bfr, \omega) = \vacmu\sqrt{\frac{\hbar \omega}{\pi}}\curl \int d \mathbf{r'}\int d\mathbf{s}\, \mathbf{G}(\mathbf{r}, \mathbf{r'}, \omega)\cdot\mathbf{K}(\mathbf{r'}, \mathbf{s}, \omega)\cdot \hat{\mathbf{f}}(\mathbf{s}, \omega). 
\end{align}
Here the tensor $\mathbf{K}(\mathbf{r},\mathbf{r'},\omega)$ is the matrix "square root" of the conductivity tensor $\mathbf{\sigma}(\mathbf{r}, \mathbf{r'}, \omega)$ such that $\boldsymbol{\sigma}(\mathbf{r}, \mathbf{r'}, \omega) =  \int d\mathbf{s} \, \mathbf{K}(\mathbf{r}, \mathbf{s}, \omega) \cdot \mathbf{K}^\dagger(\mathbf{s}, \mathbf{r'}, \omega)$. Note that this root is guaranteed to exist in media without gain due to the positivity of $\boldsymbol{\sigma}(\mathbf{r}, \mathbf{r'}, \omega)$ in absorbing media~\cite{scheel.buhmann_2009}.
The corresponding Schrödinger picture operators can be defined as follows:
\begin{align}
& \hbfE(\bfr) = \int_0^\infty d\omega \, \hbfE(\bfr, \omega) + \rm{h.c.}, \\
& \hbfB(\bfr) = \int_0^\infty d\omega \, \hbfB(\bfr, \omega) + \rm{h.c.}.
\end{align}
Here h.c. stands for the associated hermitian conjugate. 
These operators satisfy a commutator relation,
\begin{align}\label{eq:E_B_commutator_MQED}
    \left[\hbfE(\bfr), \hbfB(\bfr')\right] = -\frac{i\hbar}{\vacep}\curl\delta^{\bot}(\bfr-\bfr').
\end{align}

A remarkable feature of \ac{MQED} is that the expansion coefficients for the \ac{EM} field operators are given by the \ac{DGF}. This purely classical object can be determined analytically for simple geometries or numerically using finite element method simulations for complex setups. This provides a formulation of \ac{QED} for lossy environments and allows for a direct connection to real cavities, paving the way for realistic descriptions of the \ac{EM} environment within \textit{ab initio} methods such as \ac{QEDFT} (see Sec.~\ref{subsubsec:QEDFT}).

%
In practice, the central challenge is calculating the \ac{DGF} for the specific cavity setup of interest. Generally, the goal is to determine suitable basis functions in terms of which to expand the \ac{DGF}. For the simplest case of vacuum or a homogeneous medium, the \ac{DGF} can be expanded using spherical waves centered at the dipole position~\cite{chew_1999,novotny.hecht_2012}. 
Additionally, for certain high-symmetry geometries, such as layered media, concrete expressions for the \ac{DGF} can be derived~\cite{chew_1999}. In these cases, the \ac{DGF} can be expanded in terms of vector wave functions that reflect the symmetry of the problem, such as planar or spherically layered configurations~\cite{svendsen.kurman.ea_2021,svendsen.thygesen.ea_2024}.
Calculating the \ac{DGF} for general structured media is significantly more complex. While closed-form solutions may be unavailable in general, numerical methods can be used. 
In lossy media, the Helmholtz equation becomes non-Hermitian, requiring careful consideration when constructing the mode functions. One approach is the biorthonormal mode construction discussed in Ref.~\cite{chen.nielsen.ea_2010}. If one can restrict the necessary information to the \ac{DGF} value at certain points, for example, when the light-matter coupling is considered within the dipole approximation (see Sec.~\ref{subsec:LWA}), one can use the definition of the \ac{DGF} in Eq.~\eqref{eq:theory-nano-optics:classical_Helmholtz_dyadic_GF_particular} to connect the relevant values to dipole emission fields in the geometry of interest~\cite{bennett.buhmann_2020}.

\subsubsection{The applications of MQED and its combination with (QE)DFT}\label{subsubsec:applications-MQED-QEDFT}
\ac{MQED} has a wide range of applications, including medium-assisted atom-field interactions~\cite{buhmann_2013a}, dispersion forces~\cite{scheel.buhmann_2009,buhmann_2013a}, molecular emitters strongly coupled with surface plasmon polarition~\cite{wang.scholes.ea_2019}, polaritonic quantum-vacuum detection~\cite{lindel.bennett.ea_2021}, and excitation energy transfer between chiral molecules~\cite{franz.buhmann.ea_2023}. Recent comprehensive reviews of \ac{MQED} can be found in Ref.~\cite{scheel.buhmann_2009,buhmann_2013}. Here we highlight that the integration of \ac{MQED} with \textit{ab initio} methods like \ac{DFT}~\cite{svendsen.kurman.ea_2021} and \ac{QEDFT}~\cite{svendsen.thygesen.ea_2024}, which is a more recent advancement in the field.
For readers not focusing on \ac{DFT} or \ac{QEDFT}, this section can be skipped or revisited after reading Sec.~\ref{subsec:abinitio-matter} and~\ref{subsec:abinitio-lightmatter}, where these methods are discussed in detail.

When combining \ac{MQED} with first-principles electronic structure methods such as \ac{DFT}, the electronic system no longer couples to a lossless photon field, but instead to the polaritonic field of the lossy cavity setup. 
Ref.~\cite{svendsen.kurman.ea_2021} demonstrated such combined \ac{MQED}-\ac{DFT} framework by studying the coupling of intersubband transitions in few-layer \ac{TMD} stacks~\cite{schmidt.vialla.ea_2018} to the acoustic graphene plasmon that forms between a doped graphene sheet and a metal when these are placed in close proximity~\cite{lundeberg.gao.ea_2017}.
These setups, characterized by highly confined and dispersive EM modes with non-Lorentzian lineshapes, revealed that common approximations like the dipole approximation can fail. In such cases, the detailed shape of the electronic wave function becomes critical, with simplistic approximations leading to significant errors. Ref.~\cite{svendsen.kurman.ea_2021} highlighted the importance of accurately describing the dispersion of both \ac{EM} fields and electronic transitions for modeling confined light-matter systems.

While the \ac{MQED}-\ac{DFT} framework is effective for weak and strong coupling regimes, it faces limitations in ultra- and deep-strong coupling regimes. To address this, recent work~\cite{svendsen.thygesen.ea_2024} combined \ac{MQED} with \ac{QEDFT}, providing a self-consistent approach that incorporates both the \ac{EM} environment and the full electronic structure, with the latter now being modified by the presence of the quantized \ac{EM} fields.
The light-matter coupling $\lambda(\omega)$ originally treated as a free parameter within the \ac{QEDFT} framework (or Eq.~\eqref{eq:HPF-eff-modes} in Sec.~\ref{subsubsec:Approximation_strategies}) and in the dipole approximation can now be directly estimated from the \ac{DGF} within \ac{MQED} framework~\cite{svendsen.thygesen.ea_2024}. 
Such \ac{MQED}-\ac{QEDFT} framework was exemplified on a benzene molecule in a lossy metallic spherical cavity to examine how cavity radius and losses affect the nature of the light-matter coupling. The results demonstrated that coupling strength is intricately related to cavity geometry and material composition.
Further analysis explored larger aromatic molecules (e.g., naphthalene and anthracene) to modify coupling strength. Intuitively, larger molecules with greater transition dipole moments should increase coupling.
However, because the transition frequency also redshifts with increasing molecule size, the cavity setup has to be reoptimized for each molecule. The result is an effective reduction in the light-matter coupling for increasing molecule size. 
Ref.~\cite{svendsen.thygesen.ea_2024} thus demonstrates that one needs to be careful with simple statements about how the light-matter coupling can be altered, and that a proper description of coupled light-matter systems requires a quantitative description of \emph{both} the electronic structure and the \ac{EM} environment.

\subsection{Long-wavelength approximation (LWA) for solid-state materials coupled to a cavity}\label{subsec:LWA}
In this section, we return to the full \ac{PF} Hamiltonian [Eq.~\eqref{eq:HPF-full-continuum}] to explore a practical approximation for coupling solid-state materials with cavities: the long-wavelength approximation \ac{LWA}. While \ac{MQED} offers a first step towards the computability of the \ac{PF} Hamiltonian, the task remains daunting. Another procedure that can be employed to explore the coupling between matter, including extended systems, and cavities is to adopt the \ac{LWA}. The \ac{LWA} assumes that the momentum carried by cavity photons is negligible compared to the internal momenta of the matter system. 
The validity of the \ac{LWA} depends on the photon momentum cutoff $\bfq_{\rm{c}}^{\rm{LWA}}$, which defines the largest photon momentum of the modes affected by the cavity.
By using the \ac{LWA} and performing a consistent quantization procedure for light and matter~\cite{svendsen.ruggenthaler.ea_2023}, one can derive an effective single- or few-mode \ac{PF} Hamiltonian that avoids double counting of free space light-matter coupling and captures the essential physics of cavity-modified systems.
Importantly, the light-matter coupling strength in the effective \ac{LWA} theory is encoded in an effective interacting mode volume.

For illustrative purposes, we apply the \ac{LWA} to a planar cavity-matter setup, such as a Fabry-Pérot cavity, focusing on the electronic component of the matter system. We should keep in mind that this approximation is valid for most nano-optical setups, as long as the momentum carried by cavity modes is much smaller than the characteristic momentum scales of matter systems. Following Ref.\cite{svendsen.ruggenthaler.ea_2023}, the \ac{PF} Hamiltonian for a 2D material or thin film embedded in a Fabry-Pérot cavity simplifies under the \ac{LWA} to
\begin{equation}
\begin{aligned}\label{eq:HPF-LWA}
\hat{H}^{\rm{LWA}}_{\rm{PF}} &\simeq \hat{H}_{\rm{EM}} + \hat{H}_{\rm{el}} +\text{(coupling terms)},
\end{aligned}
\end{equation}
where $\hat{H}_{\rm{EM}}$ and $\hat{H}_{\rm{el}}$ are the free part of the \ac{EM} and electronic Hamiltonians, respectively, with their explicit forms shown in Eq.~\eqref{eq:HPF-full-continuum}. The coupling terms include the paramagnetic and diamagnetic contributions, 
\begin{equation}
\begin{aligned}\label{eq:HPF-LWA-couplings}
\text{(coupling terms)} & = \frac{1}{\sqrt{V}}\sum_{nm\sigma\bfk_{\parallel}}\hat{c}_{n\bfk_{\parallel}\sigma}^\dagger\hat{c}_{m\bfk_{\parallel}\sigma}\bar{p}_{nm\sigma\bfk\boldsymbol{0}}\sum_{\bfq_{\parallel},\lambda}^{\bfq_{\rm{c},\parallel}^{\rm{LWA}}}\sum_{q_{z},\alpha}^{\bfq_{\rm{c},z}^{\rm{LWA}}} \bar{A}_{\bfq_{\parallel}\lambda, q_{z}\alpha}(\hat{b}_{\bfq_{\parallel}\lambda,q_{z}\alpha}+\hat{b}^{\dagger}_{-\bfq_{\parallel}\lambda,q_{z}\alpha})\\
&+ \frac{1}{2V}\sum_{n\sigma\bfk_{\parallel}}\hat{c}_{n\bfk_{\parallel}\sigma}^\dagger\hat{c}_{n\bfk_{\parallel}\sigma}\sum_{\bfq_{\parallel},\lambda}^{\bfq_{\rm{c},\parallel}^{\rm{LWA}}}\sum_{q_{z},\alpha}^{\bfq_{\rm{c},z}^{\rm{LWA}}}\bar{A}_{\bfq_{\parallel}\lambda,q_{z}\alpha}^2(\hat{b}_{\bfq_{\parallel}\lambda,q_{z}\alpha} +\hat{b}^{\dagger}_{-\bfq_{\parallel}\lambda,q_{z}\alpha})^2,
\end{aligned}
\end{equation}
where the operators $\hat{c},\hat{c}^{\dagger}$ and $\hat{b},\hat{b}^{\dagger}$ denote electronic and photonic operators, respectively. 
In this section, photonic operators are denoted as $\hat{b}, \hat{b}^{\dagger}$, rather than $\hat{a}, \hat{a}^{\dagger}$ as used in Eq.~\eqref{eq:HPF-full-continuum} for convenience. 
The electronic momentum $\bfk$ and photon momentum $\bfq$ are decomposed into parallel ($\parallel$) and perpendicular ($z$) components relative to the planar direction. 
The indeces $n$ ($m$) label electronic bands, with spin index $\sigma$, while $\alpha$ indexes photon modes, with polarization index $\lambda$. 
The light-matter coupling is determined by the product of the electronic momentum matrix elements $\bar{p}_{nm\sigma\mathbf{q}\mathbf{0}}$ and the vector potential mode functions $\bar{A}_{\bfk_{\parallel}\lambda, k_{z}\alpha}$ (see Ref.~\cite{svendsen.ruggenthaler.ea_2023} for detailed definitions).

Although Eq.~\eqref{eq:HPF-LWA} is simpler than the original \ac{PF} Hamiltonian, it still involves a large number of photonic modes. 
To further simplify, the coupling terms can be reduced to an effective single-mode description by averaging the vector potential mode functions $\bar{A}_{\bfk_{\parallel}\lambda, k_{z}\alpha}$ over the relevant modes to get its average value $A_{\rm{eff},\lambda}$. This leads to the definition of a collective photon displacement coordinate, $\hat{Q}_{{\rm{eff}},\lambda}$, which captures the collective contribution of all photonic modes, 
\begin{equation}\label{eq:totdispl}
\hat{Q}_{{\rm{eff}},\lambda} \equiv \frac{1}{\sqrt{2}} \sum_{\alpha}\sum_{\bfq_{\parallel}}^{\bfq_{{\rm{c}}, \parallel}^{\rm{LWA}}}\sum_{q_{z}}^{\bfq_{{\rm{c}},z}^{\rm{LWA}}} (\hat{b}_{\bfq_{\parallel}\lambda,q_{z}\alpha}+\hat{b}^{\dagger}_{-\bfq_{\parallel}\lambda,q_{z}\alpha}),
\end{equation}
which is the photonic operator that directly appears in the paramagnetic term of Eq.~\eqref{eq:HPF-LWA-couplings}.
By combining this collective coordinate operator with the respective relative coordinates operators, a canonical transformation can be applied, and the \ac{LWA}-\ac{PF} Hamiltonian [Eq.~\eqref{eq:HPF-LWA}] simplifies to~\cite{svendsen.ruggenthaler.ea_2023},
\begin{equation}\label{eq:lwsinglefinal-simplified}
\hat{H}_{\rm{PF}}^{\rm{LWA}} \simeq 
\hat{H}_{\rm{el}}+\hat{H}_{\rm{EM}, \rm{rel}}+\sum_\lambda\left(\omega_{\rm{c}, \rm{eff}}+\frac{1}{2}\right)\hat{B}^{\dagger}_{\rm{eff}, \lambda}\hat{B}_{\rm{eff}, \lambda} + \text{(effective coupling terms)},
\end{equation}
where $\hat{H}_{\text{EM,rel}}$ groups terms associated with relative photonic coordinates, and we introduce a new set of creation and annihilation photonic operators $\hat{B}^{\dagger}_{\rm eff}$ and $\hat{B}_{\rm eff}$ associated with the collective displacement operator $\hat{Q}_{{\rm eff}, \lambda}$. The effective coupling terms, 
\begin{equation}\label{eq:lwsinglefinal-couplings}
\begin{aligned}
\text{(effective coupling terms)}&= \left(\frac{\sqrt{3}}{8L_{c}^{3}\mathcal{F}}\right)^{1/2}\hat{P}_{\rm{el}} \sum_{\lambda} A_{\rm{eff},\lambda} \left(\hat{B}^{\dagger}_{\rm{eff},\lambda}+\hat{B}_{\rm{eff},\lambda}\right) \\
& + \left(\frac{\sqrt{3}}{8L_{c}^{3}\mathcal{F}}\right)\hat{N}_{\rm{el}}\frac{A_{\rm{eff},\lambda}^2}{2}\left(\hat{B}^{\dagger}_{\rm{eff},\lambda}+\hat{B}_{\rm{eff},\lambda}\right)^{2},
\end{aligned}
\end{equation}
where we define the total electron momentum operator as $\hat{P}_{\text{el}}\equiv\sum_{nm\sigma\bfk_\parallel}\hat{c}_{n\bfk_\parallel\sigma}^\dagger\hat{c}_{m\bfk_\parallel\sigma}\bar{p}_{nm\sigma\bfk\boldsymbol{0}}$. 
The parameter $\mathcal{F}$ represents the finesse of the Fabry-Pérot cavity and $L_{c}$ the cavity length in the out-of-plane direction (the confinement direction).
In Eq.~\eqref{eq:lwsinglefinal-couplings}, the effective single-mode vector potential strength can thus be expressed as
\begin{align}\label{eq:g_eff}
    a_{\rm{eff},\lambda} \equiv \left(\frac{\sqrt{3}}{8L_{c}^{3}\mathcal{F}}\right)^{1/2} A_{\rm{eff},\lambda} \equiv \frac{1}{\sqrt{V_{\rm{eff}}}} A_{{\rm{eff}},\lambda}.
\end{align} 
This way of rewriting the effective vector potential strength highlights how regardless of the quantization volume for the original light-matter coupled problem, the cavity couples to matter only within finite mode volume $V_\mathrm{eff} \sim L_{c}^3\mathcal{F}$(and it is finite in all directions). This is remarkable because it implies that even in an arbitrarily extended cavity like the Fabry-Pérot cavity, which is open in the planar direction, the coupling with matter remains finite and it is characterized by an effective length scale which is set by the ability of the cavity mirrors to confine light.

The \ac{LWA} and dipole approximation share similarities for finite systems but differ significantly when applied to extended materials. The dipole approximation is a widely used approach in atomic and molecular physics. In the dipole approximation, the material's spatial extent is much smaller than the wavelength scale of the \ac{EM} field. This results in a single zero-momentum mode theory, where the mode volume grows with the lateral extent of the cavity. In the thermodynamic limit, as the cavity size increases, the mode volume increases, leading to the coupling vanishing and ultimately causing decoupling between light and matter within the dipole framework. Several studies~\cite{eckhardt.passetti.ea_2022,andolina.pellegrino.ea_2020} have noted that, in the dipole approximation, the effects of the cavity on the material's properties diminish as the size of the cavity-material system approaches infinity. However, this decoupling arises from the strict limitations of the dipole approximation, which overlooks the multi-mode nature of the \ac{EM} field. On the other hand, the \ac{LWA} applied to extended systems is less restrictive~\cite{amelio.korosec.ea_2021,lenk.li.ea_2022}. Excluding surface modes like surface plasmon polaritons, the momentum of modes within a Fabry-Pérot cavity is generally negligible compared to the momentum scales relevant to electronic matrix elements in extended systems. 

Compared to the dipole approximation, the \ac{LWA} effectively retains the multi-mode nature of the coupling. It assumes that the momentum of the cavity modes is negligible compared to electronic momenta, while still accounting for the oscillatory part of the mode functions, $e^{i\bfk\cdot\bfr}$. This ensures that the light-matter coupling remains finite, even in extended systems. Importantly, the \ac{LWA} avoids the issues of vanishing interactions by considering the effective volume of the cavity mode, $V_{\text{eff}}\sim L_{c}^{3}\mathcal{F}$, which remains finite irrespective of the planar extend of the cavity. The \ac{LWA} is particularly useful for examining how cavity-matter systems behave as the material's size increases. Up to the \textit{bulk limit}, where the material's size remains within the effective volume of the cavity mode, the coupling remains finite and significant. Beyond this point, the interaction weakens as the system surpasses the effective cavity volume.

\subsection{Effective model Hamiltonian approaches for light-matter coupled systems}\label{subsec:effective-H-light-matter}
Even after simplifying the \ac{PF} Hamiltonian with \ac{MQED} and the \ac{LWA}, solving the full problem remains challenging due to the complex collective behavior of multiple electronic and nuclear degrees of freedom. Condensed matter phenomena often involve the collective behavior of multiple electronic and nuclear degrees of freedom. However, by focusing on specific matter excitations relevant to the system under study, the Hamiltonian can be reduced to a manageable form with only a few degrees of freedom coupled to an effective photonic mode. This approach, while powerful, requires prior knowledge of which degrees of freedom are important and how the physical effect operates. For instance, understanding the absorption spectrum of a molecule often involves not only ground and excited states but also many off-resonant states to achieve accurate quantitative and qualitative results~\cite{yang.ou.ea_2021,konecny.kosheleva.ea_2024}. When the relevant physics is well-understood, such reduced models offer significant insights into the targeted properties of light-matter coupled systems. Below, we discuss two examples that illustrate this approach: one involving phonons and another focusing on excitons. The light-matter coupling is included in the matter-only Hamiltonians through minimal coupling, represented by the substitution of the matter momentum operator, $\hat{\bfp}\rightarrow\hat{\bfp}-Z_{p}\hat{\bfA}$ where $Z_{p}$ denotes the charge of the matter particle and $\hat{\bfA}$ is the photon field operator.

%
Consider a light-matter system consisting of an insulator with a strongly \ac{IR}-active phonon mode, whose behavior determines the stability of a material phase~\cite{latini.shin.ea_2021}. Using the \ac{LWA}, the system can be reduced to an effective \ac{QED} model Hamiltonian with a single degree of freedom for the phonon and one for the effective photon mode, 
\begin{align}
    \hat{H}^{\rm{eff,model}}_{\rm{PF}} = \omega_{c} \hat{a}^\dagger \hat{a} + \frac{1}{2M} \left[\hat{P} - \frac{A_{\rm{eff}}}{\sqrt{V_{\rm{eff}}}} Z (\hat{a}^\dagger + \hat{a})\right]^2 + V(\hat{Q}),
\end{align}
where the first term represents the free photon Hamiltonian with the photon frequency $\omega_{c}$ and the annihilation (creation) photon field operator $\hat{a}$ ($\hat{a}^\dagger$), the second term includes the phonon kinetic term and the coupling between photon and phonon, with $A_{\rm{eff}}$ as the coupling strength, $V_{\text{eff}}$ as the effective mode volume, $Z$ and $M$ the phonon mode effective charge and physical mass, respectively, and the last term is the potential energy surface $V(\hat{Q})$ felt by the phonon where $\hat{P}$ and $\hat{Q}$ are the phonon momentum and position operator, respectively. The photon mode's polarization is assumed to align with the phonon mode, and this detail has been omitted. 
These effective parameters including $\omega_c$, $A_{\rm{eff}}$, $M$, and $Z$, can be obtained from \textit{ab initio} calculations (see Sec.~\ref{subsec:abinitio-matter}) or fitted to experimental results.
The Hamiltonian can be diagonalized in a tensor product space of uncoupled states, combining grid representations for phonons with Fock states for photons. For example, in the representation $\ket{q}\otimes\ket{m}$, where $q$ represents the phonon position and $m$ the photon number, the system captures key features of the light-matter interaction, including phonon-polaritons, and its Hamiltonian can be expressed as
\begin{align}
H_{\rm{PF}, qm~q'n}^{\rm{eff,model}} & = \omega_{c}\delta_{m,n}\delta{q,q'} - \frac{\nabla_q^2}{2M} + \left(i\frac{A_{\rm{eff}}Z}{M\sqrt{V_{\rm{eff}}}}\nabla_{q} \sqrt{m}\delta_{m, n+1} + \rm{h.c.}\right) \\
& + \frac{A_{\rm{eff}}^{2}Z^{2}}{2MV_{\rm{eff}}}m(m+1)(\delta_{m, n+2} + \delta_{n+2, m}) + V(q)\delta_{q,q'},
\end{align}
where $\delta_{i,j}$ is the Kronecker delta.
This model is particularly useful for predicting cavity-induced phase transitions in systems where specific phonon modes drive macroscopic material properties (see Sec.~\ref{subsec:theory-prediction-paraferro}).
While adding more phonon or photon modes is possible, the computational cost increases exponentially, limiting the scalability of this approach for highly complex systems.

%
In semiconductors, cavity modes often couple to neutral electronic excitations, such as excitons, as the second example. For low-energy excitations near the bandgap, the system can be reduced to a few excitonic states interacting with a cavity mode~\cite{latini.ronca.ea_2019}. In 2D materials, where optical responses are dominated by band-edge excitonic states, the effective \ac{PF} Hamiltonian reduces to
\begin{equation}
\begin{split}
\hat{H}_{\text{pol}} & = \Delta |e\rangle \langle e| + \frac{\hbar^{1/2}\omega_{\textrm{d}}}{\sqrt{2 m \omega}} \left( \langle 0|\hat{\epsilon} \cdot  \hat{p} | e \rangle |0\rangle \langle e|~\hat{a}^\dagger + \langle e|\hat{\epsilon} \cdot  \hat{p} | 0 \rangle |e\rangle \langle 0|~\hat{a}\right) \\
& + \frac{\hbar \omega_{\mathrm{d}}^2}{4 m \omega}\left( \hat{a} + \hat{a}^{\dagger} \right)^2 + \hbar \omega \left(\hat{a}^\dagger \hat{a} + \frac{1}{2} \right),
\end{split}
\end{equation}
where $\Delta$ is the gap between the ground state $|0\rangle$ and the lowest excited state $|e\rangle$, $\hat{p}$ is the momentum operator, $m$ is the effective mass, $\omega$ is the photon frequency with the polarization $\hat{\epsilon}$, and $\hat{a},\hat{a}^{\dagger}$ are the photonic operators. 
The parameter $\omega_{\text{d}} = \sqrt{(e^{2} N_{e})/(m \epsilon_0 V_{\text{eff}})}$ is defined with $N_{\text{e}}=1$ and the mode volume $V_{\text{eff}}$. 
The matrix element $\mel{0}{\hat{p}}{e}$ can be computed using \textit{ab initio} methods (see Sec.~\ref{subsec:abinitio-matter}). 
This Hamiltonian highlights the connection between exciton-polaritons and the quantum Rabi model~\cite{xie.zhong.ea_2017}, capturing key features of the coupling between excitonic states and cavity photons. In more complex scenarios, where multiple excitonic states interact with the same cavity mode, the resulting composite exciton-polaritons can exhibit intricate spectral features.

In addition to including light-matter coupling via the substitution $\hat{\mathbf{p}}$ with $\hat{\mathbf{p}}-Z_{p}\hat{\bfA}$ in matter-only Hamiltonians, as demonstrated in the above two cases, another approach is starting with \ac{TB} models for matter-only systems. These model Hamiltonians are invaluable for studying the electronic structure of solids, especially when \textit{ab initio} methods become computationally prohibitive~\cite{goringe.bowler.ea_1997}. Balancing computational efficiency with physical insight, \ac{TB} models capture essential behaviors of targeted systems while allowing for calculations of material properties such as band structures, transport coefficients, and topological invariants~\cite{martin_2020}.
Unlike \textit{ab initio} methods such as \ac{DFT} (Sec.~\ref{subsec:abinitio-matter}), which aim to solve the electronic structure problem from first principles, \ac{TB} models rely on approximations and parameterization, focusing on capturing the essential physics of a system with reduced computational complexity. So far the \ac{TB} model Hamiltonians with light-matter interaction have been the main tool for studying cavity-modified material properties, which will be discussed in Sec.~\ref{sec:theory-prediction}.

Here we consider the \ac{TB} model that describes the electronic structure by considering electrons localized around atomic sites and allowing for hopping between neighboring atoms. 
%
The general form of the \ac{TB} Hamiltonian is given by
\begin{equation}\label{eq:TB-Hamiltonian-construction}
\hat{H} = \sum_{i,\alpha} \epsilon_{\alpha} \hat{c}_{i,\alpha}^\dagger \hat{c}_{i,\alpha} + \sum_{i,j,\alpha,\beta} t_{i,j}^{\alpha,\beta} \hat{c}_{i,\alpha}^\dagger \hat{c}_{j,\beta},    
\end{equation}
where $\epsilon_{\alpha}$ is the on-site energy of an electron in orbital $\alpha$ on $i$th site, $t_{i,j}^{\alpha,\beta}$ is the hopping parameter for an electron to move from orbital $\alpha$ on $i$th site to orbital $\beta$ on $j$th site, and $\hat{c}_{i,\alpha}^\dagger$ and $\hat{c}_{i,\alpha}$ are creation and annihilation operators for an electron in orbital $\alpha$ on $i$th site. The \ac{TB} Hamiltonian provides a minimal description of electronic interactions in terms of a few key parameters, which can either be derived from \textit{ab initio} calculations or fitted to experimental data. \ac{TB} models are not limited to simple electronic structures; they can also describe complex interactions in strongly correlated systems. For example, extensions like the Hubbard model incorporate onsite Coulomb repulsion (U) to capture the physics of correlated electrons. These parameters, too, can be calculated from first principles.

These \ac{TB} models can be further extended to include light-matter interactions using the Peierls substitution. In this approach, the hopping parameters in the \ac{TB} Hamiltonian [Eq.~\eqref{eq:TB-Hamiltonian-construction}] are modified to account for the interaction with classical and quantized \ac{EM} fields~\cite{li.eckstein_2020,li.golez.ea_2020,sentef.li.ea_2020,eckhardt.passetti.ea_2022,li.schamriss.ea_2022,vinasbostrom.sriram.ea_2023,arwas.ciuti_2023,passetti.eckhardt.ea_2023}. For example, the Peierls substitution incorporates photonic degrees of freedom by encoding the vector potential operator into the hopping terms. Once the light-matter coupling is introduced, the resulting Hamiltonians may then be further approximated, either by focusing on the matter Hamiltonian in specific photon Fock states~\cite{vinasbostrom.sriram.ea_2023} or by integrating them out using perturbative approaches, leading to further effective matter Hamiltonians that incorporate light-matter interaction effects implicitly~\cite{cohen-tannoudji.dupont-roc.ea_1998}.

%
In short, to build an effective \ac{QED} Hamiltonian for light-matter coupled systems, the following steps are typically followed: 1) identify relevant degrees of freedom: to determine the key electronic or phononic excitations to include in the model; 2) establish matter field approximation: to decide on the level of approximation for the matter field, such as focusing on a single excitation or neglecting weaker couplings to other degrees of freedom; 3) approximate the photon field: to choose the appropriate level of approximation for the photon field, such as the dipole approximation or \ac{LWA}, and reduce the system to an effective single mode if necessary; 4) build the minimal Hamiltonian: to construct the simplest Hamiltonian within a reduced Fock space, ensuring that it captures the essential physics of the system.
Finally, it is crucial to scrutinize the assumptions underlying these models, as they imply both explicit and implicit approximations. It is far from trivial to show that a reduced model is actually connected to the solution of a parent \textit{ab initio} theory~\cite{ruggenthaler.sidler.ea_2023}.

\subsection{\textit{Ab initio} and effective matter Hamiltonians for solid-state materials}\label{subsec:abinitio-matter}
When it comes to solving the quantum many-body problem in practice, instead of avoiding it by using simple models, one of the most employed and versatile tools is density-functional methods. They are exact reformulations of linear quantum physics in a non-linear but computationally beneficial form. In this section we provide a brief overview of \ac{DFT} methods for solid-state materials, offering a pedagogical introduction and a comparison with the \ac{QEDFT} techniques discussed in Sec.~\ref{subsec:abinitio-lightmatter}. For readers interested in exploring \ac{DFT} methods for solid-state materials in greater detail, Ref.~\cite{martin_2020} offers a good resource. For readers already familiar with \ac{DFT}, feel free to skip this section and revisit it as needed.

\subsubsection{Separation of electrons and nuclei/ions}\label{subsubsec:separation-electrons-nuclei}
Before we go into detail on \ac{DFT} let us first set the stage of the quantum theory we want to density-functionalize. A solid-state material is a crystalline solid with atoms arranged in an orderly crystal lattice, defined by a unit cell and symmetry operations like translation and rotation. The Hamiltonian of a solid-state material $\hat{H}_{\rm{M}}$ [Eq.~\eqref{eq:Matter-H-from-PF}] to describe a coupled electron-nucleus system with $N_{e}$ electrons and $N_{n}$ nuclei in the absence of any explicit photon field is rewritten as 
\begin{equation}\label{eq:matter-Hamiltonian-simplified}
\hat{H}_{\rm{M}}(\underline{\hat{\bfr}},\underline{\hat{\bfR}}) = \hat{T}_{e}(\underline{\hat{\bfr}}) + \hat{W}_{e}(\underline{\hat{\bfr}}) + \hat{T}_{n}(\underline{\hat{\bfR}}) + \hat{W}_{n}(\underline{\hat{\bfR}}) + \hat{W}_{en}(\underline{\hat{\bfr}},\underline{\hat{\bfR}}),
\end{equation}
where $\underline{\hat{\bfr}} = (\hat{\bfr}_{1},...,\hat{\bfr}_{N_{e}})$ and $\underline{\hat{\bfR}} = (\hat{\bfR}_{1},...,\hat{\bfR}_{N_{n}})$ are the electronic and nuclear coordinates, respectively, with $\hat{O}$ indicating that it is an operator. The terms on the right-hand side of the equation in order are the electronic kinetic energy operator $\hat{T}_{e}(\underline{\hat{\bfr}})$, the (repulsive) Coulomb potential operator among electrons $\hat{W}_{e}(\underline{\hat{\bfr}})$, the nuclear kinetic energy operator $\hat{T}_{n}(\underline{\hat{\bfR}})$, the (repulsive) Coulomb potential operator among nuclei $\hat{W}_{n}(\underline{\hat{\bfR}})$, and the (attractive) Coulomb potential operator between electrons and nuclei $\hat{W}_{en}(\underline{\hat{\bfr}},\underline{\hat{\bfR}})$. Explicitly, this becomes
\begin{equation}\label{eq:matter-Hamiltonian}
\hat{H}_{\rm{M}} = \sum_{i=1}^{N_{e}}\frac{\hat{\bfp}_{i}^{2}}{2m_{e}} + \frac{1}{2}\sum_{i\neq j}^{N_{e}}\frac{e^{2}}{4\pi\vacep|\hat{\bfr}_{i}-\hat{\bfr}_{j}|}   + \sum_{I=1}^{N_{n}}\frac{\hat{\bfP}_{I}^{2}}{2M_{I}} + \frac{1}{2}\sum_{I\neq J}^{N_{n}}\frac{Z_{I}Z_{J}e^{2}}{4\pi\vacep|\hat{\bfR}_{I}-\hat{\bfR}_{J}|} -  \sum_{i=1}^{N_{e}}\sum_{I=1}^{N_{n}}\frac{Z_{I}e^{2}}{4\pi\vacep|\hat{\bfr}_{i}-\hat{\bfR}_{I}|},
\end{equation}
%
In Hartree atomic units, $\hbar$, $|e|$, $m_{e}$, and $4\pi\vacep$ are set to $1$, and length and energy are measured in Bohr and Hartree units, respectively.

Solving the coupled electron-nucleus Hamiltonian $\hat{H}_{\rm{M}}$ is challenging due to the numerous degrees of freedom from both electrons and nuclei. This problem is central in solid-state and condensed matter physics~\cite{martin_2020}. Consequently, the matter Hamiltonian is typically solved using several approximations~\cite{marzari.ferretti.ea_2021}. The primary objective is to determine the eigenstates and eigenenergies (or spectrum) of the matter Hamiltonian,
\begin{equation}\label{eq:matter-Hamiltonian-eigenproblem}
\hat{H}_{\rm{M}}(\underline{\hat{\bfr}},\underline{\hat{\bfR}})\Psi_{i}(\underline{\bfr},\underline{\bfR}) = E_{i}\Psi_{i}(\underline{\bfr},\underline{\bfR}),
\end{equation} 
where $\Psi_{i}(\underline{\bfr},\underline{\bfR})$ and $E_{i}$ are the electron-nucleus coupled wave function and energy of the $i$th state, respectively. A very common approach to solving the matter Hamiltonian is to rewrite the electron-nucleus wave function in terms of an infinite sum of conditional wave functions (see below) and then to approximate the resulting coupled differential equations. To do so, we first discard the nuclear kinetic energy $\hat{T}_{n}$ in Eq.~\eqref{eq:matter-Hamiltonian} due to the heavy nuclear masses, and solve for the electronic subsystem for a given (classical) position of the nuclei. That is, we determine the electronic wave function for all possible (classical) positions of the nuclei. We call such an electronic wave function, \textit{conditional}. The matter Hamiltonian then simplifies to the so-called electronic Hamiltonian $\hat{H}_{e}(\underline{\hat{\bfr}};{\underline{\bfR}})$, where nuclear coordinates $\underline{\bfR}$ are merely parameters to be adjusted, 
\begin{equation}\label{eq:electronic-Hamiltonian}
\hat{H}_{e}(\underline{\hat{\bfr}};\{\underline{\bfR}\}) = \hat{T}_{e}(\underline{\hat{\bfr}}) + \hat{W}_{e}(\underline{\hat{\bfr}}) + \hat{W}_{en}(\underline{\hat{\bfr}};\{\underline{\bfR}\}) + \hat{W}_{n}(\{\underline{\bfR}\}),
\end{equation}
where the last term becomes a constant energy shift, given fixed nuclear coordinates $\underline{\bfR}$. We use the semicolumn and curly bracket to emphasize that the nuclear coordinates are parameters. The goal now is to solve the electronic Hamiltonian, 
\begin{equation}\label{eq:electronic-Hamiltonian-eigenproblem}
\hat{H}_{e}(\underline{\hat{\bfr}};\{\underline{\bfR}\})\psi_{j}(\underline{\bfr};\{\underline{\bfR}\}) = \epsilon_{j}(\{\underline{\bfR}\})\psi_{j}(\underline{\bfr};\{\underline{\bfR}\}),
\end{equation}
where $\psi_{j}(\underline{\bfr};\{\underline{\bfR}\})$ and $\epsilon_{j}(\{\underline{\bfR}\})$ are the (conditional) electronic wave function and energy of the $j$th electronic state at fixed nuclear coordinates $\underline{\bfR}$. The electronic eigenstates at the fixed nuclear coordinates are orthonormal, i.e., $\braket{\psi_{j}(\underline{\bfr};\{\underline{\bfR}\})}{\psi_{j'}(\underline{\bfr};\{\underline{\bfR}\})} = \delta_{j,j'}$, where $\delta_{j,j'}$ is the Kronecker delta and the integration $\langle...\rangle$ here is only over the electronic coordinates. For fixed $\underline{\bfR}$ they form a complete basis set for the electronic Hilbert subspace. Therefore, the $i$th eigenstate of the matter Hamiltonian $\Psi_{i}(\underline{\bfr},\underline{\bfR})$ can be expanded in terms of the electronic eigenstate basis set $\{\psi_{j}(\underline{\bfr};\{\underline{\bfR}\})\}$,
\begin{equation}\label{eq:matter-eigenstate-BH-expansion}
\Psi_{i}(\underline{\bfr},\underline{\bfR}) = \sum_{j=0}^{\infty}\chi_{ij}(\underline{\bfR})\psi_{j}(\underline{\bfr};\{\underline{\bfR}\}),
\end{equation}
with the $i$th nuclear wave function $\chi_{ij}(\underline{\bfR})$ based on the potential energy surface from the $j$th electronic eigenstate at the fixed nuclear coordinates $\underline{\bfR}$. This expansion is called the Born-Huang expansion and maps the original problem to a set of infinitely many coupled differential equations with their own respective Hilbert spaces. Using Eq.~\eqref{eq:matter-Hamiltonian-eigenproblem}, Eq.~\eqref{eq:matter-eigenstate-BH-expansion}, and the orthonormality of the electronic eigenstates, we derive the equation for the nuclear wave function~\cite{martin_2020}, 
\begin{equation}\label{eq:matter-nuclear-exact}
\left\{\hat{T}_{n}(\hat{\underline{\bfR}})+\epsilon_{j}(\underline{\bfR})-E_{i}\right\}\chi_{ij}(\underline{\bfR}) = -\sum_{j'}C_{jj'}\chi_{ij'}(\underline{\bfR}),
\end{equation}
where the coefficients $C_{jj'}$ couples the $j$th and $j'$th electronic eigenstate and can be computed using the associated electronic wave functions~\cite{martin_2020}.
Solving this problem seems even harder than the original coupled electron-nucleus equation. Yet, in this form we can make some approximations based on physical arguments. Taking into account that the nuclear masses are usually much larger than the electronic ones, one can often discard all off-diagonal terms of $C_{jj'}$ and absorb the diagonal terms into $\epsilon_{j}(\underline{\bfR})$. This is called the adiabatic or \ac{BOA}. In most cases (also used in the following) this approximation does also imply that we neglect the diagonal terms $C_{jj}$ as well. Consequently, the nuclear wave function within the \ac{BOA}, $\chi_{ij}^{\rm{BOA}}(\underline{\bfR})$, satisfies
\begin{equation}\label{eq:matter-nuclear-BOA}
\left\{\hat{T}_{n}(\hat{\underline{\bfR}})+\epsilon_{j}(\underline{\bfR})\right\}\chi_{ij}^{BOA}(\underline{\bfR}) = E_{i}\chi_{ij}^{\rm{BOA}}(\underline{\bfR}).
\end{equation}
This equation describes nuclei as quantum particles on a potential energy surface $\epsilon_{j}(\underline{\bfR})$ from the $j$th electronic eigenstate. The ground state of the matter Hamiltonian within the \ac{BOA} is thus approximated as $\Psi_{0}^{\rm{BOA}}(\underline{\bfr},\underline{\bfR}) \approx \chi_{00}^{\rm{BOA}}(\underline{\bfR}) \psi_{0}(\underline{\bfr};\{\underline{\bfR}\})$.

\subsubsection{Solving the electronic system}
Even after decoupling electrons and nuclei via the \ac{BOA}, solving the electronic Hamiltonian [Eq.~\eqref{eq:electronic-Hamiltonian-eigenproblem}] remains a significant challenge due to the large number of interacting electrons. Over decades, various computational methods like \ac{DFT} and Green's function methods have been developed to describe electrons in solids~\cite{marzari.ferretti.ea_2021}. Among these, \ac{DFT} emerges as a powerful and widely-used computational method for studying the electronic structure of atoms, molecules, and solids, often serving as the foundation for other methods~\cite{martin.reining.ea_2016}. 

\ac{DFT} and its variants~\cite{engel.dreizler_2011} reformulates the complex many-body problem of interacting electrons into a more tractable form using electron density as the primary variable. This approach is an example of a collective-variable theory, which simplifies a system by focusing on macroscopic quantities rather than individual particle trajectories. Similar to classical fluid mechanics, where velocity fields replace individual particle dynamics, \ac{DFT} uses the electron density $\rho(\bfr)$ to describe quantum systems. Introduced by Hohenberg and Kohn~\cite{hohenberg.kohn_1964}, \ac{DFT} predicts several material properties, such as electronic band structure, magnetic moments, and dielectric constants. The key advantage of \ac{DFT} is its ability to deal with large and complex systems while maintaining reasonable accuracy.

A key step of making \ac{DFT} practical came with the \ac{KS} scheme, which maps the complex many-body interacting electron system onto a non-interacting system, called \ac{KS} system, where electrons move in an effective (or \ac{KS}) potential $v_{\rm{KS}}(\bfr)$~\cite{kohn.sham_1965}.
This effective potential consists of
\begin{equation}
v_{\rm{KS}}(\bfr) =  v_{\rm{ext}}(\bfr) + v_{\rm{H}}(\bfr) + v_{\rm{xc}}(\bfr),
\end{equation}
where $v_{\rm{ext}}(\bfr)$ is the external (Coulomb) potential  due to the nuclei, $v_{\rm{H}}(\bfr)$ is the (classical) Hartree potential among electrons, and $v_{\rm{xc}}(\bfr)$ is the \ac{xc} potential to take exchange and correlation effects into account. The \ac{xc} potential $v_{\rm{xc}}(\bfr)$ can be expressed in terms of the electron density, yet its exact form is unknown in general. In practice, we need to use approximate \ac{xc} energy functionals to perform simulations and hence only find approximate densities. Many approximate \ac{xc} energy functionals have been developed and implemented over the past several decades~\cite{jones_2015,cohen.mori-sanchez.ea_2012}, and searching for the exact \ac{xc} energy functional is still an active research direction~\cite{verma.truhlar_2020}.
While in principle orbital-free \ac{DFT} can be used, e.g., in its simplest form that is Thomas-Fermi theory~\cite{dreizler.gross_2012}, the explicit description of the kinetic energy via the \ac{KS} scheme is the origin of the success of \ac{DFT} in practice. The \ac{KS} system is then used to reconstruct the (in principle exact) electron density of the original many-body problem using a Slater determinant of \ac{KS} orbitals, determined by the \ac{KS} equations,
\begin{equation}\label{eq:KS-Hamiltonian}
 \hat{h}_{\text{KS}}(\bfr)\phi_{i}(\bfr)= \left(-\frac{1}{2}\nabla^{2} + v_{\text{KS}}(\bfr)\right)\phi_{i}(\bfr)  =  \epsilon_{i}\phi_{i}(\bfr),
\end{equation}
%
where $\hat{h}_{\rm{KS}}(\bfr)$ is the \ac{KS} Hamiltonian, and the wave function (or \ac{KS} orbital) and energy of the $i$th eigenstate of the \ac{KS} Hamiltonian are $\phi_{i}(\bfr)$ and $\epsilon_{i}$, respectively. 
The electron density $\rho(\bfr)$ is obtained from the \ac{KS} orbitals via $\rho(\bfr) = \sum_{i=1}^{N_{e}} |\phi_{i}(\bfr)|^{2}$. By solving the \ac{KS} equations iteratively, the ground-state electronic structure of a system is determined. The energy of the system can also be computed using the energy functional~\cite{martin_2020}, once the electron density is found self-consistently from the \ac{KS} system.

%
The spatial position $\bfr$ for electrons in Eq.~\eqref{eq:KS-Hamiltonian} spans the whole crystal. However, we can use the periodicity of solid-state materials to solve the \ac{KS} Hamiltonian $\hat{h}_{\rm{KS}}$ in the primitive cell of the solid-state material. The whole crystal can be constructed by performing translations on the primitive cell via the translation vector, $\mathbf{T}_{n} = n_{1}\mathbf{a}_{1} + n_{2}\mathbf{a}_{2} + n_{3}\mathbf{a}_{3}$ where $n_{1 (2,3)}$ and $\mathbf{a}_{1 (2,3)}$ are integers and the primitive lattice vectors, respectively~\cite{ashcroft.mermin_2011}. In the \ac{KS} equations, the Hartree and \ac{xc} potential depend on the electron density, a periodic function in real space, making the effective potential also periodic. The reciprocal space, or $\bfk$ space, of the primitive cell is the first \ac{BZ}~\cite{ashcroft.mermin_2011}. According to Bloch's theorem, solutions to the Schrödinger equation in a periodic potential can be expressed as plane waves modulated by periodic functions~\cite{ashcroft.mermin_2011}. Each electronic eigenstate can thus be labeled with a wave number $\bfk$ and a band index $n$, represented as $\psi_{n\bfk}(\bfr) = u_{n\bfk}(\bfr)e^{i\bfk\cdot\bfr}$ where $u_{n\bfk}(\bfr)$ is the Bloch-periodic component of the \ac{KS} electron wave function. In terms of the Bloch functions, for each $\bfk$ point, Eq.~\eqref{eq:KS-Hamiltonian} becomes 
\begin{equation}\label{eq:KS-Hamiltonian-each-k}
 \hat{\tilde{h}}_{\text{KS}}(\bfk,\bfr)u_{n\bfk}(\bfr)= \left(-\frac{1}{2}(\nabla+i\bfk)^{2} + v_{\text{KS}}(\bfr)\right)u_{n\bfk}(\bfr) = \epsilon_{n\bfk}u_{n\bfk}(\bfr). 
\end{equation}
The Coulomb potential from the nuclei $v_{\rm{ext}}(\bfr)$ in $v_{\text{KS}}(\bfr)$ is often split into contributions from core and valence electrons. The nuclei with the core electrons become ions. The Coulomb interaction between ions and valence electrons can be computed using pseudopotentials~\cite{heine_1970,schwerdtfeger_2011} or \ac{PAW} methods~\cite{blochl_1994,kresse.joubert_1999}. After solving Eq.~\eqref{eq:KS-Hamiltonian-each-k} for each $\bfk$, the energies and wave functions can be used to compute other physical quantities and serve as the starting points for other methods such as many-body perturbation theory~\cite{martin.reining.ea_2016}. 
Standard \ac{DFT} often struggles with strongly correlated materials, where electron-electron interactions dominate. Extensions such as \ac{DFT}$+$U($+$V) methods~\cite{himmetoglu.floris.ea_2014,agapito.curtarolo.ea_2015,tancogne-dejean.oliveira.ea_2017,tancogne-dejean.rubio_2020,timrov.marzari.ea_2021}, which introduce onsite Coulomb repulsion (U) and the intersite orbital hybridizations (V), provide improved accuracy for such systems.
Among these methods, the self-consistent approach -- ACBN$0$ functional~\cite{agapito.curtarolo.ea_2015} and its extensions~\cite{tancogne-dejean.oliveira.ea_2017,tancogne-dejean.rubio_2020} -- to determine the Hubbard U term (and V term) is promising as it can serve as reliable predictive and numerical efficient tools and has been applied to many strongly correlated materials~\cite{agapito.curtarolo.ea_2015,tancogne-dejean.oliveira.ea_2017,tancogne-dejean.rubio_2020}.

\subsubsection{Construction of tight-binding models from \textit{ab initio} methods}\label{subsubsec:TB-models-from-DFT}
\ac{DFT} calculations can serve as a tool for constructing approximate light-matter or effective \ac{QED} Hamiltonians discussed in Sec.~\ref{subsec:effective-H-light-matter}. By using \ac{DFT}, one can obtain the electronic Bloch wave functions, which are essential for computing the matrix elements, for example, in an exciton-polariton-type effective \ac{QED} Hamiltonian~\cite{latini.ronca.ea_2019}. When investigating specific phonon modes coupled to photons, \ac{DFPT} can be used to determine the vibrational modes and the associated electron-phonon coupling constants. The resulting nuclear displacements of these phonon modes are then employed to construct potential energy surfaces using \ac{DFT}, which are integral to developing, for example, phonon-polariton-type effective \ac{QED} Hamiltonians~\cite{latini.shin.ea_2021}.

Furthermore, \textit{ab initio} \ac{TB} models derived from \ac{DFT} calculations provide a starting point for constructing approximate light-matter Hamiltonians. Constructing \ac{TB} models [Eq.~\ref{eq:TB-Hamiltonian-construction}] from \textit{ab initio} methods is essential for balancing accuracy and computational efficiency~\cite{goringe.bowler.ea_1997,horsfield.bratkovsky_2000}. Among the most widely used techniques, \ac{MLWFs} offer a localized basis set, ideal for generating accurate \ac{TB} Hamiltonians~\cite{marzari.mostofi.ea_2012}. \ac{MLWFs} effectively capture electronic properties, facilitating band structure interpolation and the study of phenomena like electric polarization and orbital magnetization. In addition to the \ac{MLWFs} method, the following are other methods used to construct \ac{TB} models. Density-functional tight-binding is a semi-empirical extension of \ac{DFT} that leverages a minimal basis set and two-center approximation, offering computational efficiency for large systems while maintaining reasonable accuracy~\cite{spiegelman.tarrat.ea_2020}. Downfolding techniques, another powerful approach, involve integrating out high-energy degrees of freedom to derive effective low-energy Hamiltonians~\cite{miyake.aryasetiawan_2008,miyake.aryasetiawan.ea_2009,aryasetiawan.nilsson_2022}. This is particularly useful for strongly correlated systems. \ac{ML} is emerging as a valuable tool for constructing \ac{TB} models and Hamiltonians from \textit{ab initio} calculations~\cite{wang.ye.ea_2021,kulik.hammerschmidt.ea_2022}. By training on large datasets generated from \textit{ab initio} calculations, \ac{ML} techniques can parameterize Hamiltonians, leading to accurate and computationally efficient models.

\subsubsection{Lattice vibrations: phonon modes}
Lattice vibrations in solid-state materials, known as phonon modes, provide critical insights into various materials, such as thermal conductivity, electron-phonon coupling, and optical response. With the \ac{BOA} (see Sec.~\ref{subsubsec:separation-electrons-nuclei}) and under the harmonic approximation, the nuclear Hamiltonian [Eq.~\eqref{eq:matter-nuclear-BOA}] becomes significantly simplified. In the harmonic approximation, the potential energy surface $\epsilon_{0}(\underline{\bfR})$ around the equilibrium nuclear configuration $\underline{\bfR_{0}}$ is expanded as 
\begin{equation}
\epsilon_{0}(\underline{\bfR})\approx \epsilon_{0}(\underline{\bfR_{0}}) + \frac{1}{2}\sum_{I,J,\alpha,\beta} C_{IJ}^{\alpha\beta} \Delta R_{I}^{\alpha} \Delta R_{J}^{\beta},
\end{equation}
where $\Delta R_I^\alpha = R_I^\alpha - R_{I,0}^\alpha$ is the nuclear displacement from equilibrium and $C_{IJ}^{\alpha\beta}$ is the matrix element for the inter-atomic force constant, which describes the change of the force on the $I$th nuclei/ions along the $\alpha$th Cartesian coordinate due to the change of the $J$th nuclei/ions along the $\beta$ Cartesian component,
\begin{equation}
C_{IJ}^{\alpha\beta} = \frac{\partial^{2}\epsilon_{0}(\{\bfR\})}{\partial R_{I}^{\alpha}\partial R_{J}^{\beta}} \Big|_{\underline{\bfR}=\underline{\bfR_{0}}}.
\end{equation}
Using this approximation, the nuclear Hamiltonian reduces to a system of independent harmonic oscillators, which can be described by the normal modes via the following equation, 
\begin{equation}
    \sum_{J,\beta}\left(C_{IJ}^{\alpha\beta}-M_{I}\omega^{2}\delta_{I,J}\delta_{\alpha,\beta}\right)U_{J}^{\beta} = 0,
\end{equation}
where $\omega$ is the normal mode frequency, $M_{I}$ is the mass of the $I$th atom, and $U_{I}^{\alpha}$ is the displacement patterns for the $\alpha$th Cartesian component of the $I$th atom.
Once the inter-atomic force constants are known, the normal mode frequencies and corresponding displacement patterns can be computed. 
Two main approaches are used to calculate the inter-atomic force constants. One is the finite difference method, which evaluates the force constants by calculating the changes in atomic forces due to small displacements of nuclei from their equilibrium positions~\cite{togo.chaput.ea_2023}. The other is \ac{DFPT}, which is a linear-response approach within \ac{DFT} that computes dynamical matrices (and then the force constants) and other phonon-related properties~\cite{baroni.degironcoli.ea_2001,baroni.giannozzi.ea_2010}.

%
\ac{DFPT} extends the self-consistent formalism of \ac{DFT} to include the linear response of the electronic system to nuclear displacements. In \ac{DFT}, the \ac{KS} potential $v_{\text{KS}}(\bfr)$ and the electron density $\rho(\bfr)$ are determined iteratively in a self-consistent cycle: $v_{\rm{KS}}(\bfr)\leftrightarrows\rho(\bfr)$, where $\rightarrow$ indicates using \ac{KS} potential $v_{\rm{KS}}(\bfr)$ to obtain the \ac{KS} wave functions, which are used to construct the electron density $\rho(\bfr)$, while $\leftarrow$ indicates using the electron density $\rho(\bfr)$ to update the \ac{KS} potential $v_{\rm{KS}}(\bfr)$. 
In \ac{DFPT}, a similar self-consistent cycle is used to compute the linear response of these quantities due to nuclear displacements: $\partial_{\mu}\rho(\bfr)$, together with that of the \ac{KS} potential due to the nuclear displacement: $\partial_{\mu} v_{\rm{KS}}(\bfr)\leftrightarrows \partial_{\mu}\rho(\bfr)$, where $\partial_\mu$ denotes the derivative with respect to the $\mu$th nuclear coordinate, $\rightarrow$ indicates using the linear response of the \ac{KS} potential $\partial_{\mu} v_{\rm{KS}}(\bfr)$ to obtain that of the \ac{KS} wave functions, which are used to construct the linear response of the electron density $\partial_{\mu}\rho(\bfr)$, while $\leftarrow$ means using the linear response of the electron density $\partial_{\mu}\rho(\bfr)$ to compute the linear response of the \ac{KS} potential $\partial_{\mu} v_{\rm{KS}}(\bfr)$. 
Once the linear response of the \ac{KS} potential $\partial_\mu v_{\text{KS}}(\bfr)$ and the electron density $\partial_\mu \rho(\bfr)$ is obtained, \ac{DFPT} allows the calculation of the electron-phonon coupling, a critical parameter in describing superconductivity, thermal transport, and other material properties~\cite{baroni.degironcoli.ea_2001,giustino_2017}.

The harmonic approximation and \ac{DFPT} enable the efficient calculation of phonon modes and electron-phonon interactions in solid-state materials. These methods, combined with \ac{DFT}, form a starting framework for understanding lattice dynamics, providing key insights into thermal and electronic properties of materials. By linking atomic vibrations to macroscopic phenomena, they play an essential role in materials modeling and design.
While this review focuses on the treatment of phonons within the adiabatic approximation, it is worth noting that significant advancements have been made in the study of non-adiabatic~\cite{lazzeri.mauri_2006,pisana.lazzeri.ea_2007,bauer.falter_2009,calandra.profeta.ea_2010,ponce.gillet.ea_2015,caruso.hoesch.ea_2017,verdi.caruso.ea_2017,miglio.brousseau-couture.ea_2020,hu.liu.ea_2022,girotto.novko_2023,berges.girotto.ea_2023} and anharmonic~\cite{souvatzis.eriksson.ea_2009,tadano.tsuneyuki_2015,bianco.errea.ea_2017,vanroekeghem.carrete.ea_2021,zhao.zeng.ea_2021,hellman.abrikosov.ea_2011,hellman.steneteg.ea_2013,hellman.abrikosov_2013,zhou.hellman.ea_2018,yang.hellman.ea_2020,bottin.bieder.ea_2020,erba.maul.ea_2019,zacharias.giustino_2016,zacharias.giustino_2020,zacharias.volonakis.ea_2023} phonon effects from first principles. These phenomena are crucial for understanding electron-phonon interactions, thermal transport, and dynamical stability in materials under extreme conditions.

\subsubsection{Hamiltonian for coupled electron-phonon systems}
After decoupling nuclei/ions and electrons and quantizing nuclear vibrations within the harmonic approximation, we can re-assemble the problem by introducing an effective coupled electron-phonon Hamiltonian of the form~\cite{bruus.flensberg_2004},
\begin{equation}
\hat{H}_{\rm{el-ph}} = \hat{H}_{\rm{el}} + \hat{H}_{\rm{ph}} + \hat{V}_{\rm{el-ph}},
\end{equation}
where $\hat{H}_{\rm{el}}$ is the electronic Hamiltonian, $\hat{H}_{\rm{ph}}$ is the phononic Hamiltonian, and $\hat{V}_{\rm{el-ph}}$ is the electron-phonon interaction. The electronic Hamiltonian in the Bloch basis $\{\ket{n\bfk}\}$ with the second quantization notation is 
\begin{equation}
\hat{H}_{\rm{el}} = \sum_{n\bfk} \epsilon_{n\bfk} \hat{c}_{n\bfk}^{\dagger}\hat{c}_{n\bfk},
\end{equation}
where we omit the spin index for simplicity, and $n$ and $\bfk$ are the band index and electronic crystal momentum (in the first \ac{BZ}), respectively. $\hat{c}_{n\bfk}$ ($\hat{c}_{n\bfk}^{\dagger}$) is the annihilation (creation) operator for the Bloch state $\ket{n\bfk}$ with the energy $\epsilon_{n\bfk}$, and the electronic operators follow the anticommutation relation, i.e., $\{\hat{c}_{n\bfk},\hat{c}_{m\bfk'}^{\dagger}\} = \delta_{n,m}\delta_{\bfk,\bfk'}$.
The phononic Hamiltonian in the normal mode basis $\{\ket{\bfq\nu}\}$, where $\bfq$ and $\nu$ are the phononic crystal momentum and mode index, respectively, with the second quantization notation is 
\begin{equation}
\hat{H}_{\rm{ph}} = \sum_{\nu\bfq}\hbar\omega_{\bfq\nu}\left(\hat{b}_{\bfq\nu}^{\dagger}\hat{b}_{\bfq\nu}+\frac{1}{2}\right).
\end{equation}
Here $\hat{b}_{\bfq\nu}$ ($\hat{b}_{\bfq\nu}^{\dagger}$) is the annihilation (creation) of the phonon at the crystal momentum $\bfq$ with the mode index $\nu$, and the phononic operators follow the commutation relation, i.e., $[\hat{b}_{\bfq\nu},\hat{b}_{\bfq'\nu'}^{\dagger}] = \delta_{\bfq,\bfq'}\delta_{\nu,\nu'}$.
The electron-phonon interaction is written in terms of electronic and phononic operators in the second quantization notation is~\cite{giustino_2017}
\begin{equation}
\hat{V}_{\rm{el-ph}} = \frac{1}{N_{p}^{1/2}}\sum_{\bfk,mn}\sum_{\bfq,\nu}g_{mn,\nu}(\bfk,\bfq)\hat{c}_{m\bfk+\bfq}^{\dagger}\hat{c}_{n\bfk}(\hat{b}_{\bfq\nu}+\hat{b}^{\dagger}_{-\bfq\nu}),
\end{equation}
where $N_{p}$ is the number of unit cell in the \ac{BvK} supercell and $g_{mn,\nu}(\bfk,\bfq) = \mel{u_{m\bfk+\bfq}}{\Delta_{\bfq\nu}v_{\rm{KS}}}{u_{n\bfk}}$ is the electron-phonon coupling that couples the Bloch state $\ket{n\bfk}$ to another Bloch state $\ket{m\bfk+\bfq}$ via the perturbation potential $\Delta_{\bfq\nu}v_{\rm{KS}}$ due to the phonon mode $\bfq\nu$.

\subsection{\textit{Ab initio} approaches to solve the Pauli-Fierz Hamiltonian for solid-state materials}
\label{subsec:abinitio-lightmatter}
Tackling the \ac{PF} Hamiltonian for solid-state materials presents a formidable challenge due to the intrinsic complexity of the system, which now includes not only the interactions among electrons and nuclei/ions but also the interactions among those charged particles and the quantized photon field. While solving the matter Hamiltonian alone is already a difficult task (see Sec.~\ref{subsec:abinitio-matter}), the inclusion of photons adds a new layer of complexity, requiring sophisticated approaches to capture the full spectrum of interactions accurately. In certain idealized cases, such as the \ac{pHEG} (free and non-interacting electrons in the \ac{LWA}), exact solutions of the \ac{PF} Hamiltonian can be found~\cite{schafer.buchholz.ea_2021,rokaj.ruggenthaler.ea_2022}. However, for more realistic and heterogeneous solid-state systems, one often uses adapted decoupling strategies and approximations to make the problem tractable.

\subsubsection{Separation of the Pauli-Feirz Hamiltonian into "slow" and "fast" subsystems}\label{subsubsec:separation-PF-Hamiltonian}
To solve the coupled electron-nucleus-photon system (\ac{PF} Hamiltonian), we can decouple its components as we briefly discussed in Sec~\ref{subsubsec:Approximation_strategies}, similar to the treatment of the matter Hamiltonian (Sec.~\ref{subsec:abinitio-matter}). In addition to electronic and nuclear coordinates, $\underline{\bfr}$ and $\underline{\bfR}$, respectively, we now have additional photonic coordinates $\underline{q}$. Following the matter Hamiltonian approach (see Sec.~\ref{subsec:abinitio-matter}), we can rewrite the problem in terms of conditional wave functions and then approximately decouple the different subsystems. Such an approach allows us to consider nuclear coordinates as parameters for the electronic Hamiltonian. However, the presence of photons introduces a new factor: the frequency of photons can vary across the energy scale of nuclear to electronic subsystems. Which grouping and partitioning of these coordinates is useful to devise reasonable approximations, depends on the specific physical scenario. Depending on the photon frequency, we can group photons with either nuclei/ions or electrons. For photons with frequencies comparable to nuclear vibrational energy, grouping them with nuclei/ions creates a "slow" subsystem, suitable for strong vibrational coupling~\cite{flick.ruggenthaler.ea_2017,flick.appel.ea_2017}. Conversely, for photons with frequencies comparable to electronic energy, grouping them with electrons forms a "fast" subsystem, suitable for strong electronic coupling~\cite{galego.garcia-vidal.ea_2015, feist.galego.ea_2018}. To solve the \ac{PF} Hamiltonian, we can introduce basis sets for each subsystem~\cite{schafer.ruggenthaler.ea_2018}. Analogous to the electron-nucleus wave function for the matter Hamiltonian [Eq.~\eqref{eq:matter-eigenstate-BH-expansion}], the total exact wave function for the $i$th eigenstate of the \ac{PF} Hamiltonian for a coupled electron-nucleus-photon system can be expanded as 
\begin{equation}\label{eq:exact-wfc-partition}
\Psi_{i}(\underline{\bfr},\underline{\bfR},\underline{q}) = \sum_{j=0}^{\infty}\chi_{ij}(\underline{\bfR},\underline{q})\psi_{j}(\underline{\bfr};\{\underline{\bfR},\underline{q}\}) = \sum_{j=0}^{\infty}\tilde{\chi}_{ij}(\underline{\bfR})\tilde{\psi}_{j}(\underline{\bfr},\underline{q};\{\underline{\bfR}\}).
\end{equation}
Here the first approach is the \ac{cBOA}~\cite{flick.ruggenthaler.ea_2017,flick.appel.ea_2017} where we group the nuclear and photonic coordinates, and treat these coordinates as parameters for the electronic subsystem with the eigenstates denoted as $\psi_{j}(\underline{\bfr};\{\underline{\bfR},\underline{q}\})$. The second is the polaritonic surface partition~\cite{galego.garcia-vidal.ea_2015, feist.galego.ea_2018}, where we group the photonic and electronic coordinates, and treat the nuclear coordinates as parameters for the coupled, composite electron-photon subsystem with the eigenstates denoted as $\tilde{\psi}_{j}(\underline{\bfr},\underline{q};\{\underline{\bfR}\})$. A further option is to then also decouple the photons from the electrons, which leads to the explicit-polariton partitioning~\cite{schafer.ruggenthaler.ea_2018}.

%
To derive the associated Hamiltonians for the grouped coordinates, one can use the exact wave function [Eq.~\eqref{eq:exact-wfc-partition}] and corresponding orthonormal basis sets for each subsystem, treating the other coordinates as parameters. These approaches have been applied to the \ac{PF} Hamiltonian in the \textit{length-gauge} form, which involves the total dipole moment or polarization of electrons and nuclei/ions. However, for solid-state materials or periodic systems, calculating the electronic dipole moment can be nontrivial, requiring the computation of the Berry phase~\cite{vanderbilt_2018}. Moreover, in the length-gauge formalism, the matter wave function is not purely a matter quantity but a mixed matter-photon quantity~\cite{rokaj.welakuh.ea_2018,schafer.ruggenthaler.ea_2020}. In contrast, the velocity-gauge formulation of the \ac{PF} Hamiltonian, which is the focus of this work, has the disadvantage that the coupling is described via the momenta of the charged particles. This implies that the Born-Huang expansion contains more non-adiabatic elements, making it less convenient for further approximations~\cite{schafer.ruggenthaler.ea_2018}. On the other hand, for solid-state materials under ambient conditions, the nuclei/ions typically vibrate around their equilibrium positions. Therefore, we can often focus on the nuclear ground state wave function, neglecting couplings between different eigenstates, which is the essence of the adiabatic approximation. To account for non-adiabatic effects, methods such as exact factorization~\cite{abedi.khosravi.ea_2018} or multi-trajectory approaches~\cite{hoffmann.appel.ea_2018,hoffmann.schafer.ea_2019} are then potentially more convenient in practice.

%
Given the complexity, an initial approach involves treating the nuclear subsystem classically, using dynamical equations that describe classical particles influenced by forces from the electronic and photonic subsystems~\cite{flick.narang_2018}, or from the combined electron-photon subsystem. Once we treat the nuclei/ions as classical objects and fix their coordinates, known as the \ac{CNA}, the \ac{PF} Hamiltonian with the physical electron mass and a few effective photon modes [Eq.~\eqref{eq:HPF-eff-modes}] and with the nuclear coordinates as parameters becomes 
\begin{equation}\label{eq:HPF-with-eff-modes-CNA}
\begin{aligned}
    \hat{H}_{\rm{PF}}^{\rm{eff,CNA}} & = \sum_{i=1}^{N_{e}}\left[\frac{\left(\hbfp_{i}+|e|\hbfAperp^{\rm{eff}}(\bfr_{i})\right)^{2}}{2m_{e}} + \frac{|e|\hbar}{2m_{e}}\paulimat_{i}\cdot\hbfB(\bfr_{i})-\sum_{I=1}^{N_{n}}\frac{Z_{I} e^{2}}{4\pi\vacep|\bfr_{i}-\bfR_{I}|}\right] + \frac{1}{2}\sum_{i\neq j}^{N_{e}}\frac{e^{2}}{4\pi\vacep|\bfr_{i}-\bfr_{j}|} \\
    & + \sum_{\alpha=1}^{M_{p}} \hbar \omega_{\alpha} \hat{a}_{\alpha}^{\dagger}\hat{a}_{\alpha} + E_{\rm{nuclei}}(\underline{\bfR}),
\end{aligned}
\end{equation}
where $E_{\rm{nuclei}}(\underline{\bfR})$ is a constant energy shift, depending on the nuclear coordinates $\underline{\bfR}$. Such a \ac{PF} Hamiltonian, originally constructed using full minimal coupling (where the photon and classical \ac{EM} field are spatially dependent), can be naturally reduced within the \ac{LWA}. This simplified form aligns well with standard solid-state theory for solid-state materials subjected to external \ac{EM} fields, making it a suitable choice for our analysis of the ground state of cavity-engineered solid-state materials. To go beyond \ac{LWA} for solid-state materials is non-trivial since we have now several conflicting symmetries that need to be treated consistently~\cite{svendsen.ruggenthaler.ea_2023}. For instance, to employ Bloch's theorem as done in Sec.~\ref{subsec:abinitio-matter}, we need a periodic Hamiltonian, but the enhanced modes of a cavity might not respect this symmetry. The resulting \ac{PF} Hamiltonian within the \ac{LWA}, where the magnetic field disappears, becomes
\begin{equation}\label{eq:HPF-with-eff-modes-egamma}
\begin{aligned}
    \hat{H}_{\rm{PF}}^{e\gamma,\rm{LWA}} & = \sum_{i=1}^{N_{e}}\left[\frac{\left(\hbfp_{i}+|e|\hbfAperp^{\rm{eff,LWA}}\right)^{2}}{2m_{e}}+V_{\rm{nuclei}}(\bfr_{i})\right] + \frac{1}{2}\sum_{i\neq j}^{N_{e}}\frac{e^{2}}{4\pi\vacep|\bfr_{i}-\bfr_{j}|} + \sum_{\alpha=1}^{M_{p}} \hbar \omega_{\alpha} \hat{a}_{\alpha}^{\dagger}\hat{a}_{\alpha},
\end{aligned}
\end{equation}
where we neglect the constant energy shift for the fixed nuclear coordinates and write $V_{\rm{nuclei}}(\bfr) = -\sum_{I=1}^{N_{n}}\frac{Z_{I} e^{2}}{4\pi\vacep|\bfr_{i}-\bfR_{I}|}$. The Hamiltonian presented here is an effective Hamiltonian that describes a coupled electron-photon ($e\gamma$) system within the \ac{LWA}. The electrons are subjected to an external attractive Coulomb potential from the fixed nuclei/ions. In the absence of explicit photon contributions, the Hamiltonian [Eq.~\eqref{eq:HPF-with-eff-modes-egamma}] reduces to the electronic Hamiltonian $\hat{H}_{e}$, up to a constant energy shift, as described in Sec~\ref{subsec:abinitio-matter}. Even with the \ac{CNA} and \ac{LWA}, solving the effective Hamiltonian [Eq.~\eqref{eq:HPF-with-eff-modes-egamma}] remains computationally challenging due to the numerous degrees of freedom associated with both electrons and photons. Similar to how \ac{DFT} is used to solve many-body electronic Hamiltonians, \ac{QEDFT} can be used to tackle coupled light-matter Hamiltonians. While alternative numerical methods, such as Green's functions~\cite{amelio.korosec.ea_2021} or quantum Monte Carlo~\cite{weber.vinasbostrom.ea_2023}, are available for solving light-matter interactions, we focus on \ac{QEDFT} due to its numerical efficiency and accuracy in the study of solid-state materials.

\subsubsection{Quantum-electrodynamical density-functional theory (QEDFT)}\label{subsubsec:QEDFT}
\ac{QEDFT} extends traditional density-functional methods (Sec.~\ref{subsec:abinitio-matter}) into the domain of \ac{QED}. In the early development of \ac{QEDFT}, the absence of a well-defined Hamiltonian and unique wave functions in relativistic \ac{QED} led to approaches that were largely perturbative or relied on approximations such as the Breit approximation~\cite{breit_1932}. A significant step forward came when Rajagopal~\cite{rajagopal_1978} extended the ideas of Runge and Gross to \ac{QED}, introducing a time-dependent density-functional framework. This laid the foundation for the standard form of \ac{QEDFT}, which is primarily concerned with the low-energy regime~\cite{tokatly_2013,ruggenthaler.flick.ea_2014}. The time-dependent formulation of \ac{QEDFT} highlights its relationship to collective-variable theories and is expressed as a local force-balance equation involving quantum momentum-stress and interaction tensors~\cite{ruggenthaler.flick.ea_2014}. In the coarse-grained mean-field limit, this equation reduces to a Navier-Stokes-type equation, echoing the classical Abrahams model of light-matter interactions~\cite{spohn_2004}. Additionally, this time-dependent framework can be connected to the quantum Langevin equation with memory terms, a common tool in quantum optics~\cite{haroche.raimond_2006}.

The fundamental theorems of \ac{QEDFT} assert that the memory terms in the equations of motion for the current density and vector potential, often in the Coulomb gauge, are uniquely defined. 
In the static limit, this leads to a local-force equilibrium~\cite{ruggenthaler.flick.ea_2014,jestadt.ruggenthaler.ea_2019} similar as in standard \ac{DFT}~\cite{tchenkoue.penz.ea_2019,tancogne-dejean.penz.ea_2024}, which can be reformulated as a minimal-energy principle. 
Essentially, instead of relying on the total coupled light-matter wave function, which is generally unknown, \ac{QEDFT} allows us to uniquely determine all physical quantities using only electron and photon (potentially vector-valued) densities. To perform the mapping between the external control fields and the internal fields (electron and photon densities) within \ac{QEDFT}, there are many different choices~\cite{ruggenthaler.flick.ea_2014}. Which one depends on the \ac{PF} Hamiltonian and external control fields used to formulate the corresponding \ac{QEDFT}~\cite{ruggenthaler.flick.ea_2014}. For instance, if we consider a time-dependent problem in full minimal coupling Coulomb gauge and consider all possible external four vector potentials as well as transverse external charge currents, then the corresponding collective matter variable is the full four current density and the collective photon variable is the transverse vector potential~\cite{ruggenthaler.flick.ea_2014,jestadt.ruggenthaler.ea_2019}. On the other hand, if we consider a static problem in the full minimal coupling Coulomb gauge, we usually consider only external scalar potentials and external transverse charge currents. In this case we have the zero component of the four vector current density, i.e., the charge density $\rho(\bfr)$ and the expectation value of the vector-potential operator $\langle \hbfAperp(\bfr) \rangle $~\cite{ruggenthaler_2017,penz.tellgren.ea_2023}. We discuss the static minimal-coupling case in a little more detail below. To calculate these collective variables and approximate the unknown \ac{xc} fields, \ac{QEDFT} maps the original interacting light-matter system onto an auxiliary non-interacting system that reproduces the same collective matter and photon variables, similar to the \ac{KS} scheme in standard \ac{DFT} (Sec.~\ref{subsec:abinitio-matter}). In the following, the \ac{QEDFT} mapping is applied to the electron-photon coupled system; however, the \ac{QEDFT} theorems can be extended to electron-nucleus-photon coupled systems~\cite{jestadt.ruggenthaler.ea_2019}.

\subsubsection{Build up the Kohn-Sham auxiliary system}
%
For the static case, we consider the \ac{PF} Hamiltonian in the \ac{CNA} but generalize it to arbitrary external scalar potentials $\phi_{\rm ext}(\bfr)$ and general external (transverse due to Coulomb gauge) charge currents $\mathbf{j}_{\rm ext}$. That is, we have
\begin{equation} \label{eq:HPF-continuum-static-e-gamma}
\begin{aligned}
\hat{H}_{\rm{PF}} & = \sum_{i=1}^{N_{e}}\left[\frac{1}{2\mebare} \left(\hbfp_{i}+|e|\hbfAperp(\bfr_{i})\right)^{2} + \frac{|e|\hbar}{2\mebare}\paulimat_{i}\cdot\hbfB(\bfr_{i}) -|e| \phi_{\rm ext}(\bfr)\right] + \frac{1}{2}\sum_{i\neq j}^{N_{e}}\frac{e^{2}}{4\pi\vacep|\bfr_{i}-\bfr_{j}|} \\
& + \sum_{\lambda=1}^{2}\int \hbar \omega_{\bfk} \hat{a}^{\dagger}(\bfk,\lambda)\hat{a}(\bfk,\lambda)d\bfk -\int d^{3}r\ \mathbf{j}_{\rm ext}(\bfr)\cdot\hbfAperp(\bfr).
\end{aligned}
\end{equation}
We note that by Maxwell's Eq.~\eqref{eq:MWE-BEJ}, the external classical current $\mathbf{j}_{\rm ext}(\bfr)$ corresponds to a classical static magnetic field $\mathbf{B}_{\rm ext}(\bfr)$. That is, the external current is a source term for the (quantized) \ac{EM} field and hence adding also an extra external magnetic field is physically redundant~\cite{ruggenthaler_2017,penz.tellgren.ea_2023}. 
It can then be shown that there is a bijective mapping between $(\mathbf{j}_{\rm ext}(\bfr), -|e|\phi_{\rm ext}(\bfr) =: v_{\rm ext}(\bfr)) \leftrightarrow (\langle \hbfAperp(\bfr) \rangle = \mathbf{A}_{\perp}(\bfr), \rho(\bfr) )$ and we can replace the wave function of the ground-state problem by merely the internal pair $(\mathbf{A}_{\perp}(\bfr), \rho(\bfr) )$.

A similar mapping can be established for a mean-field coupled auxiliary system of the form 
\begin{align}\label{eq:HMP-full-minimal}
\hat{H}_{\rm{MP}} & = \sum_{j=1}^{N_{e}}\left[\frac{1}{2m_{e,\rm{b}}}\left(\hbfp_{j}+|e|\bfAperp(\bfr_{j},t)\right)^{2} + \frac{|e|\hbar}{2\mebare}\paulimat_{i}\cdot \bfB(\bfr_{i})  +  v_{\rm{s}}(\bfr_{j})\right] + \int d^{3}r\ \mathbf{J}_{\rm{MP}}(\bfr)\cdot\bfAperp(\bfr) \nonumber \\
& + \sum_{\lambda=1}^{2}\int \hbar \omega_{\bfk} \hat{a}^{\dagger}(\bfk,\lambda)\hat{a}(\bfk,\lambda)d\bfk -\int d^{3}r\ \mathbf{J}_{\rm{MP}}(\bfr)\cdot\hbfAperp(\bfr) -\int d^{3}r\ \mathbf{j}_{\rm{s}}(\bfr)\cdot\hbfAperp(\bfr),
\end{align}
where $\bfB(\bfr) = \curl\bfA_{\perp}(\bfr)$, the last term in the first line avoids double counting of the mean-field interaction energy and $\mathbf{J}_{\rm{MP}}(\bfr)$ is the total charge current density calculated from the non-interacting matter subsystem. This mean-field Hamiltonian is denoted as the \ac{MP} Hamiltonian~\cite{jestadt.ruggenthaler.ea_2019}. Enforcing then that both, the fully coupled system of Eq.~\eqref{eq:HPF-continuum-static-e-gamma} and the auxiliary mean-field system of Eq.~\eqref{eq:HMP-full-minimal}, generate the same density $\rho(\bfr)$ and vector potential $\boldsymbol{A}_{\perp}(\bfr)$, we can define the corresponding \ac{KS} potential $v_{\rm s}(\bfr) = v_{\rm ext}(\bfr) + v_{\rm Mxc}(\bfr)$ as well as \ac{KS} current $\mathbf{j}_{\rm s}(\bfr)= \mathbf{j}_{\rm ext}(\bfr) + \mathbf{j}_{\rm xc}(\bfr)$.  We note that here $v_{\rm Mxc}(\bfr)$ and $\mathbf{j}_{\rm xc}(\bfr)$ depend highly-non-linearly on $\rho(\bfr)$ and $\mathbf{A}_{\perp}(\bfr)$ itself and it is those effective fields that need to be approximated in practice. We highlight that $v_{\rm Mxc}(\bfr) = v_{\rm Hxc}(\bfr) + v_{\rm pxc}(\bfr)$ contains contributions from the longitudinal Coulomb interaction (similar to standard \ac{DFT}) $v_{\rm Hxc}(\bfr)$ and transverse electron-photon contributions $v_{\rm pxc}(\bfr)$. Above, we keep the bare mass $m_{e,\rm{b}}$ and the full continuum of photon modes. We can also use the physical mass $m_{e}$ and then effective photon modes, as discussed in Sec~\ref{subsubsec:Approximation_strategies}. In the following, we use the physical mass version, but similar conclusions can be applied to the bare mass version.

Since we have a non-interacting problem to solve, we can recast it in the form of many coupled single-particle equations that are also coupled to the static Maxwell's equation, i.e., 
\begin{align}
& \label{eq:static-QEDFT-KS-EQ} \left[\frac{1}{2m_{e}}\left(-i\hbar\nabla+|e|\bfAperp(\bfr)\right)^{2}+v_{\rm{ext}}(\bfr)+v_{\rm{Mxc}}(\bfr) + \frac{|e|\hbar}{2m_{e}}\paulimat\cdot\bfB(\bfr)\right]\phi_{j}(\bfr,\tau) = \epsilon_{j}\phi_{j}(\bfr,\tau), \\ 
& \label{eq:static-QEDFT-KS-A} \qquad \qquad -\nabla^{2}\bfAperp(\bfr) = \mu_{0}\left(\mathbf{j}_{\rm{ext}}(\bfr)+\mathbf{J}_{\rm{MP},\perp}(\bfr)+\mathbf{j}_{\rm{xc}}(\bfr)\right).
\end{align}
Here the density is then constructed (assuming the \ac{KS} wave function is a single Slater-determinant) from $\rho(\bfr) = \sum_{j, \tau}|\phi_{j}(\bfr,\tau)|$, where $\tau \in \{\uparrow, \downarrow\}$ corresponds to the two spin states of the auxiliary \ac{KS} orbitals.

%
As discussed in Secs.~\ref{subsec:LWA}, for solid-state materials the \ac{LWA} is important not to run into symmetry mismatches and to make the problem more tractable. This simplifies the vector potential $\bfAperp(\bfr)\to\bfAperp^{\rm LWA}$ and the magnetic field $\bfB(\bfr)$ vanishes in the single-particle Pauli equation [Eq.~\eqref{eq:static-QEDFT-KS-EQ}]. That is, in the \ac{LWA} we have
\begin{equation}
\left[\frac{1}{2m_{e}}\left(-i\hbar\nabla+|e|\bfAperp^{\rm{LWA}}\right)^{2}+v_{\rm{ext}}(\bfr)+v_{\rm{Mxc}}(\bfr)\right]\phi_{j}(\bfr,\tau) = \epsilon_{j}\phi_{j}(\bfr,\tau).
\end{equation}
Furthermore, this approximation implies a change in the Maxwell's Eq.~\eqref{eq:static-QEDFT-KS-A}, since now each vectorial eigenmode of the (quantized) Maxwell field couples as a spatially constant vector field to the total current~\cite{ruggenthaler.flick.ea_2014}. However, if we are interested in only the density $\rho(\bfr)$, we can absorb the solution of this Maxwell's equation by a change in gauge~\cite{lu.ruggenthaler.ea_2024}. That is, since $\bfAperp^{\rm LWA}$ is just a constant we can absorb this phase in the orbitals and still have the same density. Yet, for the reconstruction of other (specifically photonic) observables, we would need to keep track of the gauge change and the corresponding $\mathbf{j}_{\rm{xc}}(\bfr)$. With this subtlety in mind, we can then recast the \ac{KS} equations for each periodic material in terms of Bloch functions and find for each $\bfk$  
\begin{equation}\label{eq:MKS-Hamiltonian-each-k}
 \left(\frac{1}{2m_{e}}(-i\hbar\nabla+\hbar\bfk)^{2} + v_{\rm{ext}}(\bfr) + v_{\rm{Hxc}}(\bfr) + v_{\rm{pxc}}(\bfr)\right)u_{n\bfk}(\bfr) = \epsilon_{n\bfk}u_{n\bfk}(\bfr). 
\end{equation}
We note that since the Hamiltonian is spin-independent, we only need to consider the spatial part of the non-linear single-particle equations. We further note that if we consider a time-dependent problem in the \ac{LWA}, we need to keep the Maxwell's equation explicitly~\cite{tokatly_2013,ruggenthaler.flick.ea_2014,lu.ruggenthaler.ea_2024}.

\subsubsection{Electron-photon exchange-correlation functionals}
%
To perform practical \ac{QEDFT} simulations, we need approximations for the nonlinear \ac{xc} fields. These approximations can be categorized into \ac{xc} potentials arising from the longitudinal Coulomb interaction and functionals for transverse matter-photon interactions. One way is to find the corresponding \ac{xc} energy functionals, the density functional derivative of which gives the \ac{xc} fields used in the \ac{KS} equations to reproduce the collective electron and photon variables, i.e., $\rho(\bfr)$ and $\bfAperp(\bfr)$, respectively, of the original system~\cite{pellegrini.flick.ea_2015,flick_2022,novokreschenov.kudlis.ea_2023}. Another way is to use the local force balance equation to find the \ac{xc} fields directly without using the energy functional~\cite{tchenkoue.penz.ea_2019,tancogne-dejean.penz.ea_2024,schafer.buchholz.ea_2021,lu.ruggenthaler.ea_2024}. Such an approach can avoid the issues such as the differentiability issue for energy functionals~\cite{lammert_2007}, the causality issue for action functionals in the time-dependent cases~\cite{vanleeuwen_2001}, and the numerical issue for \ac{OEP} procedure for orbital-dependent functionals~\cite{talman.shadwick_1976}.

%
Most existing electron-photon \ac{xc} functionals are developed within the \textit{length-gauge} form of the \ac{PF} Hamiltonian for finite systems~\cite{pellegrini.flick.ea_2015,flick_2022,novokreschenov.kudlis.ea_2023}. These functionals are primarily based on perturbation theory and are well-suited for moderate light-matter coupling regimes~\cite{pellegrini.flick.ea_2015}. However, they may not be reliable in strong light-matter coupling regimes~\cite{flick.schafer.ea_2018}. Recent efforts have focused on developing a photon-random-phase-approximation functional beyond perturbation theory, but its applicability to realistic systems remains uncertain~\cite{novokreschenov.kudlis.ea_2023}. For extended systems, the velocity-gauge form is more natural. Currently, there is only one electron-photon \ac{xc} functional available in the velocity-gauge form, specifically the electron-photon exchange functional~\cite{schafer.buchholz.ea_2021}. This functional, derived from the local-force balance equation, performs well in strong or ultrastrong light-matter coupling regimes but requires an electron-photon correlation functional in the weak coupling regime to accurately reproduce the exact electron density~\cite{lu.ruggenthaler.ea_2024}. One challenge with the local force balance approach is to unambiguously determine the electron-photon \ac{xc} energy of the original system~\cite{lu.ruggenthaler.ea_2024}. Despite the challenges, the electron-photon exchange functional has been used to investigate how field fluctuations of photons within strongly coupled cavities modify electron densities and electronic band structures in solid-state materials. The electron-photon exchange functional has been implemented in our open-source code, OCTOPUS~\cite{tancogne-dejean.oliveira.ea_2020}, and has also been shown to reproduce the vacuum Rabi splitting in the hydrogen atom~\cite{lu.ruggenthaler.ea_2024}. This functional has also been extended to the linear response regime and integrated into the \ac{DFPT} framework to study cavity-modified phonon dispersion and electron-phonon couplings, which has been used to explore cavity-modified phonon-mediated superconductivity using the anisotropic Eliashberg equations~\cite{lu.shin.ea_2024a}.

\section{Realization of cavity materials engineering}\label{sec:experiments}
%
Beyond solid-state materials, notable progress has been achieved in polaritonic chemistry, where photon field fluctuations inside cavities have been used to modify the ground states of molecular systems~\cite{ebbesen.rubio.ea_2023}. These breakthroughs offer valuable insights and methodologies that could inspire future experiments on cavity-modified solid-state materials. Section~\ref{subsec:polaritonic-chemistry} reviews these developments and explores their potential relevance to solid-state systems, though readers primarily interested in solid-state phenomena may choose to skip this section. Moving to solid-state materials, Section~\ref{subsec:quantm-Hall-effect} investigates the impact of cavity field fluctuations on integer and fractional quantum Hall effects in \ac{2D} electron gases. In contrast to non-equilibrium laser-pump or Floquet-engineered phase transitions~\cite{disa.nova.ea_2021,delatorre.kennes.ea_2021,bao.tang.ea_2022}, Section~\ref{subsec:MIT-TaS2} highlights cavity-modified metal-to-insulator phase transitions. Finally, Section~\ref{subsec:ferromagnetism} examines the enhancement of ferromagnetism in an unconventional superconductor, demonstrating the diverse potential of cavity materials engineering.

\subsection{Insights from polaritonic chemistry}\label{subsec:polaritonic-chemistry}

\subsubsection{Overview of polaritonic chemistry}
%
Polaritonic (or also QED) chemistry~\cite{ebbesen_2016,ribeiro.martinez-martinez.ea_2018,feist.galego.ea_2018,herrera.owrutsky_2020,garcia-vidal.ciuti.ea_2021,dunkelberger.simpkins.ea_2022,ebbesen.rubio.ea_2023,mandal.taylor.ea_2023, ruggenthaler.sidler.ea_2023} explores the novel phenomena arising from the strong coupling of matter and light within a cavity at room temperature. This interdisciplinary field encompasses systems ranging from individual molecules to condensed phases. The widely investigated systems are individual molecules, which are chemically highly non-trivial as they contain not only various electronic and spin excitations, but also translation, rotational, and vibrational degrees of freedom. 
Polaritonic systems can be categorized based on the type of molecular excitation coupled to the cavity. Systems where the cavity is resonant with electronic excitations are termed \textit{electronic strong coupling}, while those resonant with vibrational degrees of freedom are called \textit{vibrational strong coupling}. While QED chemistry encompasses all phases of matter, experimental studies have predominantly focused on electronic strong coupling in solids and vibrational strong coupling in liquids~\cite{garcia-vidal.ciuti.ea_2021,simpkins.dunkelberger.ea_2023}.
The number of molecules coupled to the cavity also differentiates polaritonic systems. When only a few molecules interact strongly with a micro- or nano-cavity~\cite{skolnick.fisher.ea_1998,raimond.brune.ea_2001,vahala_2003,hugall.singh.ea_2018}, this is termed \textit{few-molecule} or \textit{local strong coupling}. While this regime has been studied in several experiments~\cite{santhosh.bitton.ea_2016,chikkaraddy.denijs.ea_2016,benz.schmidt.ea_2016}, most other experiments focus on a macroscopic number of molecules coupled strongly to a cavity, so-called \textit{collective strong coupling}.
Of course, further and more detailed distinctions along these various lines are possible but those are most common.

The first question that immediately arises is whether polaritonic chemistry can be considered a distinct research field from laser-controlled chemistry~\cite{shapiro.brumer_1994,dantus.lozovoy_2004}. 
At first glance, one could argue that any changes due to coupling to the cavity modes should have a certain similarity with coupling to lasers. Unlike laser-based approaches, which typically involve strong coupling of many photons to matter, polaritonic chemistry often operates \textit{in the dark} -- the cavity is not pumped and the matter system is just coupled to the \textit{vacuum} and/or \textit{thermal} (incoherent) fluctuations of the photon field. This leads to equilibrium changes rather than transient effects. Moreover, polaritonic chemistry can potentially affect entire molecular ensembles~\cite{sidler.ruggenthaler.ea_2022,campos-gonzalez-angulo.poh.ea_2023}, challenging the assumption of independent molecules in traditional chemical kinetics. This fact brings polaritonic chemistry already very close to cavity-materials engineering. Indeed, many novel aspects in polaritonic chemistry are borrowed from solid-state physics~\cite{sidler.ruggenthaler.ea_2024} and the line between both fields is becoming more blurred with time.

%
Polaritonic chemistry demonstrates that carefully engineered and vibrationally tuned optical cavities can modify chemical processes by altering the vacuum and thermal fluctuations of the photon field. This approach offers a novel tool for chemical manipulation, comparable to traditional methods like pressure, temperature, or solvent selection. Even subtle changes in reaction rates can have significant implications for industrial processes, where many complex steps are needed to achieve a specific product via chemical synthesis. Beyond its potential applications, polaritonic chemistry pushes the boundaries of our understanding of light-matter interactions~\cite{ebbesen_2016,ruggenthaler.tancogne-dejean.ea_2018,garcia-vidal.ciuti.ea_2021}. Traditionally, chemical properties are attributed to individual molecules or their immediate nanoscale environment. However, polaritonic chemistry challenges this local perspective by demonstrating that macroscopic properties of the ensemble can influence molecular behavior~\cite{ebbesen_2016,ruggenthaler.sidler.ea_2023,sidler.ruggenthaler.ea_2022}. Furthermore, polaritonic chemistry reveals that even weak light-matter interactions in complex systems can lead to strong effects at equilibrium and entangled light-matter systems, challenging the notion of equilibrium states governed by perturbative approaches. \ac{QED} chemistry offers a new avenue to explore the intricacies of \ac{QED} beyond traditional scattering theory~\cite{ruggenthaler.tancogne-dejean.ea_2018,ruggenthaler.sidler.ea_2023}.

\subsubsection{Theoretical methods and models for polaritonic chemistry}
To well describe a molecular system coupled to a cavity, a ground state for the matter system coupled to a continuum of photon modes is essential for a theoretical description. The \ac{PF} Hamiltonian, expressed in full minimal coupling~\cite{spohn_2004}, serves as a suitable starting point for describing coupled light-matter systems. 
Recent studies have established its mathematical consistency~\cite{bach.frohlich.ea_1995,fefferman.frohlich.ea_1997,hiroshima_2002} and demonstrated its ability to support ground states~\cite{hiroshima.spohn_2002,loss.miyao.ea_2007,hidaka.hiroshima_2010} (see also discussion in Sec.~\ref{subsubsec:PF-Hamiltonian}).
Semi-relativistic extensions of the \ac{PF} Hamiltonian are available~\cite{hidaka.hiroshima_2015,miyao.spohn_2009}, allowing for the inclusion of relativistic effects such as spin-orbit coupling using established mathematical methods~\cite{thirring_2013,spohn_2004}. The \ac{PF} Hamiltonian, while exact in the semirelativistic limit, often necessitates approximations. The \ac{LWA}, or dipole approximation, is commonly used (see Sec.~\ref{subsec:LWA}). An important but often overlooked advantage of the \ac{LWA} is that it permits flexible manipulation of the photon field, unlike the full \ac{QED} framework where stringent gauge conditions restrict arbitrary modification of the \ac{EM} field mode structure necessary for cavity emulation~\cite{ruggenthaler.tancogne-dejean.ea_2018}. Unlike the solid-state cases, a unitary equivalent form, the so-called \textit{length-gauge} form, is used in polaritonic chemistry~\cite{faisal_2013,tokatly_2013}. However, the length gauge form intertwines light and matter degrees of freedom, obscuring the inherent periodicity of the matter subsystem within a combined matter-displacement-field coordinate~\cite{rokaj.welakuh.ea_2018,schafer.ruggenthaler.ea_2020}. While the length gauge complicates the connection to Bloch states due to its mixed representation, it proves convenient for most QED chemistry systems lacking periodicity. This gauge enables a Born-Huang-like decomposition, allowing for different groupings of particles (electrons, nuclei/ions, and photons), as for the \ac{BOA} used in solid-state cases discussed in Sec.~\ref{subsec:abinitio-matter}. These groupings facilitate the development of tailored approximation strategies aligned with specific physical and experimental conditions (see Sec.~\ref{subsubsec:separation-PF-Hamiltonian}). For vibrational strong coupling, coupling photons to nuclei/ions leads to the \ac{cBOA}~\cite{flick.ruggenthaler.ea_2017,flick.appel.ea_2017,bonini.ahmadabadi.ea_2024}, while for electronic strong coupling, coupling photons to electrons results in \ac{PoPES}~\cite{galego.garcia-vidal.ea_2015, feist.galego.ea_2018}. While other approaches exist, such as the explicit-polariton method where the photons are solved analytically and a Born-Huang expansion akin to Floquet theory arises~\cite{schafer.ruggenthaler.ea_2018}, the \ac{cBOA} and \ac{PoPES} frameworks are predominantly used.

Building upon the \ac{cBOA} or \ac{PoPES} partitioning, diverse theoretical methods from chemistry and electronic structure theory can be employed to address the resulting coupled nucleus/ion-photon and/or electron-photon systems~\cite{mandal.taylor.ea_2023,ruggenthaler.sidler.ea_2023}. Full configuration-interaction simulation (or exact diagonalization in physics terminology) is feasible for small systems~\cite{sidler.ruggenthaler.ea_2020,sidler.ruggenthaler.ea_2023,flick.ruggenthaler.ea_2015}, while path integral~\cite{li.nitzan.ea_2018,mandal.huo_2019,mandal.krauss.ea_2020,li.nitzan.ea_2022}, hierarchical equations of motion~\cite{lindoy.mandal.ea_2023,wu.cerrillo.ea_2024}, and molecular dynamics methods~\cite{hoffmann.schafer.ea_2019,chen.li.ea_2019,fregoni.corni.ea_2020,antoniou.suchanek.ea_2020} are applicable to \ac{cBOA} problems. \ac{QEDFT}~\cite{ruggenthaler.flick.ea_2014,jestadt.ruggenthaler.ea_2019,ruggenthaler.sidler.ea_2023,penz.tellgren.ea_2023}, \ac{CC-QED}~\cite{mordovina.bungey.ea_2020,haugland.ronca.ea_2020,fregoni.haugland.ea_2021}, and density matrix approaches~\cite{buchholz.theophilou.ea_2019,nielsen.schafer.ea_2019,buchholz.theophilou.ea_2020} are commonly used for \ac{PoPES} problems. Notably, a comparison of results from these methods to simplified few-level approximations like Jaynes-Cummings or Tavis-Cummings reveals significant qualitative discrepancies, even in the weak coupling regime~\cite{schafer.ruggenthaler.ea_2018,mandal.taylor.ea_2023, ruggenthaler.sidler.ea_2023, foley.mctague.ea_2023}. While these simplified models are designed to accurately capture certain spectroscopic features, they often fail to describe complex matter observables. Consequently, the development of novel and reliable models remains an active area of research in \ac{QED} chemistry.

\subsubsection{Progress in experiments and open questions}
%
The coupling of molecular systems to photonic structures has yielded a diverse range of effects. While early studies focused on modifying electronic and excitonic properties in disordered J and H aggregates~\cite{lidzey.bradley.ea_1998,lidzey_2003,tischler.scottbradley.ea_2007}, the field gained significant momentum with the observation of altered photochemical reactions under electronic strong coupling~\cite{hutchison.schwartz.ea_2012,hutchison.liscio.ea_2013, ebbesen_2016}. Subsequent research explored the change in chemical properties using cavities~\cite{ebbesen_2016, garcia-vidal.ciuti.ea_2021}, such as the manipulation of electronically excited states~\cite{stranius.hertzog.ea_2018}, suppression of photo-oxidation and photo-degradation~\cite{munkhbat.wersall.ea_2018,peters.faruk.ea_2019}, and modification of energy transfer processes~\cite{zhong.chervy.ea_2016}. The focus gradually shifted from photochemistry to ground-state chemistry and vibrational strong coupling~\cite{dunkelberger.spann.ea_2016,thomas.lethuillier-karl.ea_2019,vergauwe.thomas.ea_2019,hirai.takeda.ea_2020,xiang.ribeiro.ea_2020,pang.thomas.ea_2020,ahn.triana.ea_2023}, often conducted under ambient conditions (room temperature and pressure) and in a "dark" cavity (no external pumping). For example, experimental studies have demonstrated the ability to modify reaction pathways~\cite{thomas.lethuillier-karl.ea_2019} and control crystallization processes~\cite{hirai.ishikawa.ea_2021} through strong coupling. Strong coupling phenomena are not exclusively linked to the target molecules; effects arising from solvent coupling have also been documented. To differentiate these indirect coupling effects, the term \textit{cooperative strong coupling} has been coined~\cite{garcia-vidal.ciuti.ea_2021}.

\begin{figure}[t]
    \centering
    \includegraphics[width=1.0\textwidth]{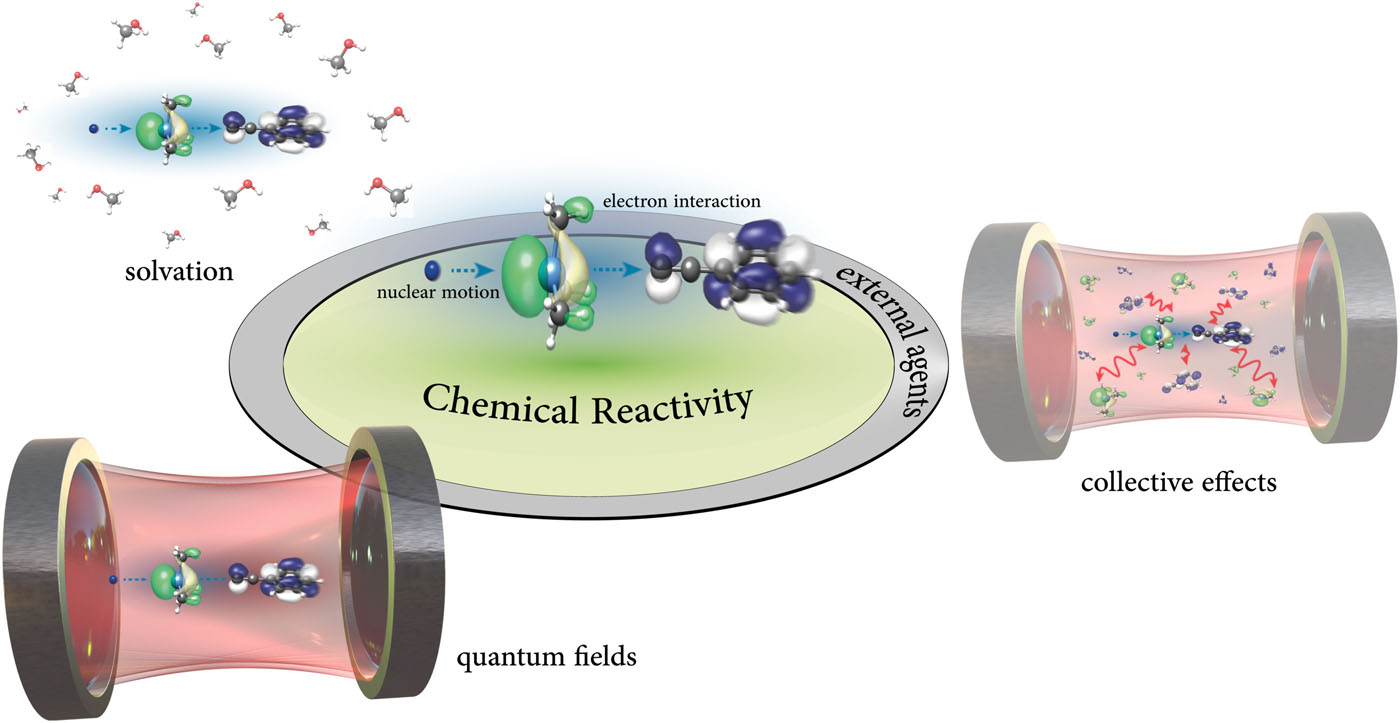}
    \caption{
    \textbf{Illustration of complexity landscape of polaritonic chemistry.}
    Beyond the inherent complexity of chemical systems (denoted as chemical reactivity) and solvent effects, the introduction of quantum fields and collective molecular behavior within optical cavities significantly increases the challenge of theoretical modeling.
    Reprinted from D. Sidler \textit{et al.}, J. Chem. Phys., 156, 230901, 2022~\cite{sidler.ruggenthaler.ea_2022}; licensed under a Creative Commons Attribution (CC BY) license.
    }\label{fig:exp-polaritonic-chemistry}
\end{figure}

%
Despite numerous experimental observations, a comprehensive understanding of the underlying mechanisms in polaritonic chemistry remains elusive~\cite{ebbesen.rubio.ea_2023}. The difficulty in reproducing certain experimental results highlights the complexity of the system and the presence of subtle~\cite{imperatore.asbury.ea_2021,wiesehan.xiong_2021}, yet critical factors. Theoretically, accurately modeling these systems requires a delicate balance between molecular detail and collective effects, while also considering temperature, solvation, and the cavity environment. The complexity of this problem is overwhelming, as illustrated in Fig.~\ref{fig:exp-polaritonic-chemistry}, particularly due to symmetry breaking and the interplay of multiple length scales. The introduction of the cavity breaks fundamental symmetries, mixes different length scales~\cite{sidler.ruggenthaler.ea_2022,ruggenthaler.sidler.ea_2023,svendsen.ruggenthaler.ea_2023}, and necessitates the treatment of molecules as a coupled ensemble. This can lead to emergent phenomena, such as frustration effects resembling spin glasses~\cite{sidler.schnappinger.ea_2024}, challenging traditional thermodynamic descriptions. In addition to these non-perturbative micro-macro relationships, the inclusion of the full continuum of modes in a chemically detailed description demands addressing the ultraviolet divergences inherent in \ac{QED} and the associated mass renormalization~\cite{spohn_2004,ruggenthaler.sidler.ea_2023,welakuh.rokaj.ea_2023,svendsen.ruggenthaler.ea_2023}. As chemical systems typically employ free-space renormalized masses, i.e., the \textit{physical masses}, incorporating the full continuum of modes requires either reverting to \textit{bare masses} or excluding free-space contributions (see Sec.~\ref{subsubsec:PF-Hamiltonian}). Consequently, even equilibrium systems encounter the fundamental complexities of interacting quantum field theories~\cite{ruggenthaler.sidler.ea_2023}. 

\subsection{Cavity-modified quantum Hall effect (QHE)}\label{subsec:quantm-Hall-effect}

\subsubsection{Introduction to the quantum Hall effect}
The classical Hall effect occurs when a current-carrying conductor is subjected to a perpendicular magnetic field, generating a transverse Hall voltage due to the Lorentz force on charge carriers~\cite{ashcroft.mermin_2011}. In a \ac{2DEG} at low temperatures and high magnetic fields, the \ac{QHE} arises, exhibiting quantized Hall resistance in integer multiples of the von Klitzing constant ($h/e^{2}$)~\cite{klitzing.dorda.ea_1980}. The \ac{IQHE} is robust against impurities and stems from the topological properties of the electronic system~\cite{shen_2013}, with the quantized conductance linked to the Chern number, a topological invariant~\cite{vanderbilt_2018}. At even lower temperatures and stronger magnetic fields, electron-electron interactions in the \ac{2DEG} give rise to the \ac{FQHE}, characterized by fractional filling factors and quasiparticles with fractional charge~\cite{tsui.stormer.ea_1982}. The behaviors of the \ac{IQHE} and \ac{FQHE} -- the former explained by single-particle physics, the latter by intricate many-body interactions -- make the \ac{2DEG} an ideal platform for probing the effect of field fluctuations of photons on matter systems. Advances in cavity platforms (Sec.~\ref{subsubsec:cavity-platforms}), such as split-ring resonators, enable strong coupling between \ac{2DEG}s and quantum light, allowing photon-mediated modifications of \ac{QHE} properties. This section reviews experimental and theoretical insights into how cavity fields influence the topological protection of the \ac{IQHE} and the fractional states of the \ac{FQHE}.

\subsubsection{Experimental realization of cavity-modified QHE}

\begin{figure}[t]
    \centering
    \includegraphics[width=1.0\textwidth]{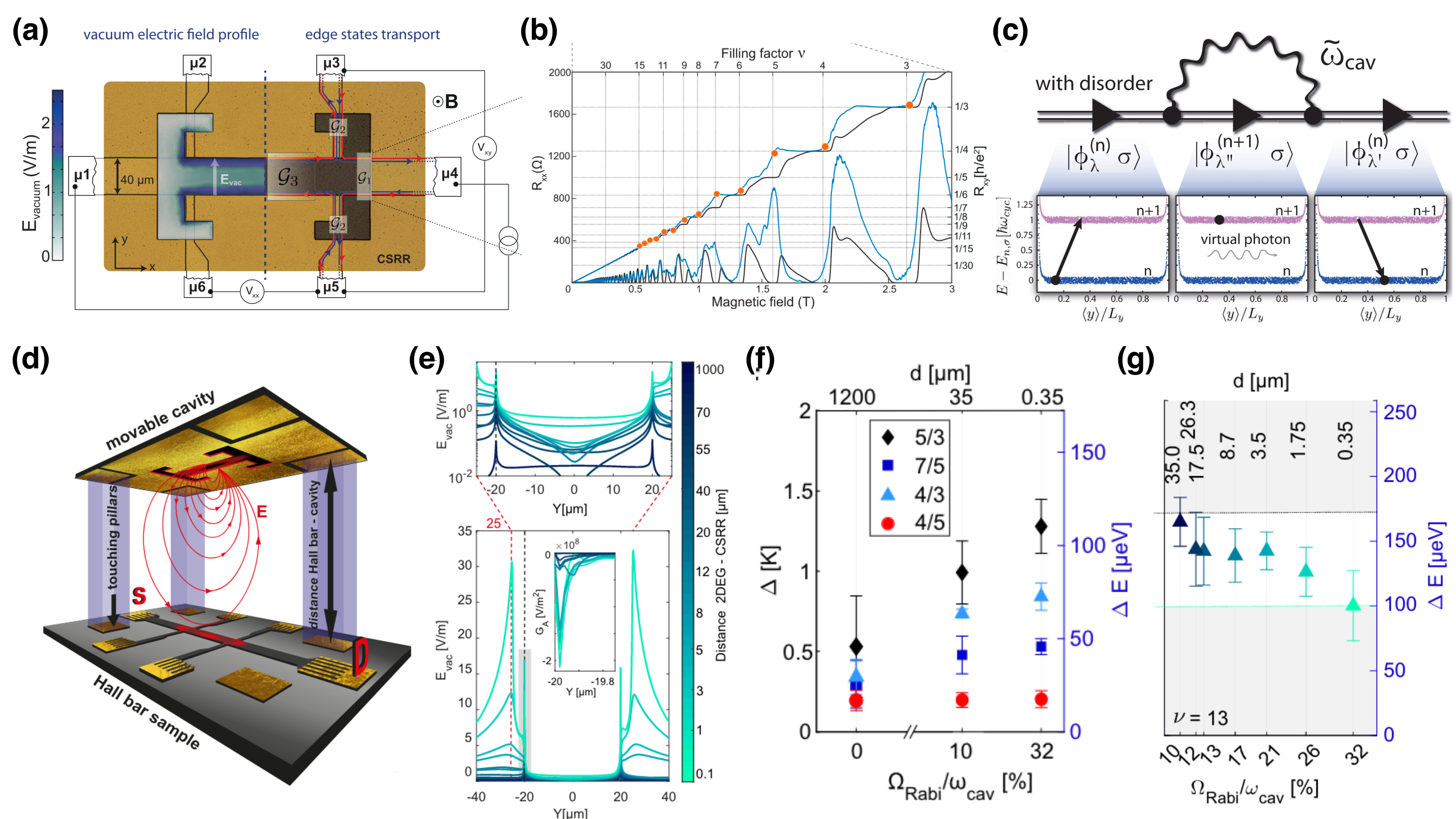}
    \caption{
    \textbf{Cavity-modified integer and fractional Quantum Hall effect of two-dimensional electron gas.}
    (a) The setup of the quantum Hall bar embedded inside a split-ring resonator where the vacuum electric field $E_{\text{vacuum}}$ is localized at the edges of the Hall bar sample and is polarized along the $y$-axis. The color bar represents the intensity of the field.
    (b) The longitudinal resistance $R_{xx}$ and Hall resistance $R_{xy}$ as a function of the external magnetic field. The black (blue) lines are for the Hall bar outside (inside) the cavity, while the orange points are the simulated results. The results show the breakdown of topological protection for the \ac{IQHE} inside the cavity.
    (c) The theoretical explanation of the effective cavity-mediated long-range hopping between two disordered eigenstates at the same Landau band via the exchange of a virtual cavity photon, leading to the scattering channels for the electronic states. 
    (a)-(c) are from F. Appugliese \textit{et al.}, Science 375, 1030-1034 (2022)~\cite{appugliese.enkner.ea_2022}, reprinted with permission from AAAS.
    (d) The schematic diagram of the experimental setup of the movable split-ring resonator.
    (e) The simulated spatial profile of $E_{\text{vacuum}}$ via tuning the distance between the resonator and the quantum Hall bar. The color bar represents the distance between the movable cavity and the quantum Hall bar. The inset in the lower panel shows the gradient of the electric field near the edge of the Hall bar. 
    (f) The enhancement of the activation energy of the fractional quantum Hall $\Delta$ for a wide range of fractional filling factors as a function of the effective light-matter coupling $\Omega_{\text{Rabi}}/\omega_{\text{cav}}$, where $\Omega_{\text{Rabi}}$ is the Rabi frequency and $\omega_{\text{cav}}$ is the photon frequency. $d$ is the distance between the cavity and the Hall bar.
    (g) The effective spin splitting $\Delta$ decreases as a function of the effective light-matter coupling.
    (d)-(e) are reprinted from J. Enkner \textit{et al.}, arXiv, arxiv.org/abs/2405.18362, 2024~\cite{enkner.graziotto.ea_2024}. Copyright 2024 Authors, licensed under Creative Commons Attribution License 4.0 (CC BY).
    }\label{fig:exp-hall-effect-2D-gas}
\end{figure}

Recent experiments~\cite{appugliese.enkner.ea_2022,enkner.graziotto.ea_2024} demonstrated the coupling of a quantum Hall bar to a split-ring resonator, generating intense vacuum electric field fluctuations localized at the Hall bar's edges [Fig.~\ref{fig:exp-hall-effect-2D-gas}(a)]. These fluctuations, significantly stronger than those in free space, led to notable deviations in the Hall resistance [Fig.~\ref{fig:exp-hall-effect-2D-gas}(b)], particularly in odd-numbered plateaus of the \ac{IQHE}. Key observations include the breakdown of integer quantization in \ac{IQHE} and the enhanced stability of fractional states in \ac{FQHE}~\cite{appugliese.enkner.ea_2022,enkner.graziotto.ea_2024}. The breakdown of integer quantization is suggested to be attributed to photon-mediated long-range electron hopping between edge and bulk states~\cite{ciuti_2021}. This process involved virtual photon exchanges between Landau levels, enabling electron scattering into bulk states, which impacts the overall conductivity of the \ac{2DEG} [Fig.~\ref{fig:exp-hall-effect-2D-gas}(c)]. The cavity-\ac{2DEG} coupling was tuned by varying the distance between the split-ring resonator and the Hall bar [Fig.~\ref{fig:exp-hall-effect-2D-gas}(d)]. As the resonator approaches the Hall ball, the effective light-matter coupling strength -- denoted as $\Omega_{\text{Rabi}}/\omega_{\text{cav}}$ where $\Omega_{\text{Rabi}}$ is the Rabi frequency and $\omega_{\text{cav}}$ is the cavity frequency -- increases as well as the vacuum electric field gradients near the Hall bar edges [Fig.~\ref{fig:exp-hall-effect-2D-gas}(e)]. Consequently, the activation energy gap for fractional fillings such as $5/3$ increased significantly [Fig.~\ref{fig:exp-hall-effect-2D-gas}(f)], enhancing the \ac{FQHE}'s stability. These findings highlight the interplay between cavity-induced modifications and the intrinsic robustness of the \ac{QHE}.

\subsubsection{Mechanism of cavity-induced modifications in quantum Hall effects}

The robustness of the \ac{IQHE} stems from edge states immune to backscattering~\cite{shen_2013}. However, photon-mediated processes in a cavity can undermine this protection. For example, virtual photon exchanges in a split-ring resonator setup enable an electron at the $n$th Landau level to transit to the $n+1$th Landau level via a virtual photon emission, followed by its return to a different state within the original $n$th Landau level~\cite{ciuti_2021}. Simulations incorporating these photon-mediated processes reproduced experimental deviations in transverse resistance [orange points in Fig.~\ref{fig:exp-hall-effect-2D-gas}(b)]~\cite{appugliese.enkner.ea_2022}.

In the \ac{IQHE} cases, the longitudinal resistance minima increases due to the light-matter coupling breaking the integer quantization~\cite{enkner.graziotto.ea_2024} [see Fig.~\ref{fig:exp-hall-effect-2D-gas}(b)]. This behavior suggests a reduction in the effective spin splitting $\Delta_{s} = \mu_{\rm{B}} g^{*}/k_{\rm{B}}$, where $g^{*}$  is the effective gyromagnetic factor (or $g$-factor) including exchange energy corrections among electrons, $\mu_{\rm{B}}$ is the Bohr magneton, and $k_{\rm{B}}$ is the Boltzmann constant. For integer filling of $\nu=13$, as light-matter coupling increases, the effective $g$-factor decreased from $6.5$ to $4.4$, estimated from the spin splitting $\Delta_{s}$ in Fig.~\ref{fig:exp-hall-effect-2D-gas}(g), indicating a cavity-induced modification of electron-electron interactions.

An effective \ac{QED} Hamiltonian was constructed to explain the decrease in the effective $g$-factor, describing electron interactions in an external magnetic field coupled to a single cavity mode~\cite{enkner.graziotto.ea_2024}. This Hamiltonian incorporates the spatial gradient of the electric field, extending beyond the \ac{LWA}. A perturbative treatment, accounting for up to two virtual photon exchanges, yielded a further effective matter Hamiltonian with an attractive cavity-mediated electron-electron interaction. This attractive interaction reduces the exchange splitting, countering the repulsive Coulomb interaction and leading to an estimated effective $g$-factor reduction of approximately $2$ ($\Delta g^{*}\sim-2$), consistent with the experimental value ($-2.1$)~\cite{enkner.graziotto.ea_2024}.

In the \ac{FQHE} regime, the enhanced activation energy gap observed in the fractional fillings [Fig.~\ref{fig:exp-hall-effect-2D-gas}(f)] can be attributed to the above cavity-mediated attractive electron-electron interaction. The activation energy gap is estimated using magneto-roton theory~\cite{girvin.macdonald.ea_1986,girvin.yang_2019}, which involves calculating the density-density correlation function for the Laughlin ground state and the so-called projected oscillator strength~\cite{enkner.graziotto.ea_2024}. The projected oscillator strength, a linear functional of the electron-electron interaction (including both the repulsive Coulomb and the attractive cavity-mediated interactions), is key in this estimation. For the $1/3$-Laughlin state, its enhanced activation energy gap is predicted to align well with experimental observations~\cite{enkner.graziotto.ea_2024}.

While the above analysis indicated it may be necessary to go beyond the \ac{LWA} to include the gradient of the electric field to explain the observed cavity-modified, another study~\cite{rokaj.wang.ea_2023} that uses the \ac{LWA} suggests another mechanism -- it is the softened lower Landau polariton that leads to the weakening of topological protection of the \ac{IQHE}. It also indicates that the \ac{FQHE} can be modified when the lower polariton becomes softer. Remaining in the \ac{LWA}, another theoretical study~\cite{rokaj.penz.ea_2022} suggests the von Klitzing constant can also be modified; however, a recent experiment has not been able to approach the regime where the change can be observed~\cite{enkner.graziotto.ea_2024a}.

Overall, these studies highlight the significant influence of cavity-mediated interactions on the quantum Hall effect, demonstrating how light-matter coupling reshapes fundamental electronic properties. Cavity-induced effects, such as enhanced fractional states and altered topological protection, provide powerful tools for exploring and controlling quantum materials. Diverse mechanisms, from virtual photon processes to Landau polariton softening, reveal the richness and complexity of these interactions, emphasizing the need for further experimental and theoretical research to deepen our understanding of cavity-modified quantum phenomena in solid-state materials.

\subsection{Cavity-controlled metal-to-insulator phase transition}\label{subsec:MIT-TaS2}
\subsubsection{Experimental insights into cavity-induced metal-to-insulator transition of 1T-TaS$_{2}$}
The \ac{MIT} is a fundamental phenomenon in solid-state physics, offering insights into fundamental electronic interactions and promising applications in advanced devices~\cite{imada.fujimori.ea_1998}. This transition, characterized by a dramatic change in electrical conductivity, can be induced by external stimuli such as temperature, pressure, gate field, or light, making them a versatile platform for tunable electronic properties. Among various materials, 1T-TaS$_{2}$ stands out as a model system for studying \ac{MIT} due to its sensitivity to temperature, doping, and pressure~\cite{fazekas.tosatti_1979,sipos.kusmartseva.ea_2008,yu.yang.ea_2015}. Its rich phase diagram reflects the complex interplay between electronic and structural degrees of freedom, leading to various phases from commensurate (C-CDW), nearly commensurate (NC-CDW), and incommensurate charge density wave (I-CDW) phases as temperature increases. At low temperatures below $150$ K, the insulating C-CDW phase, characterized by the Star-of-David pattern, dominates. This transitions to a metallic NC-CDW phase between $150$ K and $350$ K and eventually to an I-CDW metallic state above $350$ K.
The ability to control these \ac{CDW} phases offers exciting possibilities for device applications. For instance, memory storage devices could exploit the reversible transition between insulating and metallic states. Moreover, observing light-induced \ac{MIT} in 1T-TaS$_{2}$ opens up new avenues for ultrafast optical control of electronic properties~\cite{stojchevska.vaskivskyi.ea_2014}.
Unlike the above methods relying on strain, doping, or external light, recent research has explored optical cavities to modify the critical temperature of the \ac{MIT} in 1T-TaS$_{2}$, providing a new dimension for phase control~\cite{jarc.mathengattil.ea_2023}.

A recently developed cryogenic tunable terahertz Fabry-Pérot cavity (see Sec.~\ref{subsubsec:cavity-platforms}) offers a mechanically controlled platform for light-matter coupling in crystalline solids by adjusting mirror spacing and alignment~\cite{jarc.mathengattil.ea_2022}. In a notable experiment, this cavity was shown to alter the \ac{MIT} temperature between the C-\ac{CDW} and NC-\ac{CDW} phases in 1T-TaS$_2$~\cite{jarc.mathengattil.ea_2023}. Contactless terahertz spectroscopy detected conductive charges indicative of the metallic state, with low-frequency transmission changes marking the \ac{MIT}. While the transition occurs at $150$K in free space, it shifts to $120$K inside an $11.5$ GHz cavity~\cite{jarc.mathengattil.ea_2023}. Further, by varying the tilt angle and spacing of the cavity mirrors, the critical transition temperature was tuned, demonstrating the cavity’s capacity to manipulate the electronic properties of 1T-TaS$_2$. This work highlights the potential of cavity photonics for driving phase transitions and controlling material properties.

%
\subsubsection{Theoretical perspectives on mechanisms}
Two primary microscopic mechanisms have been proposed to explain the cavity-induced shift in \ac{MIT} temperature in 1T-TaS$_2$: 1) the renormalization of the free energy of the two phases by the cavity mode, or 2) the thermodynamic or temperature alterations from the cavity's modulation of the \ac{EM} density of states~\cite{jarc.mathengattil.ea_2023}. In the first scenario, the cavity stabilizes the metallic phase with the long wavelengths and the insulating phase with shorter wavelengths, respectively. The free energy in the insulating and metallic phases can be modified inside the cavity by strong interaction between photons and materials. The modified free energy is related to the change in the critical temperature of the \ac{MIT}. In the second scenario, on the other hand, the thermal Purcell effect~\cite{chiriaco_2024}, which modifies the exchange rate of heat radiation with the photon environment via the cavity structure, is suggested. 
One of the main differences between the cavity and the vacuum is the modified photon density of states inside the cavity. Unlike the vacuum, the distance between two optical cavity mirrors leads to the quantum confinement condition for photons.
Ref.~\cite{pannir-sivajothi.yuen-zhou_2024,fassioli.faist.ea_2024} show that the modification in photon density of states is significant with THz range fundamental cavity frequency. On the other hand, the above experiment on 1T-TaS$_2$ shows a maximized modification in the GHz range while minimized in the THz range. Regarding discrepancy, Ref.~\cite{chiriaco_2024} suggests that spherical cavity conditions can describe the thermodynamical equilibrium as the experimental observations.
Unlike the Fabry-Pérot cavity, the spherical configuration provides higher photon density of states in the GHz range than the THz range. The low(high)-frequency photons show higher(lower) photon density of states than in a vacuum. This behavior predicts consistent temperature modification in the experiment. Despite the predictive consistency of spherical cavities, the experiment employed a Fabry-Pérot cavity. While the insights from spherical cavities offer valuable perspectives, further investigation is necessary to fully understand the role of Fabry-Pérot cavities in such systems.

The experimental study reviewed in this section highlights the impact of cavity environments on the stability of matter phases, raising intriguing questions about their relationship to traditional thermodynamic equilibrium principles. In conventional matter-only systems, equilibrium states are dictated by intrinsic particle-particle interactions (e.g., electron-electron, electron-phonon) and external parameters such as temperature, pressure, and chemical potential. In cavity quantum materials, however, the equilibrium state is co-determined by the interaction between the material and the quantized \ac{EM} field. Coupling between electronic, vibrational, or other material degrees of freedom and photonic modes alters the whole system’s Hamiltonian, potentially stabilizing entirely new ground states. For example, experiments in polaritonic chemistry have shown that cavity vacuum fields can renormalize interactions, create hybrid light-matter quasi-particles, or stabilize phases otherwise energetically unfavorable in conventional systems (see Sec.~\ref{subsec:polaritonic-chemistry}). Such changes can also be rationalized theoretically~\cite{sidler.ruggenthaler.ea_2020,sidler.ruggenthaler.ea_2023}. Theoretical studies further reveal that the matter subsystem in strongly coupled systems behaves differently from its uncoupled equilibrium counterpart, driven by photon-mediated correlations and fluctuations absent in purely matter-based systems. These findings highlight the hybrid nature of cavity quantum materials and the need to extend conventional matter-only equilibrium concepts to fully capture the physics of these coupled systems. The reviewed experiment emphasizes the need for further theoretical and experimental efforts to understand how thermodynamics equilibrium in cavity-coupled systems compares to that in purely material-based systems.

\subsection{Cavity-controlled ferromagnetism and superconductivity of an unconventional superconductor, YBCO}\label{subsec:ferromagnetism}

%

In addition to cavity-modified \ac{MIT} phase transitions, optical cavities can also influence other phases, such as ferromagnetic and superconducting states. YBa$_2$Cu$_3$O$_{7-x}$ (YBCO), an unconventional superconductor with a superconducting transition temperature of $92$ K, exhibits antiferromagnetic behavior in its bulk form. However, in nanoparticle form, YBCO demonstrates room-temperature ferromagnetism due to intrinsic oxygen vacancies, rather than magnetic impurities, although its magnetic moment is typically small~\cite{sundaresan.rao_2009,sharma.gupta.ea_2003}. A recent experiment revealed that coupling cavity photons \textit{indirectly} to the phonon modes of YBCO nanoparticles can enhance this ferromagnetic phase in the normal state, offering new ways to explore light-matter interactions in unconventional superconductors~\cite{thomas.devaux.ea_2021}.

\begin{figure}[t]
    \centering
    \includegraphics[width=1.0\textwidth]{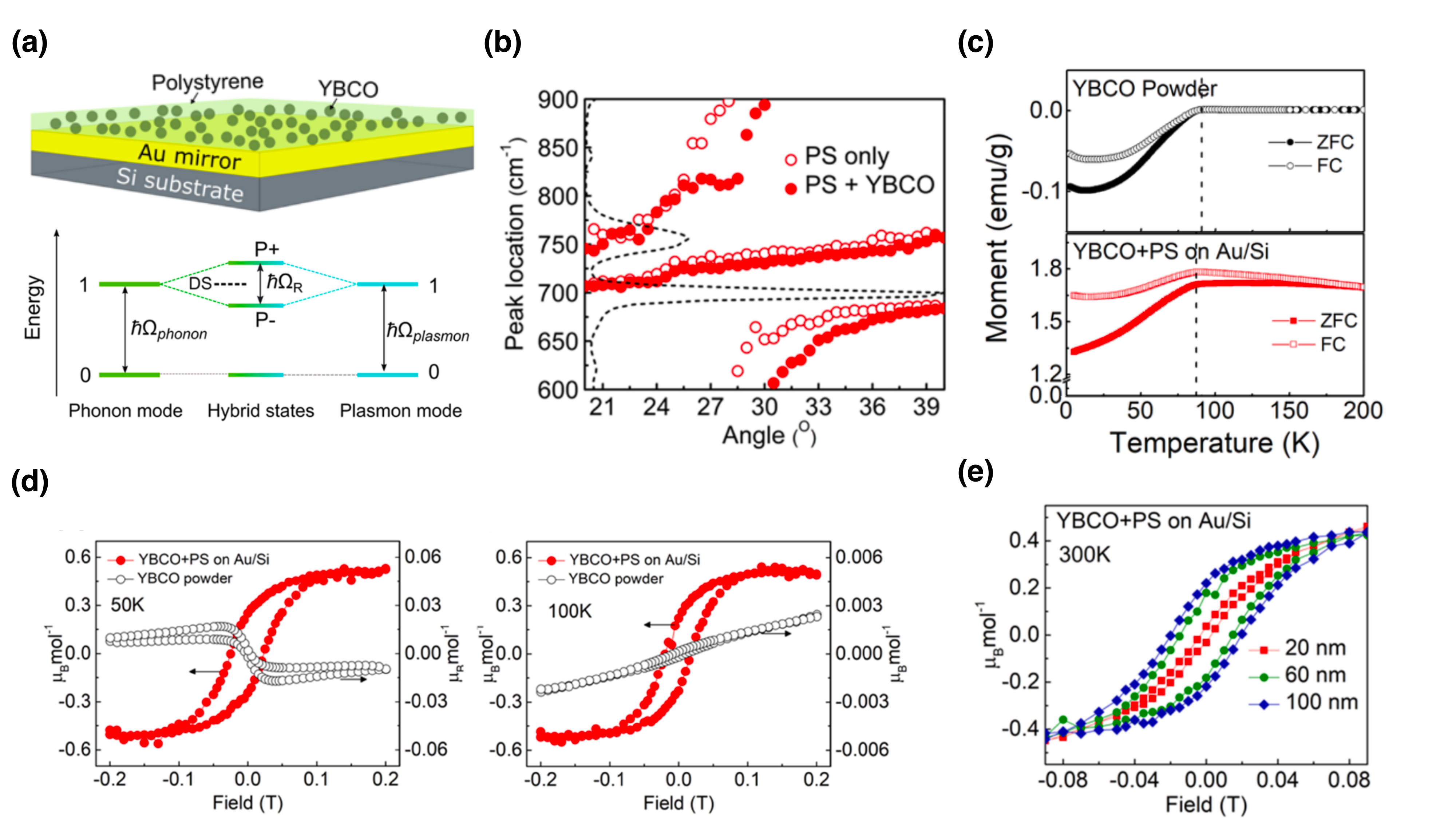}
    \caption{\textbf{Cavity-modified ferromagnetism and superconducting transition temperature of an unconventional superconductor, YBCO, via field fluctuations.}
    (a) The schematic of the cavity and material setup in the top panel. YBCO is embedded inside the polystyrene (PS) polymer matrix, which is deposited on a gold (Au) mirror with a thickness of $10$ nm supported by a silicon (Si) substrate. The Au mirror provides the surface plasmon mode with a frequency of $\Omega_{\rm{plasmon}}$, which is coupled to the phonon mode of YBCO with a frequency of $\Omega_{\rm{phonon}}$ to form lower (P-) and upper (P+) polaritons with a Rabi splitting $\hbar\Omega_{\rm{R}}$, shown in the bottom panel. Here DS stands for dark states. 
    (b) Dispersion curves are shown in red dots and obtained using attenuated total reflectance FTIR spectroscopy. The transmission spectrum of PS is shown in the black dashed curve.
    (c) Temperature dependence of magnetization during zero-field cooling (ZFC) and field cooling (FC) for the YBCO powder and the YBCO+PS coupled to the surface plasmon. 
    (d) Magnetization curves for the YBCO powder and the YBCO+PS coupled to the surface plasmon mode at $50$ and $100$ K.
    (e) Magnetization curves for YBCO+PS on the Au with different thicknesses at 300 $K$.
    Reprinted with permission from A. Thomas \textit{et al.}, Nano Letters 21, 4365 (2021)~\cite{thomas.devaux.ea_2021}. Copyright 2021 American Chemical Society.
    }\label{fig:exp-cavity-modified-ferromagnetism}
\end{figure}

%
In the experiment~\cite{thomas.devaux.ea_2021}, YBCO nanoparticles were dispersed in a \ac{PS} polymer matrix and spin-coated onto a $10$ nm-thick gold (Au) film on a silicon (Si) substrate. The Au film, serving as a cavity, supports surface plasmon modes that resonate with the absorption peaks of both PS and YBCO near $700$ cm$^{-1}$ [Fig.~\ref{fig:exp-cavity-modified-ferromagnetism}(a)]. The phonon mode band of the YBCO powder overlaps with the vibrational mode of the PS matrix, which has a stronger absorption at this wavenumber. The \ac{IR} spectrum of YBCO shows the two \ac{IR}-active phonon peaks ($692$ and $856$ cm$^{-1}$) are weakly coupled to a plasmonic mode of the metal. In contrast, the \ac{IR} spectrum of PS has a strong peak (around $700$ cm$^{-1}$), which overlaps with one of the phonon modes of YBCO. The coupling of the plasmon mode with the YBCO phonon band via the PS vibrational mode leads to the formation of the upper and lower polaritons near $700$ cm$^{-1}$ [Fig.~\ref{fig:exp-cavity-modified-ferromagnetism}(b)]. This coupling alters the YBCO phonon band, potentially affecting the magnetic and superconducting properties of the nanoparticles.

Superconducting behavior in the YBCO+PS mixture was analyzed using field-cooled (FC) and zero-field-cooled (ZFC) magnetization curves, measured with a superconducting quantum interference device magnetometer. A shift in the superconducting transition temperature (T$_{\text{c}}$) from $92$ K in bulk YBCO to approximately $87$ K in the YBCO+PS mixture was observed [Fig.~\ref{fig:exp-cavity-modified-ferromagnetism}(c)]. This T$_{\text{c}}$ shift suggests indirect coupling of cavity photons to YBCO via the vibrational modes of PS, as no shift was detected when PS was replaced with polymethylmethacrylate (PMMA), which lacks a matching vibrational mode~\cite{thomas.devaux.ea_2021}.

In addition to the T$_{\text{c}}$ shift, enhanced ferromagnetic was evident in the YBCO+PS mixture. Magnetization curves at low ($50$ K) and high ($100$ K) temperatures revealed a ferromagnetic phase in the YBCO+PS mixture, with the magnetic moment increasing by two orders of magnitude compared to YBCO alone at $100$ K [Fig.~\ref{fig:exp-cavity-modified-ferromagnetism}(d)]. This enhancement is attributed to the strong coupling between the plasmon mode and the YBCO phonon band via the PS vibrational mode.

To further explore the impact of light-matter coupling on ferromagnetism, the quality of the Au film can be enhanced by increasing its thickness [Fig.~\ref{fig:exp-cavity-modified-ferromagnetism}(e)]. Increasing the thickness of the Au film enhances the cavity's plasmonic properties, leading to larger hysteresis and a larger net magnetic moment [Fig.~\ref{fig:exp-cavity-modified-ferromagnetism}(e)]. These effects are saturated when the Au film exceeds $100$ nm in thickness. Under collective strong coupling, the enhanced ferromagnetic phase could compete with and even disrupt the superconducting state of YBCO, lowering its T$_{\text{c}}$.

While these findings illustrate the potential of cavity engineering to manipulate ferromagnetism and superconductivity in unconventional superconductors, the underlying microscopic mechanisms of superconductivity in YBCO remain unresolved. In addition to YBCO, the superconductivity behaviors of other materials seem to be able to be controlled via cavities~\cite{thomas.devaux.ea_2019,song.seo_2017}. This highlights the need for further theoretical and experimental investigations to fully understand the interplay between light-matter interactions and the intrinsic properties of these materials.

\section{Theoretical predictions for cavity-engineered solid-state materials}\label{sec:theory-prediction}
This section builds on the experimental evidence of cavity vacuum field fluctuations discussed earlier by reviewing recent theoretical proposals for cavity-modified equilibrium ground states of solid-state materials in dark cavities. However, the feasibility of experimentally realizing these predictions depends significantly on the maturity of the required supporting technologies. This section aims to identify and prioritize theoretical proposals that are most amenable to near-term experimental validation while acknowledging the challenges that remain for more speculative ideas. While prior studies have explored cavity-modified phenomena in 1D systems, including electronic topology~\cite{mendez-cordoba.mendoza-arenas.ea_2020, mendez-cordoba.rodriguez.ea_2023, perez-gonzalez.platero.ea_2023}, quantum transport~\cite{arwas.ciuti_2023,nguyen.arwas.ea_2024}, quantum Floquet engineering~\cite{eckhardt.passetti.ea_2022}, and strongly correlated systems~\cite{kiffner.coulthard.ea_2019}, this section primarily focuses on 2D and 3D materials. These higher-dimensional platforms are more extensively studied in experimental settings (see Sec.~\ref{sec:experiments}) and offer greater immediate potential for practical realization. The discussion is organized as follows: Section~\ref{subsec:theory-prediction-paraferro} reviews predictions of cavity-modified para-to-ferroelectric phase transitions. Section~\ref{subsec:cavity-modified-superconductivity} examines a wide range of mechanisms of cavity-controlled superconductivity. Section~\ref{subsec:cavity-modified-strongly-correlated-systems} explores cavity-modified strongly correlated systems. Section~\ref{subsec:cavity-modified-electronic-topology} highlights advances in cavity-modified electronic topology and quantum geometry. Finally, Section~\ref{subsec:feasibility-prediction} evaluates the experimental feasibility of these theoretical predictions, offering guidance for future research directions.

\subsection{Cavity-controlled para-to-ferroelectric phase transition}\label{subsec:theory-prediction-paraferro}
%
 
The para-to-ferroelectric phase transition, marked by the emergence of spontaneous polarization, is typically described by Landau theory. Here, a decrease in temperature leads to free energy changes that stabilize the ferroelectric order, often driven by the softening of a lattice vibration mode, known as the "soft mode"~\cite{rabe.ahn.ea_2007}. This mode signals lattice instability, ultimately breaking symmetry and enabling ferroelectricity.

Unlike conventional ferroelectrics, quantum paraelectrics such as strontium titanate (SrTiO$_{3}$) remain in a paraelectric state even at very low temperatures, as nuclear quantum fluctuations prevent long-range ferroelectric order~\cite{muller.burkard_1979}. The quantum paraelectric state of SrTiO$_{3}$ has been studied using effective models~\cite{fujishita.kitazawa.ea_2016} and first principles calculations~\cite{shin.latini.ea_2021,esswein.spaldin_2022}. Inducing a ferroelectric phase in SrTiO$_{3}$ has been the subject of extensive research, with methods such as (epitaxial) strain engineering~\cite{haeni.irvin.ea_2004,xu.huang.ea_2020}. In recent years, the concept of utilizing light-matter interactions to control the phases of materials via external \ac{EM} fields has emerged as a promising avenue~\cite{li.qiu.ea_2019,nova.disa.ea_2019,shin.latini.ea_2022,fechner.forst.ea_2024,orenstein.krapivin.ea_2024}. A novel approach to controlling quantum matter emerges, which involves modifying the field fluctuations of its \ac{EM} environment, rather than relying on classical radiation. Using a two-level dipole model, it has been demonstrated that ferroelectric phases can be stabilized through coupling with photons~\cite{schuler.debernardis.ea_2020,pilar.bernardis.ea_2020}. By confining light within a cavity, more refined theoretical studies showed that it is possible to manipulate material properties and induce~\cite{ashida.imamoglu.ea_2020,latini.shin.ea_2021} or suppress~\cite{curtis.michael.ea_2023} ferroelectricity via the confined photon field fluctuations. 


\begin{figure}[t]
    \centering
    \includegraphics[width=1.0\textwidth]{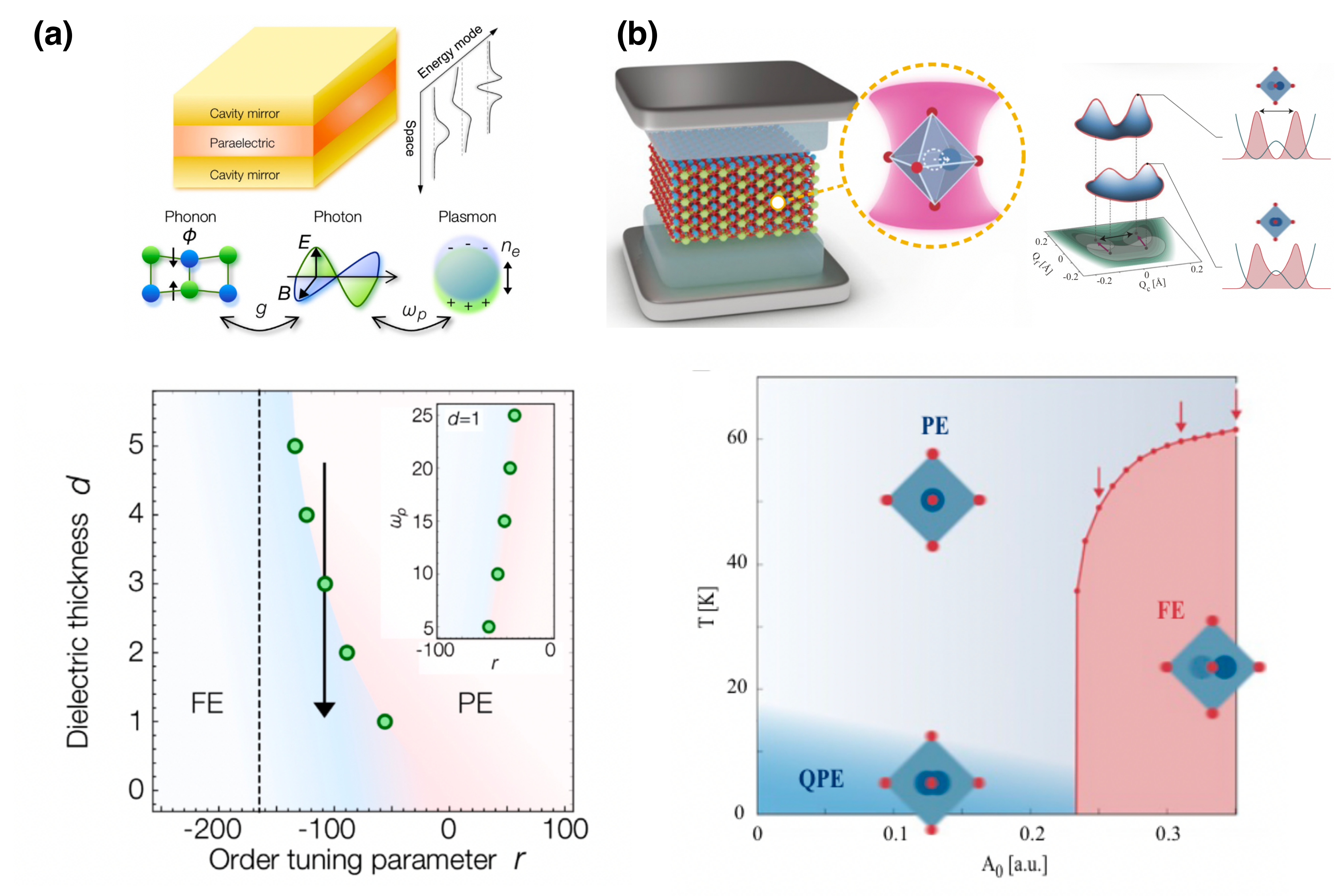}
    \caption{\textbf{Proposed cavity-induced para- to ferro-electric phase transition via quantum vacuum fluctuations in a dark cavity.}
    (a) A paraelectric material with a thickness $d$ is embedded inside a Fabry-Pérot cavity with the spatial profile of the energy modes shown next to it (top panel). The light-matter system is modeled by coupling between phonons and photons (with the coupling strength $g$), and between photons and plasmon (with the coupling strength $\omega_{p}$, plasma frequency). As the tuning parameter $r$ (for example, pressure) decreases, the phase transits from the paraelectric phase (PE) to the ferroelectric phase (FE). The black dashed line denotes the phase transition point for the bulk geometry. The inset shows the dependence of the phase transition point on the plasma frequency at $d=1$. 
    Reprinted from Ashida \textit{et al.}, Phys. Rev. X 10, 041027 (2020)~\cite{ashida.imamoglu.ea_2020}. Copyright 2020 Authors, licensed under Creative Commons Attribution License 4.0 (CC BY).
    (b) SrTiO$_{3}$ (STO), a quantum paraelectric material, is placed inside a Fabry-Pérot cavity with the cavity photons coupled to the ferroelectric soft mode $Q_{f}$ and a second lattice vibration $Q_{c}$ in STO (left at the top panel). Tuning the photonic environment can potentially realize the para-to-ferroelectric phase transition (right at the top panel). The phase diagram as a function of the effective light-matter coupling $A_{0}$ shows the ground states of STO at different temperatures and $A_{0}$ (bottom panel).
    Reprinted from Latini \textit{et al.}, Proc. Natl. Acad. Sci. 118, e2105618118 (2021)~\cite{latini.shin.ea_2021}. Copyright 2021 Authors, licensed under Creative Commons Attribution License 4.0 (CC BY).
    }\label{fig:theory-cavity-modified-para-to-ferro-electric}
\end{figure}

A range of theoretical and computational studies highlight the potential of cavities to control para-to-ferroelectric transitions, particularly in SrTiO$_{3}$. Early models incorporated the interaction between phonons, photons, and plasmons within the cavity~\cite{ashida.imamoglu.ea_2020}, demonstrating that the hybridization of cavity photons with optical phonons can soften the phonon modes, driving the system toward a ferroelectric phase transition [Fig.~\ref{fig:theory-cavity-modified-para-to-ferro-electric}(a)]. A separate study~\cite{latini.shin.ea_2021}, employing first-principles (\ac{DFT}) calculations within an effective \ac{QED} framework, demonstrated the ability of strong light-matter coupling to induce a ferroelectric phase transition in SrTiO$_{3}$ [Fig.~\ref{fig:theory-cavity-modified-para-to-ferro-electric}(b)]. This "photo-groundstate" is stabilized across a wide range of cavity photon energies, distinguishing it from conventional approaches reliant on resonant excitation. The resulting ferroelectric phase shares similarities with the dynamically localized ferroelectric phase observed in laser-induced systems~\cite{latini.shin.ea_2021}. Additional support comes from studies coupling two-level dipoles on 2D lattices to surface plasmon-polaritons, solved beyond mean-field theory using \ac{DMFT}, which further reinforced the feasibility of cavity-induced ferroelectricity~\cite{lenk.li.ea_2022}. Such findings highlight the potential of cavity-modified \ac{EM} environments to manipulate matter phase transitions by nuclear quantum fluctuations.

However, not all theoretical work supports the stabilization of the ferroelectric phase inside a cavity. Recent work incorporating a full continuum of transverse photon modes suggests that such modes can suppress ferroelectric phases, particularly near material surfaces, due to screening effects at cavity boundaries~\cite{curtis.michael.ea_2023}. This nuanced finding points out the importance of model assumptions, such as the treatment of bare versus physical masses, which can significantly influence predictions (as described in Sec.~\ref{subsubsec:PF-Hamiltonian}). Models using physical masses implicitly account for photon contributions, potentially requiring a reduced set of effective photon modes to avoid double counting (as done in this review).


The studies reviewed here highlight both the potential and challenges of cavity-modified ferroelectricity. From a practical standpoint, the ability to induce or suppress ferroelectricity without applying external strain or \ac{EM} fields could revolutionize the design of tunable quantum devices. Theoretical insights into the interplay between photons, phonons, and material properties provide a foundation for cavity materials engineering. Experimental feasibility remains a significant challenge. Although models elucidate mechanisms for cavity-induced phase transitions, scaling these effects to macroscopic systems or achieving the required coupling strengths demands major advancements in cavity design and fabrication. Additionally, conflicting predictions about the stabilization or suppression of ferroelectric phases underscore the need to better understand photon-mode contributions and their interplay with material boundaries. Future research must balance computational sophistication with experimental practicality to realize cavity-modified quantum phases.

\subsection{Cavity-modified superconductivity of solid-state materials}\label{subsec:cavity-modified-superconductivity}

Unlike conventional approaches that rely on intense laser fields to transiently induce superconductivity~\cite{fausti.tobey.ea_2011,hu.kaiser.ea_2014,mitrano.cantaluppi.ea_2016,cantaluppi.buzzi.ea_2018,cremin.zhang.ea_2019,buzzi.nicoletti.ea_2020,budden.gebert.ea_2021,buzzi.nicoletti.ea_2021,isoyama.yoshikawa.ea_2021,rowe.yuan.ea_2023} (for comprehensive review, see Ref.~\cite{cavalleri_2018}), cavity-based methods use quantum vacuum fluctuations of \ac{EM} fields to enhance light-matter coupling without the need for strong laser intensities. This gentle yet effective approach enables the exploration of cavity-induced modifications to superconducting critical temperatures (T$_{\text{c}}$) and pairing mechanisms, offering new tools for controlling and engineering quantum materials.
While some work~\cite{gao.schlawin.ea_2020,chakraborty.piazza_2021} demonstrates using a cavity to tune the superconductivity of a cavity-coupled material with the assistance of an external laser to tune the effective electron-electron interaction or light-matter coupling, in this section we mainly focus on the dark cavity cases, where superconducting properties are modified without external laser pumping.
Recent studies highlight diverse mechanisms, such as cavity-modified electron-phonon coupling and phonon frequency via phonon-polaritons~\cite{sentef.ruggenthaler.ea_2018,hagenmuller.schachenmayer.ea_2019} or the change in electron density~\cite{lu.shin.ea_2024a}, quantum Eliashberg effect~\cite{curtis.raines.ea_2019}, cavity-modified band structure~\cite{kozin.thingstad.ea_2024}, cavity-induced superconductivity in 2D electron gas~\cite{schlawin.cavalleri.ea_2019,andolina.depasquale.ea_2024}, and controlling the competing matter phases~\cite{li.eckstein_2020}. 


%
\subsubsection{Phonon-polaritons and electron-phonon coupling}

Cavity photons can hybridize with phonons to form phonon-polaritons, modifying electron-phonon coupling and phonon frequencies. For instance, Ref.~\cite{sentef.ruggenthaler.ea_2018} studied a monolayer of FeSe on an SrTiO$_{3}$ substrate, focusing on the interaction between a cross-interfacial phonon mode between these two materials and cavity photons within a Fabry-Pérot cavity. A photon mode is polarized along the $z$-axis, parallel to the phonon dipoles. The phonons and cavity photons hybridize to form (phonon-)polaritons, which are then coupled to the electrons in the FeSe/SrTiO$_{3}$ system. The hybrid system was modeled using an electron-polariton Hamiltonian within the Migdal-Eliashberg framework with a self-consistent Nambu Green’s function approach and the bare polaritonic Green's function to compute the superconducting gap and then superconducting transition temperature. The dimensionless electron-phonon coupling strength is then extracted from the electron self-energy. The key finding of Ref.~\cite{sentef.ruggenthaler.ea_2018} is that while the cavity-induced modification enhances the electron-phonon coupling strength at all temperatures, this enhancement is offset by a reduction in the effective frequency, leading to a slight decrease in the superconducting transition temperature T$_{\text{c}}$. These results highlight that cavity coupling can enhance electron-phonon interactions, but the overall superconducting transition temperature depends on the balance between electron-phonon coupling strength and phonon frequency. Similarly, Ref.~\cite{hagenmuller.schachenmayer.ea_2019} explored hybrid plasmon-phonon modes in metallic crystals, demonstrating significant enhancements in electron-phonon scattering under specific conditions, such as thin crystal geometries and resonant phonon frequencies.

\begin{figure}[t]
\centering
\includegraphics[width=1.0\textwidth]{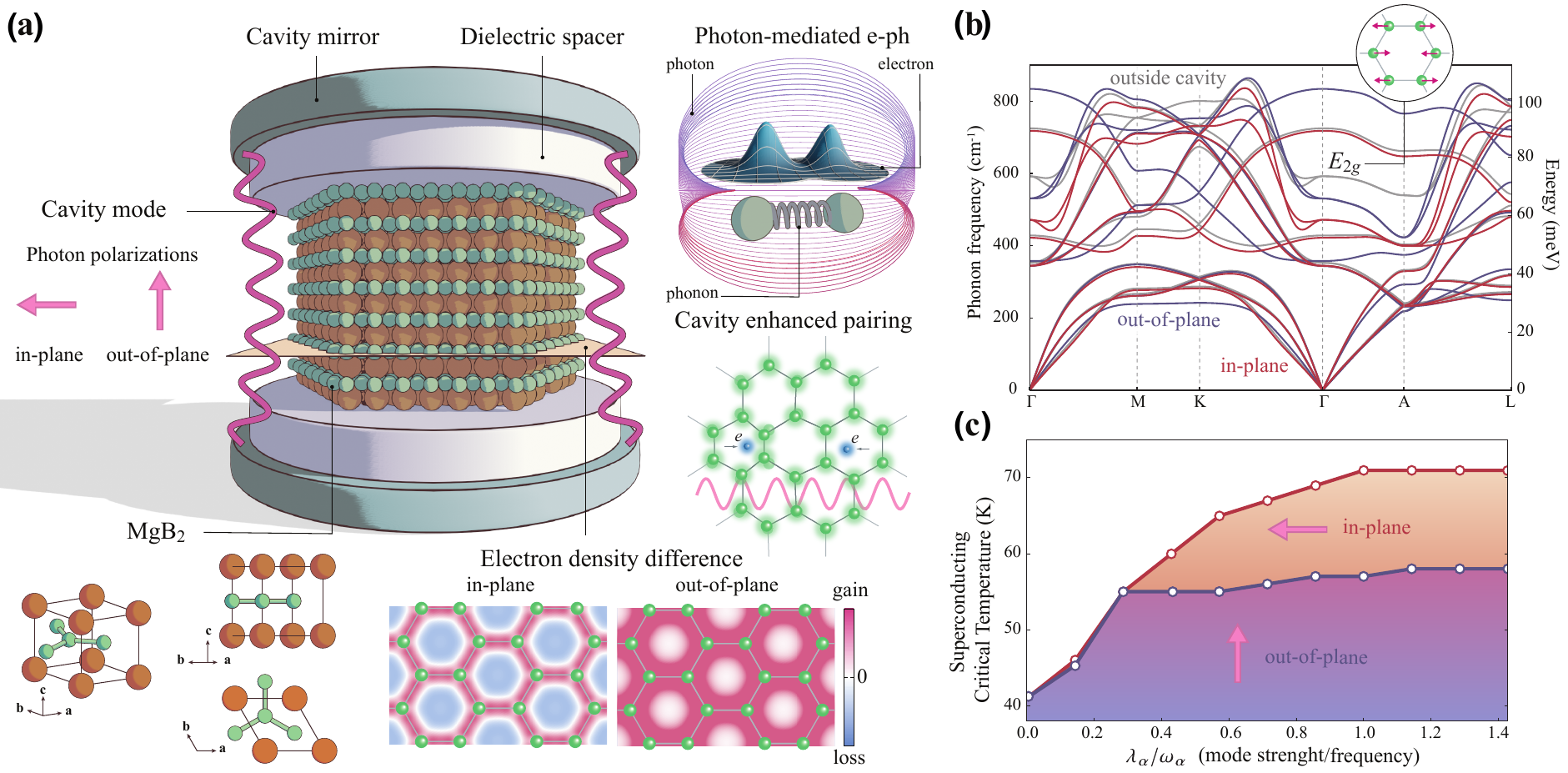}
\caption{
\textbf{Cavity-modified superconductivity using \ac{QEDFT}.} 
(\textbf{a}) MgB$_{2}$ (Magnesium atoms in orange and Boron atoms in green) is placed inside an optical cavity.
Here we explore two cavity setups, an out-of-polarized cavity with one "effective" photon mode perpendicular to the Boron planes and an in-plane polarized cavity with two effective modes parallel to the planes.
The electronic ground state and phonon properties are computed using \ac{QEDFT} and \ac{DFPT}, respectively.
Cavity-induced change in electron density on the Boron plane in MgB$_2$ modifies the phonon-mode $E_{2g}$, which mainly drives the superconductivity.
The electron-phonon coupling is enhanced due to the light-matter coupling, leading to enhanced superconductivity. 
%
%
(\textbf{b}) Calculated phonon dispersion of MgB$_{2}$ along high symmetry lines outside and inside the in-plane and the out-of-plane polarized cavity. 
The $E_{2g}$ mode softens due to the screening of the enhanced electron density between the Boron-Boron bonds inside a cavity, reducing the Coulomb repulsion between Boron atoms. 
The ratio of the mode strength (light-matter coupling) and photon frequency ($\lambda_{\alpha}/\omega_{\alpha}$) is $1.0$ for both cavity setups.
(\textbf{c}) Superconducting transition temperature as a function of the mode strength $\lambda_{\alpha}$ and photon frequency $\omega_{\alpha}$ ratio.
$\lambda_\alpha/\omega_\alpha$ can reach up to around $0.1$ for a polaritonic cavity setup. 
Reprinted from Lu \textit{et al.}, Proc. Natl. Acad. Sci. 121, e2415061121 (2024)~\cite{lu.shin.ea_2024a}. Copyright 2024 Authors, licensed under Creative Commons Attribution License 4.0 (CC BY).
}
\label{fig:theory-cavity-modified-superconductivity}
\end{figure}

\subsubsection{Cavity-modified electron density and electron-phonon coupling}
In contrast to coupling phonon and photon to modify the electron-phonon interaction, another approach to modifying electron-phonon coupling involves directly coupling electrons to cavity photons. This interaction can alter the electron density, indirectly affecting phonon frequencies and electron-phonon coupling. Recent advancements in \ac{QEDFT} (see Sec.~\ref{subsubsec:QEDFT}) have enabled the simulation of these coupled systems, providing insights into the effects of cavity-mediated interactions on superconducting properties. Ref.~\cite{lu.shin.ea_2024a} applied this approach to MgB$_{2}$ and predicted an enhancement of the superconducting transition temperature by up to $5$ K in a realistic cavity (see Fig.~\ref{fig:theory-cavity-modified-superconductivity}). This enhancement was attributed to softening the $E_{2g}$ Raman phonon mode, a key mediator of Cooper pairing in MgB$_{2}$, and to the increase of the electron-phonon coupling. The softening of this phonon mode resulted from the redistribution of electrons within the material, screening the Coulomb repulsion between boron atoms. Ref.~\cite{lu.shin.ea_2024a} emphasized the importance of self-consistent first-principles calculations for cavity-modified systems, combining \ac{QEDFT} with phonon simulations. Such methods provide a quantitative understanding of cavity effects on superconducting materials. Note that in Ref.~\cite{lu.shin.ea_2024a}, the light-matter coupling was treated as a free parameter but can be potentially computed using \ac{MQED} as discussed in Sec.~\ref{subsec:describe-photonic-env}.

\subsubsection{Cavity-induced Ampearean superconductivity}
Compared to the above BCS-type Cooper pairing mechanism with zero center-of-mass momentum, cavities can potentially mediate non-zero center-of-mass momentum pairing, known as Amperean superconductivity. Ref.~\cite{schlawin.cavalleri.ea_2019} investigated a 2D electron gas on a square lattice coupled to virtual cavity photons, revealing effective attractive interactions between electrons with parallel momenta via current-current interactions. The study estimated a critical temperature in the low-kelvin range in sub-wavelength cavities. However, Ref.~\cite{andolina.depasquale.ea_2024} argues that Amperean superconductivity cannot be achieved in deep sub-wavelength cavities, which induce mainly density fluctuation interactions rather than current-current interactions. Ref.~\cite{andolina.depasquale.ea_2024} estimated the superconducting transition temperature in a Fabry-Pérot cavity to be around $10^{-6}$ K for mirrors separated by approximately $1 \mu$m. Further analysis revealed a power-law dependence of the cavity-induced critical temperature on the light-matter coupling constant, contrasting the exponential dependence of the electron-phonon coupling constant in phonon-mediated superconductivity. While planar optical cavities support Amperean superconductivity, subwavelength nanoplasmonic cavities, as analyzed using a dyadic Green's function approach~\cite{andolina.depasquale.ea_2024}, were found to induce density-density interactions rather than the current-current interactions necessary for Amperean pairing.

\subsubsection{Cavity-induced quantum Eliashberg effect}
The Eliashberg effect, which describes the enhancement of superconductivity through the redistribution of quasiparticles, has been traditionally studied under external pumping~\cite{demsar_2020,sobolev.lanz.ea_2023}. 
Ref.~\cite{curtis.raines.ea_2019} proposed that fluctuating \ac{EM} fields within a resonant cavity can redistribute Bogoliubov quasiparticles (BQPs) in an $s$-wave superconductor, modifying the superconducting gap. Through a perturbative analysis, three types of BQP scattering processes were identified: scattering among BQPs, pair breaking, and pair recombination. These processes contribute differently to the superconducting gap depending on the photon frequency. At lower frequencies, scattering processes dominate, cooling existing BQPs and leading to a buildup near the gap edge. Conversely, at higher frequencies, recombination processes become more prevalent, reducing the BQP population and enhancing the superconducting gap. These findings suggest that cavity photons can potentially redistribute the quasiparticles, modifying superconducting properties without requiring external radiation using intrinsic quantum fluctuations instead.

\subsubsection{Cavity-enhanced superconductivity via band engineering}
Cavity photons can modify the electronic band structure of materials, leading to enhanced superconducting properties. Ref.~\cite{kozin.thingstad.ea_2024} demonstrated this effect in a 2D superconducting electron system described by an attractive Hubbard model. Coupling to cavity photons via a Peierls phase squeezed the electronic band structure, effectively renormalizing the electron mass~\cite{rokaj.ruggenthaler.ea_2022,li.schamriss.ea_2022,welakuh.rokaj.ea_2023}. This mass renormalization increased the density of states, resulting in a larger superconducting gap within BCS theory. The study showed a linear relationship between the gap and light-matter coupling strength, offering a pathway for engineering superconductivity through band structure control.

\subsubsection{Competing phases and superconductivity}
Coupling cavity photons to materials can shift the balance between competing phases -- the matter phases with similar free energies -- such as \ac{CDW} and superconducting phases. Therefore, controlling the competing phases is another approach to control superconductivity. 
Ref.~\cite{li.eckstein_2020} used a quantum Floquet formalism~\cite{sentef.li.ea_2020} to demonstrate how cavity-mediated interactions can selectively enhance either \ac{CDW} or superconducting states. For example, in the absence of photons, increasing the light-matter coupling strength promotes \ac{CDW} order while suppressing the superconducting state. This behavior contrasts with classical Floquet driving, where blue-detuned light enhances superconductivity. By tuning the cavity environment, the energy landscape of materials can be manipulated to favor desired quantum phases.

\subsubsection{Challenges and opportunities}
While cavity-modified superconductivity offers exciting possibilities, several challenges remain. For example, accurate predictions require self-consistent modeling among electrons, phonons (or nuclei/ions), photons within light-matter coupled systems, such as the first-principles methods with \ac{QED} frameworks. Another significant challenge is to extend these methods to complex, real-world materials without exponential computational costs. Finally, demonstrating cavity-induced, especially, cavity-enhanced superconductivity experimentally in dark cavity setups demands precise control over cavity parameters and material systems.

\subsection{Cavity engineering of strongly correlated systems}\label{subsec:cavity-modified-strongly-correlated-systems}
Strongly correlated electron systems exhibit a rich array of exotic phenomena, including high-temperature superconductivity, colossal magnetoresistance, and metal-insulator transitions~\cite{chen.wang.ea_2024,imada.fujimori.ea_1998}. Unlike conventional materials, these systems defy description by standard band theory due to the dominant role of electron-electron interactions. To capture these effects, theoretical models beyond mean-field approximations such as the Hubbard model~\cite{arovas.berg.ea_2022,qin.schafer.ea_2022} are essential~\cite{martin.reining.ea_2016}. Recent advances in \ac{cQED} provide tools to manipulate these systems by introducing photon-mediated interactions and stabilizing novel quantum phases. This section explores the interplay between strong correlations and cavity fields, focusing on theoretical progress in modifying Hubbard models, magnetic phases, and the Kondo effect.

\subsubsection{Introduction to strongly correlated systems}
The Hubbard model provides a foundational framework for describing strongly correlated systems, capturing phenomena like metal-to-insulator transitions and magnetism~\cite{khomskii_2010}.
The Hubbard model explicitly incorporates the strong repulsion between electrons on the same atomic site, allowing for a more accurate description of these systems. 
The Hamiltonian for the Hubbard model in the simplest form is given by:
\begin{equation}
\hat{H}_{\rm{Hubbard}}=-t\sum_{\langle i,j\rangle,\sigma}(\hat{c}_{i\sigma}^{\dagger}\hat{c}_{j\sigma}+\hat{c}_{j\sigma}^{\dagger}\hat{c}_{i\sigma})+U\sum_{i}\hat{n}_{i\uparrow}\hat{n}_{i\downarrow},   
\end{equation}
where $t$ is the hopping amplitude between nearest-neighbor sites $i$ and $j$, $U$ is the on-site Coulomb repulsion, $\hat{c}_{i\sigma}^{\dagger}$ and $\hat{c}_{i\sigma}$ are creation and annihilation operators for an electron with spin $\sigma$ at $i$th site, and $\hat{n}_{i\sigma}$ is the number operator for electrons with spin $\sigma$ at $i$th site. The first term in the Hamiltonian describes the kinetic energy of electrons hopping between lattice sites, while the second term represents the potential energy due to the Coulomb repulsion between electrons on the same site. While the Hubbard model provides a foundational framework for understanding strongly correlated electron systems, it represents a simplified representation of real materials. To capture more complex phenomena, various extensions have been developed, including extended Hubbard models that incorporate long-range Coulomb interactions and multi-orbital models that consider multiple electronic orbitals per site. Techniques like \ac{DMFT}~\cite{pavarini.koch.ea_2018} further refine these models by mapping the lattice problem onto an effective impurity problem to address the challenges of strong correlations. More complex phenomena require extensions like the $t$-$J$ model, Kondon, Anderson impurity, or Heisenberg models, tailored to specific materials~\cite{khomskii_2010}. 

The interplay between hopping $t$ and on-site repulsion $U$, often quantified by the $U/t$ ratio, determines electronic and magnetic phases~\cite{khomskii_2010}. Experimental techniques such as strain engineering, twisting in \ac{2D} materials, or applying external fields have traditionally been used to tune these parameters~\cite{sclauzero.dymkowski.ea_2016,kim.liu.ea_2018,wu.lovorn.ea_2018,tancogne-dejean.sentef.ea_2018,baykusheva.jang.ea_2022,granas.vaskivskyi.ea_2022,xu.kang.ea_2022}. However, recent advances in \ac{cQED} offer an additional avenue for controlling Hubbard model parameters, potentially leading to novel phases and properties. For the discussion later, when solving the Hubbard model in the large $U$ limit at half-filling (one electron per site), where $U \gg t$, the model is treated perturbatively up to the doubly occupancy using the Schrieffer-Wolff transformation~\cite{khomskii_2010}. This approach results in an effective low-energy Hamiltonian, known as the spin-$\frac{1}{2}$ Heisenberg model, given as 
\begin{equation}
H = J_{\text{ex}} \sum_{i} \hat{\mathbf{S}}_{i} \cdot \hat{\mathbf{S}}_{i+1},
\end{equation}
where the exchange interaction is $J_{\rm{ex}} = 4t^2/U$ and $\hat{\mathbf{S}}_{i}$ represents the spin-$\frac{1}{2}$ operator.

\subsubsection{Cavity-modified Hubbard model and its extensions}

The extended Hubbard model with electron-photon coupling is proposed by incorporating photon field operators via the Peierls substitution~\cite{sentef.li.ea_2020}, building on a minimal gauge-invariant lattice model~\cite{li.golez.ea_2020}:
\begin{equation} \label{eq:H-QF-TB}
\hat{H}_{\rm{QF}}=-t\sum_{\langle i,j\rangle,\sigma}(\hat{c}_{i\sigma}^{\dagger}\hat{c}_{j\sigma}e^{i\hat{A}}+\hat{c}_{j\sigma}^{\dagger}\hat{c}_{i\sigma}e^{-i\hat{A}})+U\sum_{i}\hat{n}_{i\uparrow}\hat{n}_{i\downarrow} + \Omega\hat{a}^{\dagger}\hat{a}.
\end{equation}
Similar to the Hubbard model, the Hamiltonian can be reduced to the so-called \textit{spin-photon} Heisenberg model under the limit $U\gg t$ at the half-filling~\cite{sentef.li.ea_2020,vinasbostrom.sriram.ea_2023}, 
\begin{equation}
\hat{H} = \sum_{\langle rs\rangle}\left(\hat{\mathbf{S}}_{r} \cdot \hat{\mathbf{S}}_{s}-\frac{1}{2}\right)\hat{\mathcal{J}}_{\langle rs\rangle}[\hat{a}^{\dagger},\hat{a}] + \Omega\hat{a}^{\dagger}\hat{a},
\end{equation}
where $\hat{\mathcal{J}}_{\langle rs\rangle}$ is determined by the exchange operator of the two-side model with the hopping $t$ and light-matter coupling $g$ for a given bond~\cite{sentef.li.ea_2020}; the cavity-modified exchange interaction thus deviates from the bare value $J_{\rm{ex}}$. 
This spin-photon Heisenberg model has been recently applied to an AA-stacked honeycomb and solved numerically using quantum Monte Carlo simulations to explore its size scaling behavior of the magnetic structure factor, susceptibility, and energy~\cite{weber.vinasbostrom.ea_2023}.
In the high-frequency limit, the resulting exchange operator in the spin-photon Hamiltonian becomes diagonal in the photon Fock space. Thus, for example, the spin-photon Heisenberg Hamiltonian for two-site Hubbard dimer becomes~\cite{sentef.li.ea_2020,vinasbostrom.sriram.ea_2023}
\begin{equation}
\hat{H} = \sum_{n}\ket{n}\bra{n}\left(\hat{H}_{n}^{\rm{spin}}+n\Omega\right),
\end{equation}
with $\hat{H}_{n}^{\rm{spin}} = (\hat{\mathbf{S}}_{1}\cdot\hat{\mathbf{S}}_{2}-\frac{1}{2})J_{\rm{ex}}^{(n)}$, and the exchange interaction $J_{\rm{ex}}^{(n)}=\mel{n}{\hat{\mathcal{J}}_{0}}{n}$.
Thus, this photon-coupled Hubbard model reveals two main effects: the renormalization of hopping (dynamical localization) and cavity-modified spin exchange interaction. Additionally, Ref.~\cite{sentef.li.ea_2020} shows that Floquet-like engineering can be achieved with relatively weak photon amplitude in the strong light-matter coupling regime.
The above spin-photon Hamiltonian only includes the short-range interaction between nearest neighbor spins, i.e., $\hat{\mathbf{S}}_{1}\cdot\hat{\mathbf{S}}_{2}$, because of the approximation up to the second-order perturbation terms to eliminate the charge excitation. Ref.~\cite{fadler.schmidt.ea_2024} further extended the quantum Floquent model [Eq.~\eqref{eq:H-QF-TB}] to fourth-order perturbation terms, obtained long-range cavity-mediated interactions between pairs of spins, which cannot be connected by a hopping process, and pointed out the limitation of the spin-photon approach.

In contrast to the above extended Hubbard model that uses Peierls substitution for photon fields, Ref.~\cite{masuki.ashida_2024} uses full minimal coupling within the light-matter coupled Hamiltonian, which is then reduced to an effective low-energy Hamiltonian with cavity-modified band dispersion. Although Ref.~\cite{masuki.ashida_2024} assumes a constant Hubbard $U$, the appendix in Ref.~\cite{masuki.ashida_2024} demonstrates that the change of the Hubbard U via cavity is similar to the changes in band dispersion caused by the cavity. A recent theoretical proposal~\cite{tancogne-dejean.sentef.ea_2018} and experimental evidence~\cite{baykusheva.jang.ea_2022,granas.vaskivskyi.ea_2022} also suggest that the Hubbard U term can be strongly renormalized using external lasers. The modification of the Hubbard U inside a cavity has not been fully explored, which might be one of the future directions. In addition to modifying the hopping term (and the Hubbard U term) in the Hubbard model via electron-photon coupling, coupling photons with phonons to form phonon-polariton can also effectively modify the Hubbard U term for electrons via electron-phonon(-polariton) coupling~\cite{de.eckhardt.ea_2022}.

\subsubsection{Cavity-modified magnetic phases}

The ground state of the Hubbard model is highly related to the magnetic phases. Recent studies have demonstrated the potential of manipulating the magnetic properties of materials using photon field fluctuations. By coupling matter to the photons within a cavity, it is possible to induce significant changes in the electronic and magnetic structure.

\begin{figure}[t]
    \centering
    \includegraphics[width=1.0\textwidth]{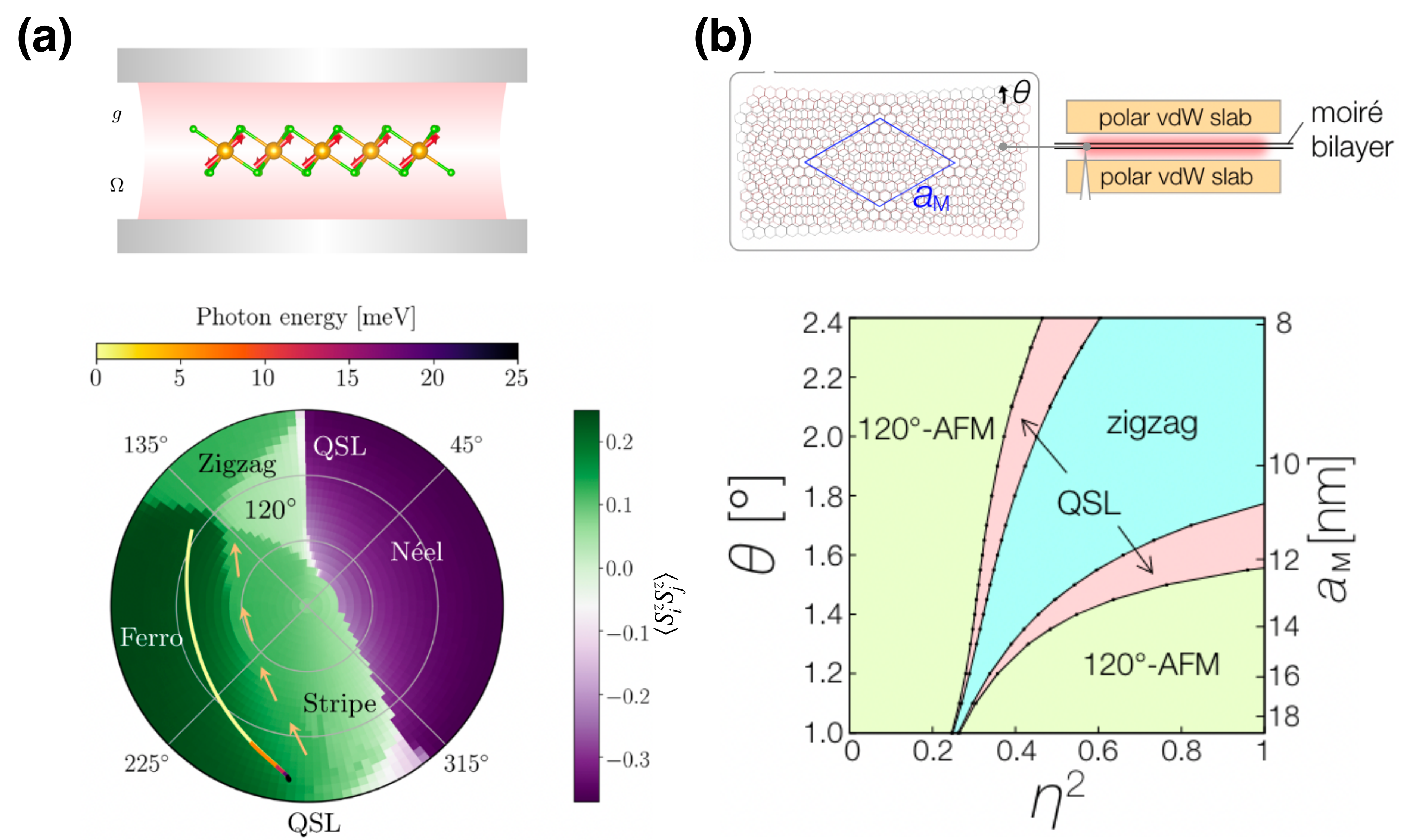}
    \caption{\textbf{Theoretical proposed cavity-modified magnetic phases of electronic systems.}
    (a) The ground state of $\alpha$-RuCl$_{3}$ is a zigzag antiferromagnetic phase outside a cavity (top panel); here Ru atoms are shown in organe, while Cl atoms are shown in green. The red arrows indicate the spin direction of each Ru atom. The schematic of $\alpha$-RuCl$_{3}$ inside a cavity with a bare light-matter coupling $g$ and photon frequency $\Omega$. The magnetic phase diagram of $\alpha$-RuCl$_{3}$ inside a cavity (bottom panel) is computed using the extended Kitaev Hamiltonian coupled to effective photon modes with the materials parameters obtained from the first principles calculations. The paths on the diagram are traced out and colored according to the cavity photon energy as the effective light-matter coupling $g_{\rm{eff}}$ increases from $0$ to $0.5$.
    Reprinted from Viñas Boström \textit{et al.}, Npj Comput. Mater. 9, 202 (2023)~\cite{vinasbostrom.sriram.ea_2023}. Copyright 2023 Authors, licensed under Creative Commons Attribution License 4.0 (CC BY).
    (b) A twisted 2D Moiré \ac{TMD} lattice is embedded in a phonon-polariton cavity (top panel). The magnetic phase diagram as a function of small twisted angles and (integrated dimensionless) light-matter coupling $\eta^{2}$ includes the zigzag, the \ac{AFM}, and the quantum spin liquid (QSL) phase (bottom panel).
    Reprinted figure with permission from Masuki et al., Phys. Rev. B 109, 195173 (2024)~\cite{masuki.ashida_2024}. Copyright 2024 by the American Physical Society. 
    }\label{fig:theory-cavity-modified-magnetic-phase}
\end{figure}

%
Ref.~\cite{vinasbostrom.sriram.ea_2023} demonstrated control over the magnetic ground state of a strongly correlated system, $\alpha$-RuCl$_{3}$, using \ac{cQED} model Hamiltonian. $\alpha$-RuCl$_{3}$ has several competing magnetic orders near its ground state, serving as a good candidate for exploring cavity-controlled phase transitions. By incorporating photon interactions via a Peierls substitution into a Hubbard-like model~\cite{winter.li.ea_2016,sriram.claassen_2022} and employing a downfolding technique to eliminate electronic degrees of freedom, Ref.~\cite{vinasbostrom.sriram.ea_2023} derived an effective spin-photon model from the Kitaev model, introducing cavity-modified magnetic interactions (or spin parameters). The original magnetic interaction parameters, including exchange couplings and Hubbard U, were obtained from \ac{DFT} calculations (as discussed in Sec.~\ref{subsec:abinitio-matter}). The magnetic phase transition is driven by a complex interplay of factors, including the suppression of ligand-mediated hopping and the enhancement of direct Ru-Ru hopping. Below a critical photon energy of approximately $10$ meV, the system undergoes a transition from a zigzag antiferromagnetic to a ferromagnetic ground state as light-matter coupling strengthens [Fig.~\ref{fig:theory-cavity-modified-magnetic-phase}(a)]. Furthermore, the model predicts the possibility of accessing a Kitaev \ac{QSL} phase through cavity pumping~\cite{vinasbostrom.sriram.ea_2023}, demonstrating the potential for engineering exotic quantum states using light-matter interactions. However, the practical realization of such engineered phases relies on precise control over cavity parameters and photon number, posing challenges for experimental implementation. Similarly, another study~\cite{chiocchetta.kiese.ea_2021} uses a long-range Heisenberg model on a square lattice to show that quantum field fluctuations can stabilize \ac{QSL}.

In addition to $\alpha$-RuCl$_{3}$, Ref.~\cite{masuki.ashida_2024} used a Moiré \ac{TMD} heterobilayer, which is coupled to a phonon-polariton cavity composed of polar van der Waals slabs such as hexagonal boron nitride, for exploring light-matter interactions and their impact magnetic phases. By tuning the twist angle $\theta$ and light-matter coupling strength $\eta$, a range of magnetic phases, from antiferromagnetic states to exotic \ac{QSL}s, can be accessed [Fig.~\ref{fig:theory-cavity-modified-magnetic-phase}(b)]. Using an effective triangular Heisenberg model, Ref.~\cite{masuki.ashida_2024} demonstrated that virtual interband transitions drive significant band energy renormalization, with pronounced effects in systems with flatter electronic bands. While this work sheds light on the interplay between electronic structure and cavity-mediated interactions, it neglects the potential impact of cavity-induced modifications to the Hubbard $U$ term. Incorporating such effects could provide a more comprehensive understanding of energy renormalization and extend the applicability of this approach.

\subsubsection{Cavity-enhanced Kondo effect}
%
The Kondo effect is another example of strong correlations in electron systems, highlighting how localized magnetic impurities can significantly influence the behavior of conduction electrons. It illustrates the intricate interplay between localized spins and the conduction electron sea in strongly correlated materials. The characteristic energy scale associated with this phenomenon is the Kondo temperature, $T_{\rm{K}}$, which dictates the temperature range where the Kondo effect becomes prominent. Below $T_{\rm{K}}$, resistivity increases logarithmically due to enhanced electron scattering by magnetic impurities. Ref.~\cite{mochida.ashida_2024} investigates the potential for controlling the Kondo effect through cavity confinement, starting with an extended single-impurity Anderson model that incorporates cavity photons coupled to conduction electrons in the \ac{LWA}. By applying an asymptotically decoupling unitary transformation~\cite{ashida.imamoglu.ea_2021,ashida.yokota.ea_2022}, the model is reduced to an effective Hamiltonian for the cavity-induced Kondo effect via perturbation theory. This Hamiltonian includes an additional nonlocal electron-electron interaction mediated by the cavity field, which competes with the electrons' kinetic energy. The cavity-mediated interaction effectively increases the electron mass, leading to an enhanced density of states at the Fermi surface, $\rho_{\rm{F}}$, and subsequently raising the Kondo temperature $T_{\rm{K}}\propto \exp(-1/(J\rho_{\rm{F}}))$, where $J$ represents the Kondo exchange interaction. These findings highlight the potential for engineering electronic correlations and localized spin behaviors using cavity fields.

\subsection{Theoretical predicted cavity-modified electronic topology of solid-state materials}\label{subsec:cavity-modified-electronic-topology}


Topological materials exhibit unique behaviors that distinguish them from trivial systems~\cite{xiao.chang.ea_2010}. Their topological properties, including topological surface states, Chern numbers, and the quantum anomalous Hall effect, profoundly impact the electronic and optical characteristics of solid-state materials~\cite{xiao.chang.ea_2010,thouless.kohmoto.ea_1982,haldane_1988,bernevig.hughes.ea_2006}. 
These properties are closely linked to the quantum geometry of electronic structures in materials, providing critical fundamental electronic behavior in the \ac{BZ}. For instance, the Chern number -- $C = \frac{1}{2\pi} \int_{\text{BZ}} \Omega(\mathbf{k}) \, d^2\mathbf{k}$, where $\Omega(\mathbf{k})$ is the Berry curvature at $\mathbf{k}$ in the \ac{BZ} -- is associated with the quantum Hall effect, which leads to quantized electrical conductance in two-dimensional materials under strong magnetic fields~\cite{thouless.kohmoto.ea_1982}. Investigating and exploiting these properties can unravel new states of matter, such as topological insulators, which facilitate efficient, dissipationless electron transport on their surfaces, which are the boundary between the non-trivial states inside the material and the trivial states outside~\cite{haldane_1988,murakami.nagaosa.ea_2003}. For example, the non-zero Chern number can be achieved in magnetic material without an external magnetic field, which is known as a quantum anomalous Hall insulator. Additionally, these topological properties hold promise for quantum computing, spintronics, and topological photonics applications, driving technological advancements with significant societal impact~\cite{gilbert_2021,liu.hersam_2019}. 
The topology of a material can be altered through strong light-matter coupling in quantum light-matter systems. When a laser pump excites carriers or distorts the lattice, it can trigger topological phase transitions due to various factors, such as the altered symmetry group from lattice distortion and the Floquet-engineered electronic structure~\cite{mciver.schulte.ea_2020,vaswani.wang.ea_2020,disa.nova.ea_2021,shin.rubio.ea_2024}. Similarly, the following theoretical studies suggest that coupling confined photons with a material in a cavity can modify its topological phase transitions. 

\subsubsection{Cavity-modified electronic topological properties of 2D materials}

\begin{figure}[t]
    \centering
    \includegraphics[width=1.0\textwidth]{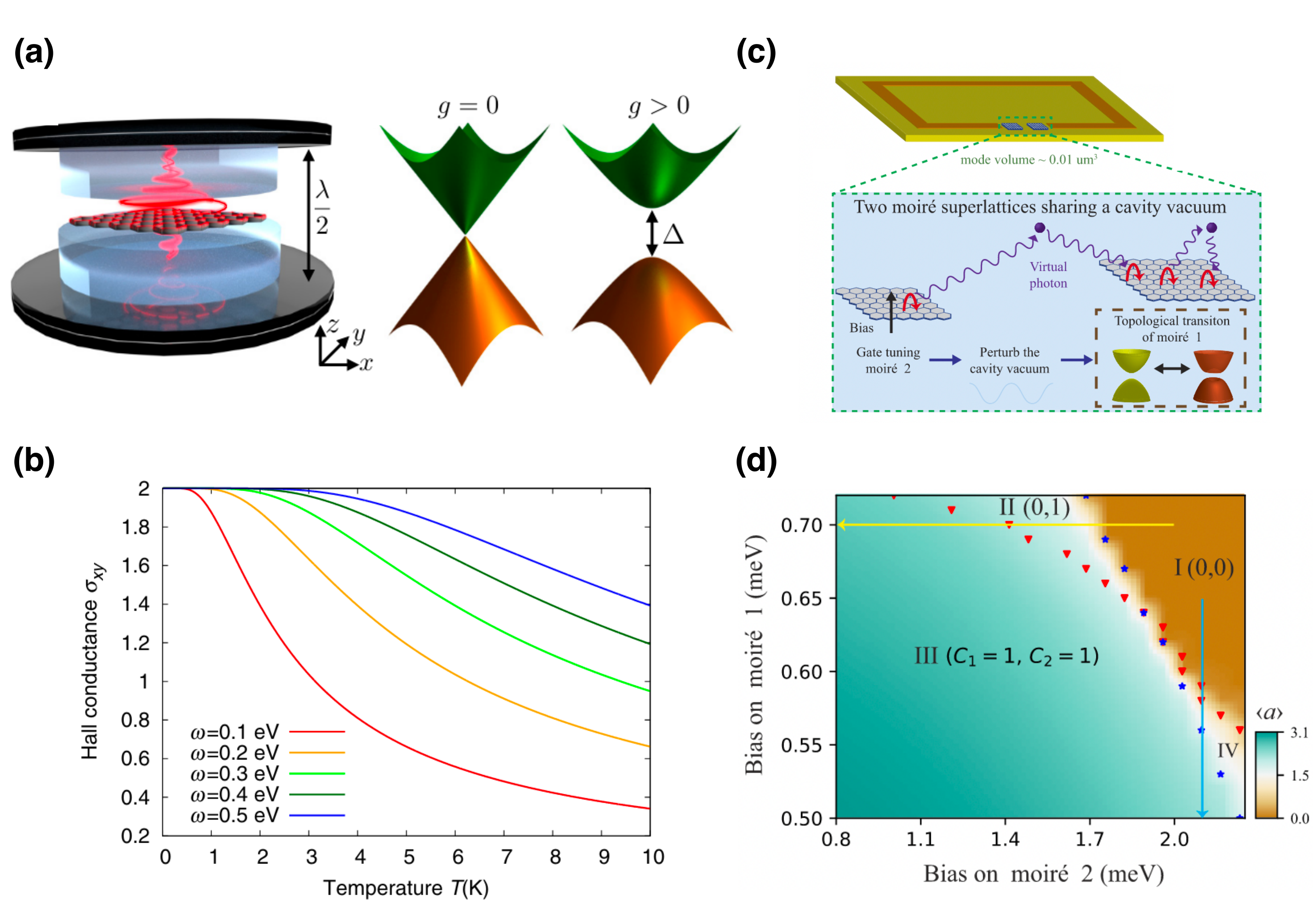}
    \caption{\textbf{Proposed electronic topology via quantum vacuum fluctuations.}
    (a) Modified Dirac points of graphene through a chiral photon mode of the optical cavity. Here $\lambda/2$ is the cavity mirror distance, $g$ is the electron-photon coupling strength, and $\Delta$ is the band gap at Dirac points.
    (b) Chiral cavity-induced Hall conductivity $\sigma_{xy}$ of graphene with respect to temperature and photon frequency $\omega$.
    (a) and (b) are reprinted figures with permission from Wang \textit{et al.}, Phys. Rev. B 99, 235156 (2019)~\cite{wang.ronca.ea_2019}. Copyright 2019 by the American Physical Society.  
    (c) Schematic scheme for gate control of the topology of two Moiré lattices, which are inside a metallic split-ring THz \ac{EM} resonator, via a cavity photon mode. Applying a bias on the Moiré lattice 2 perturbs the vacuum field inside the cavity, leading to the topological transition in the Moiré lattice 1. 
    (d) Phase diagram of twisted Moiré lattice with respect to gate bias. $\langle\hat{a}\rangle$ is the ground-state expectation value of the field operator $\hat{a}$.
    (c) and (d) are reprinted from Lin \textit{et al.}, Proc. Natl. Acad. Sci. 120, e2306584120 (2023)~\cite{lin.xiao.ea_2023}. Copyright 2023 Authors, licensed under under Creative Commons Attribution-NonCommercial-NoDerivatives License 4.0 (CC BY-NC-ND).  
    }\label{fig:theory-cavity-modified-topology}
\end{figure}

Quantum vacuum fluctuations of circularly polarized photon modes were demonstrated to open the band gap of graphene at the Dirac cone~\cite{kibis.kyriienko.ea_2011,wang.ronca.ea_2019,dag.rokaj_2024} due to the time-reversal symmetry breaking by the circularly polarized photon modes [see Fig.~\ref{fig:theory-cavity-modified-topology}(a)]. In addition to the modification of the band gap, the quantum vacuum fluctuations have been suggested and explored to modify the electronic topological properties~\cite{hubener.degiovannini.ea_2021,wang.ronca.ea_2019,masuki.ashida_2023,nguyen.arwas.ea_2023,dag.rokaj_2024}. The computed topological invariant for \ac{2D} materials is primarily the Chern number, which can be computed using the electronic wave functions/eigenstates~\cite{vanderbilt_2018} or Green's function method~\cite{wang.qi.ea_2010}. With the latter method, a circularly polarized photon mode in a cavity was shown to induce the quantum anomalous Hall effect, which is deeply connected to the electronic topology, in graphene~\cite{wang.ronca.ea_2019} [see Fig.~\ref{fig:theory-cavity-modified-topology}(b)]. By evaluating temperature dependency on Hall conductance, cavity-induced quantized anomalous Hall conductivity can be achieved at low-temperature conditions (less $1$ K). This result indicates that the circularly polarized photon in a cavity can induce the topological phase transition of graphene to a Chern insulator from a Dirac semimetal. 

In addition to monolayer materials such as graphene, the framework of cavity-induced topological phase transitions has been extended to \ac{2D} Moiré materials and twisted bilayer systems~\cite{nguyen.arwas.ea_2023}. These systems introduce new opportunities due to their tunable electronic structures, governed by twist angles and superlattice periodicities. By computing the so-called electron-photon Chern number, Ref.~\cite{nguyen.arwas.ea_2023} demonstrated that coupling Moiré superlattices to cavity fields enables topological phases transitions~\cite{nguyen.arwas.ea_2023}. This approach represents a significant advance over prior methods, which relied solely on electronic wave functions or Green's function techniques~\cite{wang.ronca.ea_2019,dag.rokaj_2024}, as it directly incorporates the effects of light-matter hybridization.


Interestingly, superradiant phase transitions in cavity-coupled electronic systems have also been proposed as a mechanism for generating topological transitions, including Fermi surface lifting and the emergence of bands with nonzero Chern numbers~\cite{guerci.simon.ea_2020}. This highlights the versatility of light-matter interactions in enabling exotic quantum states, though experimental verification of these predictions remains an open challenge.

Quantum light-matter interaction under the optical cavity can also remotely induce topological phase transition of \ac{2D} Moiré materials via a gate bias~\cite{lin.xiao.ea_2023}.
It is experimentally observed that ferromagnetism emerges in a supercell of twisted MoTe$_2$ bilayer with one hole carrier~\cite{anderson.fan.ea_2023}. Similar to this observation, this theoretical work~\cite{lin.xiao.ea_2023} demonstrates that gate bias leads to the topological phase transition by evaluating the system's Chern number ($C_{1}$ and $C_{2}$) in the twisted \ac{TMD} homobilayer, as shown in Fig.~\ref{fig:theory-cavity-modified-topology}(c). Here, a \ac{TB} model Hamiltonian to describe the Moiré superlattice is constructed, and the light-matter interaction is included via a Peierls substitution. With the gate bias, the topological phase transitions toward the Chern insulator for both are achieved, as shown in Fig.~\ref{fig:theory-cavity-modified-topology}(d). In this phase transition, the band inversion is achieved at non-symmetry points modified by an optical cavity. Because the expectation value of the cavity field $<\hat{a}>$ can be non-negligible with interband coherence, it can be an indicator for topological phase transition in this system. Notably, the single-layer phase transition can be formed depending on the ratio between gate biases for each layer. For example, the red (blue) dots indicate the phase transition boundary for the 1st (2nd) layer, respectively, as shown in Fig.~\ref{fig:theory-cavity-modified-topology}(d). This result indicates that the vacuum field fluctuations contribute to the topological phase transition with gate-biased conditions.

\subsubsection{Cavity-controlled quantum geometry of materials}
The quantum geometric tensor, comprising the Berry curvature and quantum metric, plays a crucial role in understanding topological phenomena and material properties~\cite{ma.grushin.ea_2021}. Direct measurements of both the Berry curvature and quantum metric can also be achieved in experiments~\cite{gianfrate.bleu.ea_2020}. Recently, the broader implications and open problems of quantum geometry in quantum materials have been discussed~\cite{torma_2023}. For example, it is demonstrated that light can modulate quantum geometry through Floquet engineering, such as anomalous Hall effect and higher order non-local responses~\cite{oka.aoki_2009,tai.claassen_2023,sato.giovannini.ea_2020,mciver.schulte.ea_2020}. These studies suggest strong light-matter interaction can control the quantum geometry and related properties. Therefore, exploring how quantum vacuum fluctuations of photons affect the quantum geometry in quantum materials by the ground state modification is a new direction away from the light-induced non-equilibrium state. Among quantum materials, flat-band systems, such as Moiré materials, are particularly intriguing due to their vanishing band velocity and curvature. In these systems, the linear paramagnetic contribution from the gauge-invariant minimal coupling Hamiltonian is typically proportional to the electronic current density, which depends on the band velocity, while the quadratic diamagnetic contribution is proportional to electron density and inversely related to electron mass. Based on this, one might expect the light-matter coupling in flat-band systems to vanish. However, this is not the case. Recent work using a \ac{TB} model, which incorporates photon effects through the Peierls substitution (an exponential function), has revealed exceptional behavior~\cite{topp.eckhardt.ea_2021}. When the exponential term associated with the Peiers substitution is expanded, the matrix elements of light-matter coupling between different band indices emerge, introducing quantum geometry into the equation. This was demonstrated using two model systems: (i) a sawtooth quantum chain with a single flat band and (ii) a tight-binding model for twisted bilayer graphene. The first model showed that despite the flat band, there is a nonvanishing diamagnetic light-matter coupling due to quantum geometry. The second model explored how the twist angle in twisted bilayer graphene influences various light-matter coupling matrix elements. This indicates that the light-matter coupling plays a role in flat-band systems via quantum geometry.

\subsection{Discussion of experimental feasibility of theoretical predictions}\label{subsec:feasibility-prediction}

Among the above theoretical predictions, several appear more realistic to implement experimentally given current technology. For example, the induction of para-to-ferroelectric phase transitions within cavities has garnered significant theoretical attention. Numerous models suggest the feasibility of such transitions, yet experimental validation remains elusive. Key challenges include the design and fabrication of cavities with the requisite quality factors and mode volumes, as well as identifying the appropriate photon energy regimes that allow effective photon penetration and coupling to targeted phonon modes within materials.

In the domain of superconductivity, theoretical predictions for cavity-mediated enhancements vary in their experimental feasibility. Enhancing electron-phonon coupling, a well-established mechanism in conventional superconductivity, has been theoretically demonstrated for MgB$_{2}$ in realistic cavity settings. However, the metallic nature of MgB$_{2}$ presents a significant challenge: its plasma frequency causes photon reflection, preventing photons with lower frequencies from penetrating sufficiently deep into the material. If the material thickness exceeds the penetration depth, the photon fields cannot effectively couple with the electrons inside. Similarly, Amperean pairing—a proposed mechanism involving virtual photon-mediated electron pairing with non-zero center-of-mass momentum—faces practical limitations in deep sub-wavelength cavities, as indicated by recent theoretical studies.

The exploration of cavity-controlled magnetism is at an earlier stage, with theoretical efforts focused on modifying hopping parameters in Hubbard models and renormalizing the Hubbard $U$ term. While promising, these approaches require more detailed studies to account for material-specific effects, refine the models, and address practical challenges, such as achieving the necessary light-matter coupling strength and ensuring spatial uniformity within cavity environments.

In summary, the field of cavity materials engineering is evolving rapidly. While theoretical predictions continue to inspire new experimental directions, overcoming practical challenges will require close collaboration between theorists and experimentalists. Addressing these challenges will be crucial to controlling material properties through quantum vacuum fluctuations.\\

\section{Conclusion and outlook}\label{sec:outlook}

In this review, we have discussed the field of cavity materials engineering, which some of us initiated and pioneered. This area has since attracted numerous theoretical and experimental groups worldwide, who are now making key contributions and broadening the scope of the research. Let us first summarize what we have presented and then discuss future directions.

\subsection{Conclusion} 

We began with Sec.~\ref{sec:intro} by emphasizing that the \ac{EM} interactions among charged constituents govern most properties of matter. We argued that these interactions can be manipulated in innovative ways, particularly through the use of optical cavities, which enable profound modifications even under equilibrium conditions, without requiring external driving. Recognizing the complexity of real materials and the diverse range of effects, we highlighted the necessity of adopting a comprehensive \textit{ab initio} framework rather than relying solely on simplified model approaches. In Sec.~\ref{sec:methodoloy}, we introduced the Pauli-Fierz Hamiltonian, detailing its mathematical structure and theoretical underpinnings as the foundation for describing light-matter interactions in cavities. We addressed critical nuances, such as mass renormalization and the mismatches between the periodicity of light and matter, proposing methods to circumvent these inconsistencies using the long-wavelength approximation. For the quantum-optics community, we also provided an overview of key concepts from solid-state physics, with a focus on density-functional theory. This discussion naturally extended to a cavity-modified framework based on the Pauli-Fierz Hamiltonian, enabling the non-perturbative treatment of electrons, nuclei/ions, and photons. In Sec.~\ref{sec:experiments}, we reviewed various experimental advancements where optical cavities have successfully modified the equilibrium properties of solid-state matter. These studies highlight the potential of cavity quantum electrodynamics for solid-state material design and control. Lastly, in Sec.~\ref{sec:theory-prediction}, we explored the state-of-the-art theoretical predictions, identifying promising phenomena yet to be experimentally realized, which offer significant opportunities for future research. By integrating insights from theoretical, computational, and experimental perspectives, this review highlights the immense potential of optical cavities to reshape the fundamental properties of matter at equilibrium. Further advances in this field are poised to reveal unprecedented ways to engineer materials and quantum systems, bridging gaps between quantum optics, condensed matter physics, and materials science.

\subsection{Outlook}

\subsubsection{Experimental efforts and future directions from our perspective}
%
While theoretical proposals for cavity-engineered solid-state materials under dark-cavity conditions are plentiful (Sec.~\ref{sec:theory-prediction}), experimental validation remains scarce (Sec.~\ref{sec:experiments}). Key predictions, such as cavity-induced para-to-ferroelectric or anti-to-ferromagnetic phase transitions, have yet to be conclusively demonstrated in experiments. Experimental investigations are not only necessary to verify these phenomena but also to distinguish between competing mechanisms. For example, the cavity-modified phase transitions observed in 1T-TaS$_{2}$ (Sec.~\ref{subsec:MIT-TaS2}) require clarification on whether they arise from the thermal Purcell effect or changes to the ground-state energy landscape across different phases. Rigorous experimental benchmarks and systematic approaches are essential for testing these models and refining our understanding of cavity-induced phase transitions.

Achieving strong light-matter coupling in realistic experimental setups presents significant challenges. Issues such as cavity losses, limited tunability, and insufficient coupling strength often hinder the realization of predicted phenomena like cavity-induced superconductivity or ferroelectric transitions. Experimental setups could address factors such as cavity design and photon penetration depth into materials. High-quality cavities with minimal losses and uniform coupling are critical, yet scaling up these systems to accommodate larger, more complex materials while maintaining coherence and coupling strength remains a major challenge.

Chiral cavities represent another frontier for experimental exploration, offering a way to manipulate material properties through photon polarization-induced symmetry breaking~\cite{hubener.degiovannini.ea_2021}. These systems hold particular promise for materials where symmetry-breaking drives phase transitions, enabling control over quantum states and potentially unlocking previously inaccessible phases. Beyond dark cavities, the ability to regulate photon numbers within a cavity presents opportunities for dynamic control of material phases. Injecting photons into a cavity could stabilize phases like superconductivity or access exotic quantum spin liquids~\cite{li.eckstein_2020,vinasbostrom.sriram.ea_2023}. 
The reciprocal nature of light-matter interactions in such scenarios opens the door to fine-tuned control, where changes in the material also modify the properties of the cavity field. Similarly, driving cavities with external light fields -- coherent, squeezed, or otherwise -- offers another layer of control, potentially enabling non-equilibrium phases and Floquet-engineered states~\cite{zhou.liu.ea_2024}. This approach could mitigate heating issues inherent to intense laser pulses while providing precise control over electronic bands and topological phases in quantum materials.

The experimental challenges in cavity materials engineering are closely tied to the inherent complexity of real-world systems. In contrast to idealized theoretical models, experiments are subject to a host of uncontrolled variables, including imperfections in materials, environmental conditions, and cavity geometry. For example, factors such as defects, impurities, and substrate interactions can significantly alter the response of the material to cavity fields, making it difficult to predict experimental outcomes. Furthermore, real cavities do not behave like the idealized models typically assumed in theory. Cavity losses, photon penetration depth, and imperfect coupling between the material and the cavity modes introduce additional challenges in aligning theory with experiments. Additionally, different models are often used to simulate various material systems, each with varying levels of complexity in how photon environments are treated. The lack of a unified approach to simulating both the material and its photonic environment in a self-consistent way adds another layer of complexity, making it challenging to draw direct connections between theoretical predictions and experimental results. Moving forward, advancing experimental techniques and refining theoretical models to address these challenges will be critical for bridging the gap between theory and experiment in cavity materials engineering.

Identifying material-cavity systems where theory and experiment naturally converge can provide a platform for mutual validation. Certain material classes stand out as promising candidates for significant cavity-induced modifications. Materials with strong intrinsic dipole moments or high polarizabilities, such as ferroelectrics and charge density wave systems, are particularly sensitive to light-matter interactions and are likely to exhibit pronounced effects. Resonant coupling to cavity frequencies can amplify energy exchange between the cavity and material, making materials with electronic, vibrational, or magnetic modes at or near resonance highly attractive for experimental validation. Nonresonant interactions, however, should not be overlooked, as they can still lead to substantial modifications through effective potential renormalization or changes to key material parameters. Additionally, materials with nearly degenerate or competing ground states, such as those found in strongly correlated systems, are especially promising for cavity engineering. Their sensitivity to external perturbations makes them ideal platforms for stabilizing new phases or enhancing exotic properties through cavity-induced effects.

Finally, addressing the technical hurdles of cavity engineering is essential for advancing the field. Plasmonic nanocavities, while offering subwavelength confinement and strong coupling, suffer from optical losses that limit their practicality, particularly at room temperature. Fabry-Pérot cavities provide ease of implementation and control but may struggle with large mode volumes that dilute coupling strength. Innovative cavity designs (e.g., cavity geometry and materials) tailored to specific applications, along with advanced in-situ tunability and characterization techniques, will be crucial. Furthermore, mitigating environmental factors such as temperature fluctuations and external light sources will help bridge the gap between theoretical predictions and experimental outcomes, driving the field toward practical applications.

\subsubsection{General challenges for theoretical developments}
Theoretical advances are also crucial to bridge the gap between existing models and the behavior of real solid-state materials in cavities. This requires integrating diverse approaches from fields, such \ac{DFT}, \ac{QEDFT}, and correlated methods (e.g., coupled cluster, configuration interaction, etc), to provide a holistic description of the coupled light-matter system. Keeping both the intrinsic complexity of the quantized light field and the quantized many-body matter system provides a large scientific playground for future discoveries. 

While significant progress has been made, as highlighted in this review, several challenges lie ahead. One specific issue is accurately capturing collective coupling effects, which are critical for understanding macroscopic interactions between solids and cavity fields. Similar to polaritonic chemistry, artificially enhanced coupling constants are often used in simulations to replicate experimentally observed phenomena. This approach highlights the challenge: the interaction between the solid and the cavity occurs on a macroscopic scale~\cite{svendsen.ruggenthaler.ea_2023}, but its effects feed back non-trivially on the microscopic constituents of the material~\cite{sidler.schnappinger.ea_2024}. Addressing this requires simulating very large supercells, a numerically intensive task. While existing first-principles methods for light-matter systems have been validated for local strong-coupling regimes, where individual electrons and nuclei/ions exhibit strong interactions with the cavity field, their applicability to collective coupling phenomena remains an open question. For instance, such methods could be tested in the context of recently proposed polarization-glass phase transitions, which provide a theoretical framework for explaining collective effects in polaritonic chemistry~\cite{sidler.ruggenthaler.ea_2024}.

Another critical frontier is the self-consistent inclusion of nuclear and ionic contributions. Beyond the dipole approximation, these effects become especially significant when investigating systems in the thermal or time domain. Explicitly simulating multiple modes in such cases is necessary to capture dissipation into the far field from first principles~\cite{flick.welakuh.ea_2019,svendsen.thygesen.ea_2024,konecny.kosheleva.ea_2024}. Moreover, complex cavity structures and collective coupling effects demand simulations that go beyond the \ac{LWA}, incorporating retardation effects to ensure the locality of interactions. Implementing these effects at the wave-function level is particularly challenging, as discussed in Secs.~\ref{subsubsec:PF-Hamiltonian} and~\ref{subsubsec:Approximation_strategies}. Approximation strategies must therefore be applied with precision and caution.

In parallel with first-principles developments, there is a pressing need for simplified, consistent models that can distill the fundamental physics emerging from these complex systems. Such models would provide a conceptual framework for understanding the new phenomena in cavity materials engineering and guide experimental and computational efforts. These combined advancements in theory and modeling will be instrumental in unlocking the full potential of cavity-engineered quantum materials.

\subsubsection{Future directions for QEDFT development for solid-state materials}
To address the challenges outlined above, advancements in the framework of \ac{QEDFT} are essential. For collective light-matter coupling effects, several points must be undertaken. Currently available \ac{QEDFT} functionals, which have been tested extensively for local strong coupling, require validation and potential adaptation for collective coupling scenarios. Encouragingly, in molecular systems, mean-field terms account for much of the effect, suggesting that existing functionals may already capture essential physics in collective cases~\cite{sidler.ruggenthaler.ea_2024}. Moreover, the interplay between nuclei/ions and electrons -- identified as significant in polaritonic chemistry -- needs to be integrated into \ac{QEDFT} for solid-state systems. Initial coupled \ac{QEDFT}+Ehrenfest simulations for nuclei have been successful in molecular cases~\cite{flick.narang_2018,jestadt.ruggenthaler.ea_2019,sidler.schnappinger.ea_2024}, but their extension to solids is still under development.

Incorporating dissipation mechanisms is another critical area of focus. Explicit photon mode treatments~\cite{flick.welakuh.ea_2019,svendsen.thygesen.ea_2024,konecny.kosheleva.ea_2024} and radiation-reaction approaches~\cite{schafer_2022,bustamante.gadea.ea_2021} have shown promise in accurately capturing dissipative effects. How to devise approximated \ac{QEDFT} functionals beyond the \ac{LWA} has already been discussed theoretically in, e.g., Ref.~\cite{schafer.buchholz.ea_2021}. Developing beyond-\ac{LWA} charge-current density functionals is another priority. Progress in this area can use insights from current-density-functional theory~\cite{ruggenthaler.flick.ea_2014,penz.tellgren.ea_2023} to handle more complex photonic interactions and include retardation effects. To further expand the capabilities of \ac{QEDFT}, integrating it with \ac{MQED} provides a natural pathway for describing general photonic environments, including lossy cavities. The dyadic Green's function, central to \ac{MQED}, offers a powerful tool for capturing the spatially varying \ac{EM} response of realistic cavity structures~\cite{scheel.buhmann_2009,rivera.kaminer_2020,svendsen.kurman.ea_2021,svendsen.thygesen.ea_2024}. By incorporating these Green's functions into \ac{QEDFT}, the framework can account for dissipative processes, complex geometries, and nonuniform light-matter interactions. This combination not only enhances the modeling of general photonic environments but also bridges the gap between idealized theoretical models and experimental conditions, enabling the study of cavity-modified material properties in realistic and technologically relevant settings.

Photonic observables remain an underexplored frontier in \ac{QEDFT} for solids. Yet, much could be learned from such photon functionals~\cite{lu.ruggenthaler.ea_2024,schafer.buchholz.ea_2021}; we are at present in the development of \ac{QEDFT} as we were in five decades ago with \ac{DFT} functionals. The detailed behavior of the photonic collective variables in \ac{QEDFT} is important specifically for time-dependent and thermal simulations. Furthermore, existing approximations, typically based on linearly polarized modes, need refinement to accommodate circular or chiral modes, especially when using the length gauge~\cite{ruggenthaler.sidler.ea_2023}.

Integrating \ac{QEDFT} with other \textit{ab initio} techniques offers promising directions for practical applications. For instance, combining \ac{QEDFT} with Wannier function methods~\cite{marzari.mostofi.ea_2012} can facilitate the construction of cavity-modified tight-binding (\ac{TB}) models. Additionally, extending methods like self-consistent \ac{DFT}$+$U(+V)~\cite{agapito.curtarolo.ea_2015,tancogne-dejean.oliveira.ea_2017,tancogne-dejean.rubio_2020} to \ac{QEDFT}$+$U(+V) could provide a way to compute cavity-modified Hubbard $U$ terms within this framework. Another potential advance is extending many-body perturbation theory methods to the \ac{QEDFT} framework, similar to the connection between GW+BSE to \ac{DFT} functional development~\cite{onida.reining.ea_2002}. Similarly, refining pseudopotentials~\cite{heine_1970,schwerdtfeger_2011} and \ac{PAW} methods~\cite{blochl_1994,kresse.joubert_1999} for \ac{QEDFT} will broaden its applicability.

With these theoretical advancements, diverse applications in cavity-modified materials science could be pursued. This includes studying cavity-induced phase transitions, ion diffusion in solids for energy storage, and cavity-modified electron or nuclear dynamics. Future efforts could also explore material properties such as heat capacities, conductivities, and dielectric responses under cavity conditions. These developments provide promising insights into material behavior within general photonic environments, potentially unlocking new avenues in materials design and functionality.

\begin{backmatter}
\bmsection{FUNDING}
This work was supported by the Cluster of Excellence ‘CUI:Advanced Imaging of Matter’ of the Deutsche Forschungsgemeinschaft (DFG) (EXC 2056 and SFB925), and the Max Planck-New York City Center for Non-Equilibrium Quantum Phenomena. I-T.L. thanks Alexander von Humboldt-Stiftung for the support from Humboldt Research Fellowship. D.S. was supported by the National Research Foundation of Korea (NRF) grant funded by the Korea government (MSIT) (No. RS-2024-00333664 and RS-2023-00218180) and the Ministry of Science and ICT(No. 2022M3H4A1A04074153). M.K.S. is supported by the Novo Nordisk Foundation, Grant number NNF22SA0081175, NNF Quantum Computing Programme.

\bmsection{ACKNOWLEDGMENTS}
The Flatiron Institute is a division of the Simons Foundation. We thank Dominik Sidler, Jérôme Faist, Dimitri Basov, Thomas Ebbesen, Tal Schwartz, Emil Viñas Boström, and Lukas Grunwald for the fruitful discussions.

\bmsection{DISCLOSURES}
The authors declare no conflicts of interest.

\bmsection{DATA AVAILABILITY}
All data needed to evaluate the conclusions are present in the paper. Additional data related to this paper may be requested from the authors.

\end{backmatter}

\bibliography{review-AOP-reference-new}

\newpage

\begin{wrapfigure}{l}{32mm} 
\includegraphics[width=32mm,clip,keepaspectratio]{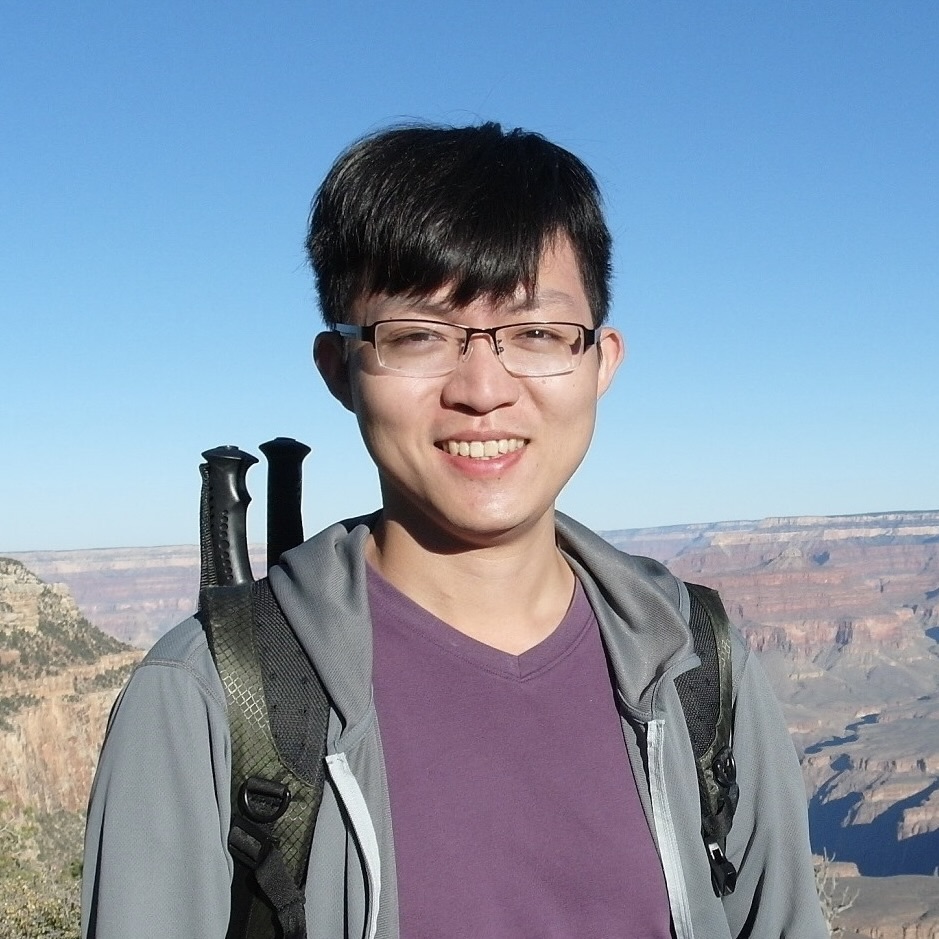}
\end{wrapfigure}
\noindent\textbf{I-Te Lu} received his BS degree in Materials Science and Engineering and MS degree in Applications of Synchrotron Radiation on Materials from National Chiao Tung University, Hsinchu, Taiwan, in 2010 and 2012, respectively. He received his PhD degree in Materials Science from the California Institution of Technology, Pasadena, the United States, in 2020. In 2021, he moved to the Theory department, Max Planck Institute for the Structure and Dynamics of Matter, Hamburg, Germany, as a postdoc researcher, supported by the Alexander von Humboldt Research Fellowship. His research interests include quantum electrodynamical density-functional development, cavity QED materials, and \textit{ab initio} methods for solid-state materials. \\

\begin{wrapfigure}{l}{32mm} 
\includegraphics[width=32mm,clip,keepaspectratio]{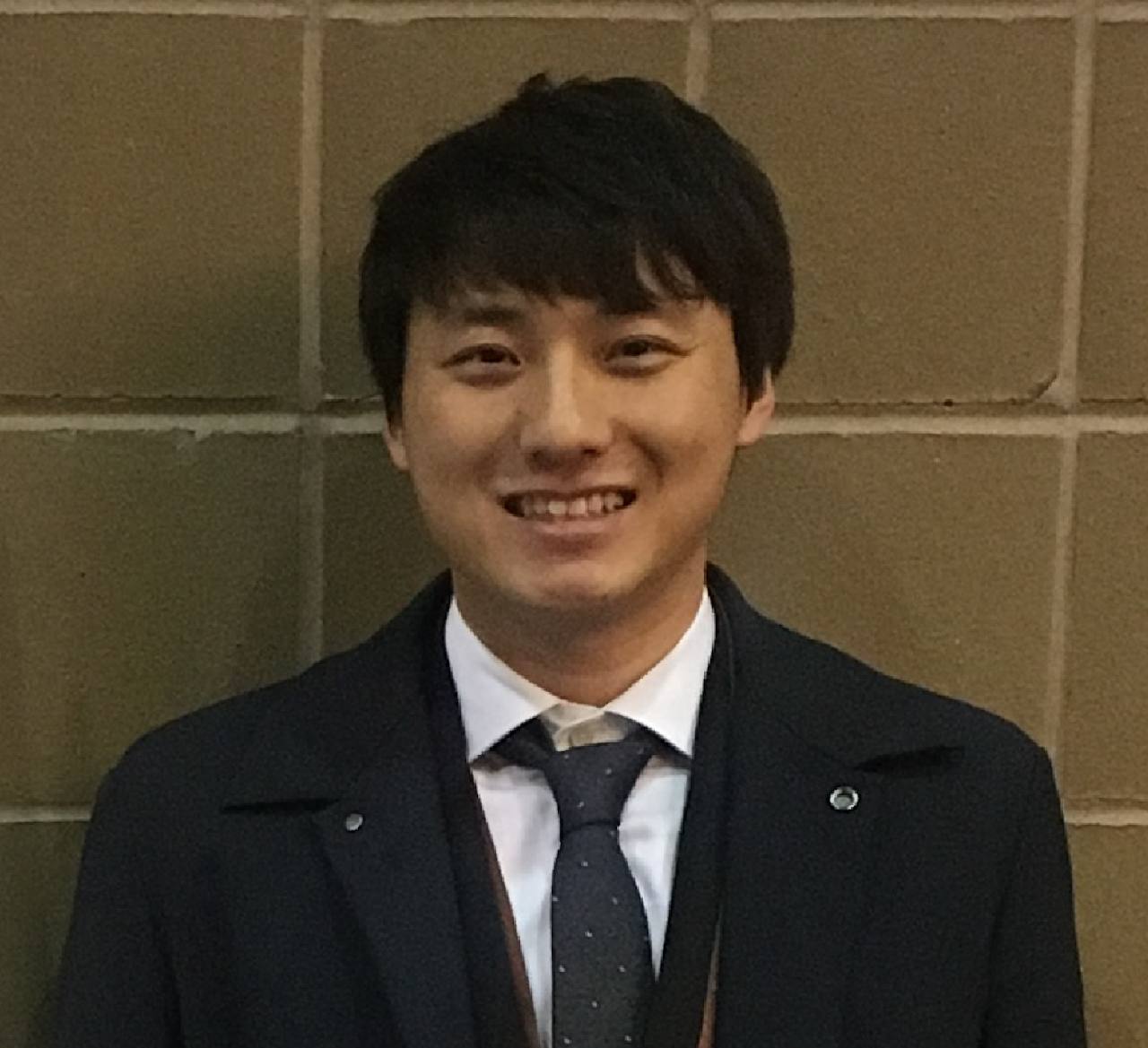}
\end{wrapfigure}
\noindent \textbf{Dongbin Shin} received his BS and PhD degrees from the Ulsan National Institute of Science and Technology, South Korea, in 2013 and 2019, respectively. He subsequently held a Humboldt Research Fellowship as a postdoctoral researcher in the Theory Department at the Max Planck Institute for the Structure and Dynamics of Matter in Hamburg, Germany. Since 2022, he has been an Assistant Professor at the Gwangju Institute of Science and Technology, South Korea. His research focuses on light-induced phase transitions and non-equilibrium dynamics in condensed matter systems, employing \textit{ab initio} methods. \\

\begin{wrapfigure}{l}{32mm} 
\includegraphics[width=32mm,clip,keepaspectratio]{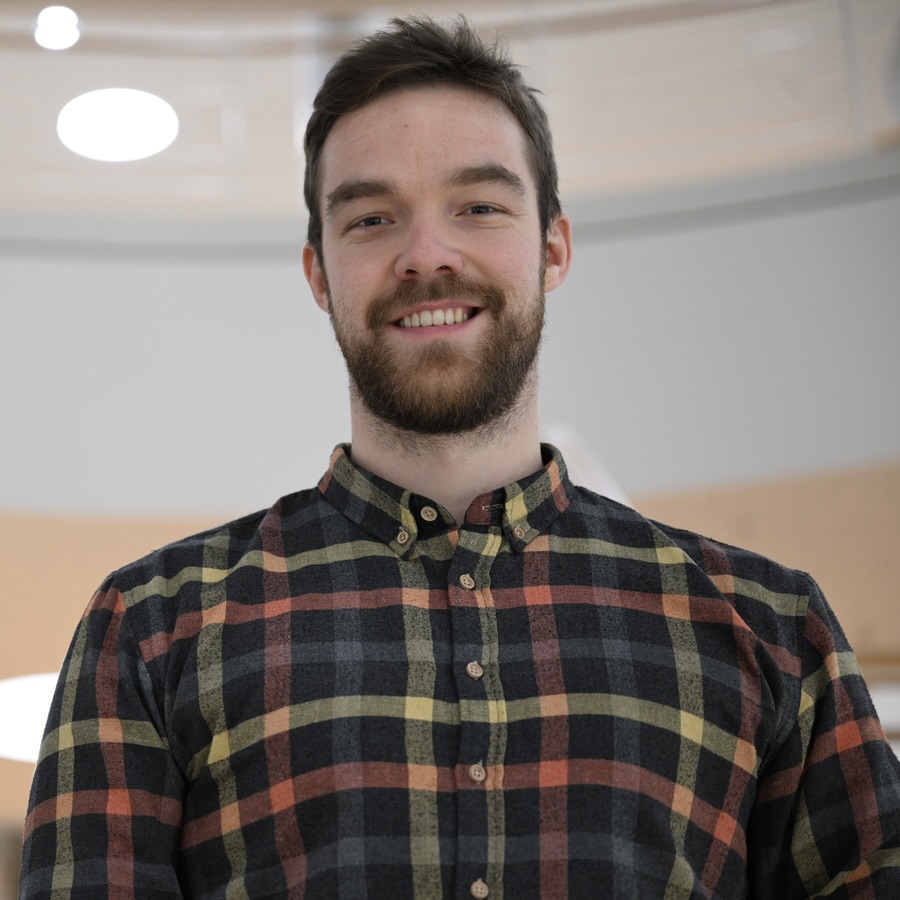}
\end{wrapfigure}
\noindent \textbf{Mark Kamper Svendsen} received his BSc in Physics and Nanotechnology from the Technical University of Denmark (DTU) in 2017 and his MSc degree from the 1:1 program in Physics and Nanotechnology between DTU and the Technical University of Munich in 2019. This was followed by a PhD in Physics from DTU in 2022 for which he won a Young Researcher Award. He subsequently held a postdoctoral researcher position in the Theory Department at the Max Planck Institute for the Structure and Dynamics of Matter in Hamburg, Germany. Since 2024, he has served as an Assistant Professor at the Niels Bohr Institute, University of Copenhagen, Denmark. His research focuses on the development of first-principles methodology to describe the electronic and optical properties of condensed matter systems and the description of electromagnetic environments in ab initio QED. \\

\begin{wrapfigure}{l}{32mm} 
\includegraphics[width=32mm,clip,keepaspectratio]{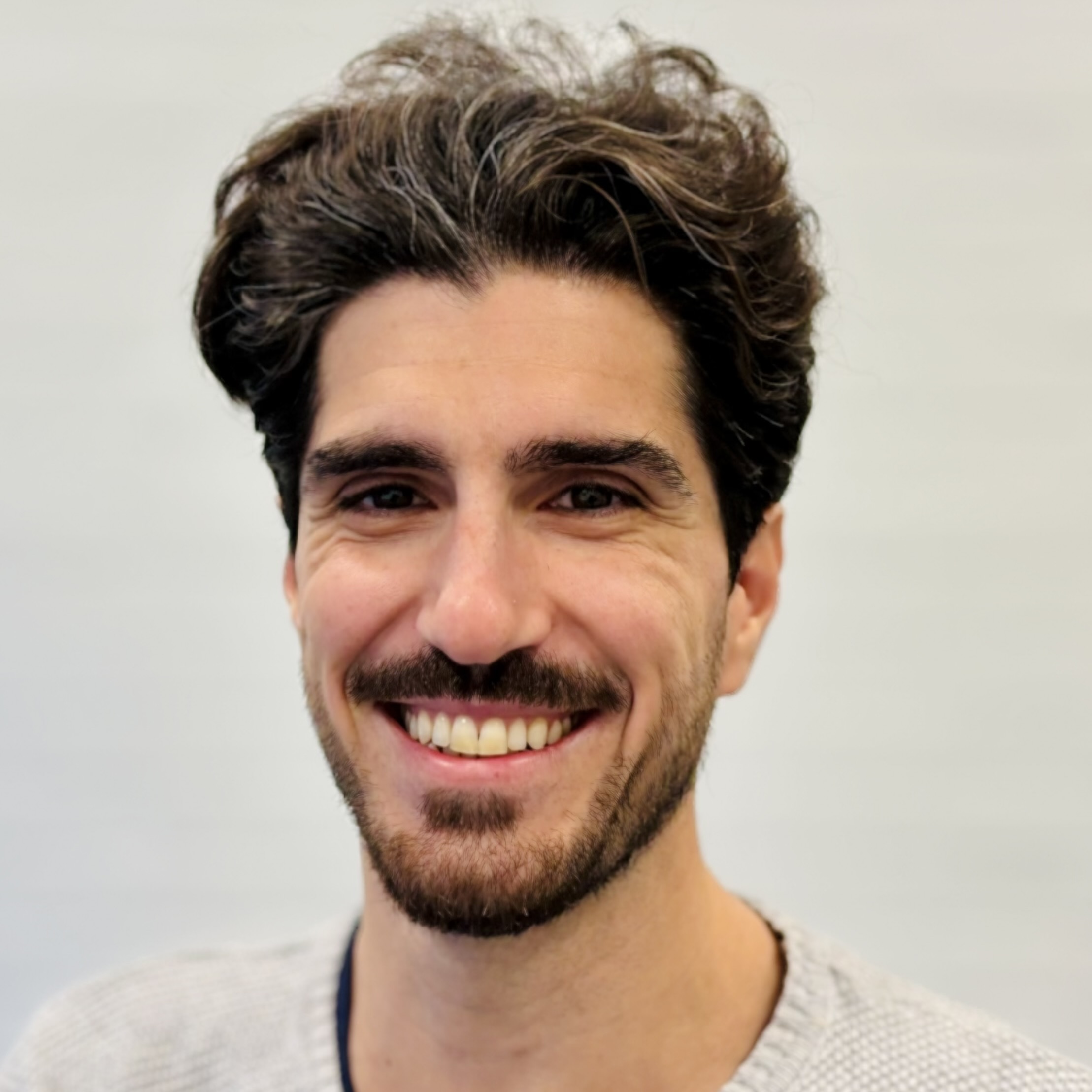}
\end{wrapfigure}
\noindent \textbf{Simone Latini} received his Bachelor and MSc degree in Material Science from the University of Rome Tor Vergata in 2013. Afterward, in 2016, he obtained a PhD in physics from the Technical University of Denmark. The PhD thesis was awarded with the best thesis prize from the Danish Academy of Natural Sciences. Following the PhD he moved to the Theory Department of the Max Planck Institute for the Structure and Dynamics of Matter in Hamburg, Germany. There, he first worked as a postdoctoral researcher, supported by the prestigious Alexander von Humboldt fellowship, and later in 2021 became group leader in the same department. Since the end of 2022, he is an Assistant Professor at the physics department of the Technical University of Denmark. His research is rooted in the development of first-principles frameworks to investigate non-equilibrium electron dynamics, optical properties of two-dimensional materials and strong-light matter interaction in cavity-matter systems. \\

\begin{wrapfigure}{l}{32mm} 
\includegraphics[width=32mm,clip,keepaspectratio]{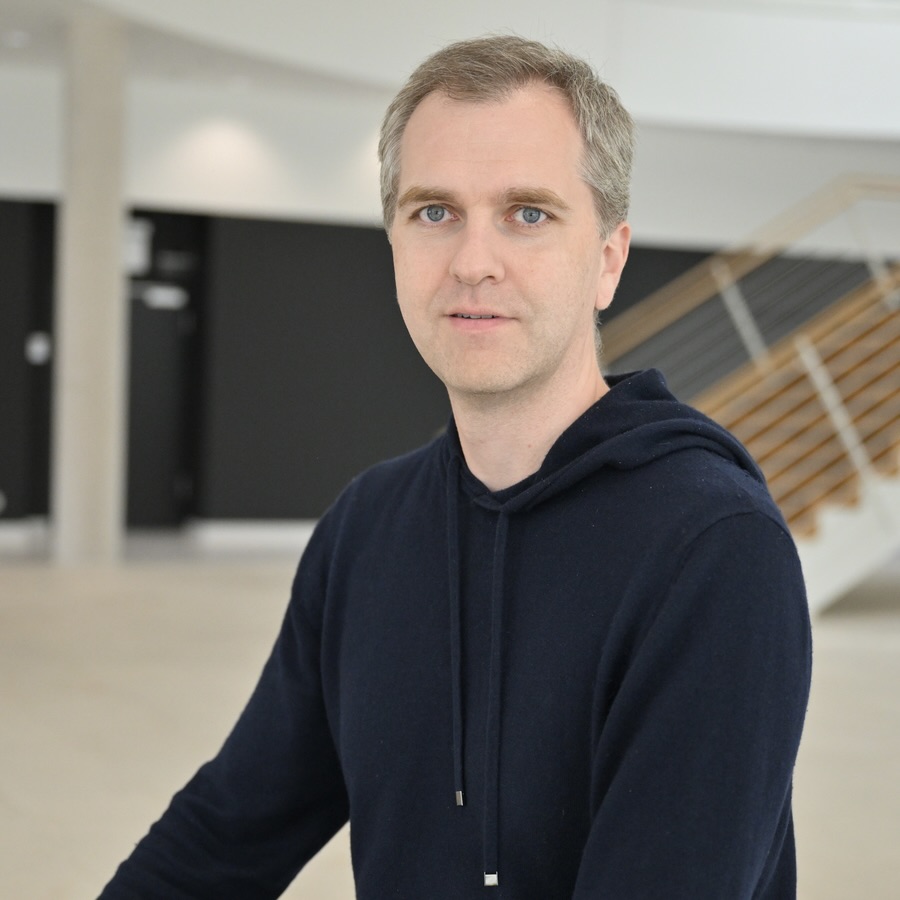}
\end{wrapfigure}
\noindent \textbf{Hannes Hübener} obtained his degree in physics at the University of Hamburg in 2007 and his doctorate at the École polytechnique de Paris in France in 2010. He then spent four years as a Research Fellow in the Materials Modelling Laboratory at the University of Oxford. In 2014, he moved to San Sebastian, Spain, as a Marie Curie Fellow. Since 2017 he heads a research group in the Theory Department at the MPI for the Structure and Dynamics of Matter in Hamburg. \\

\begin{wrapfigure}{l}{32mm} 
\includegraphics[width=32mm,clip,keepaspectratio]{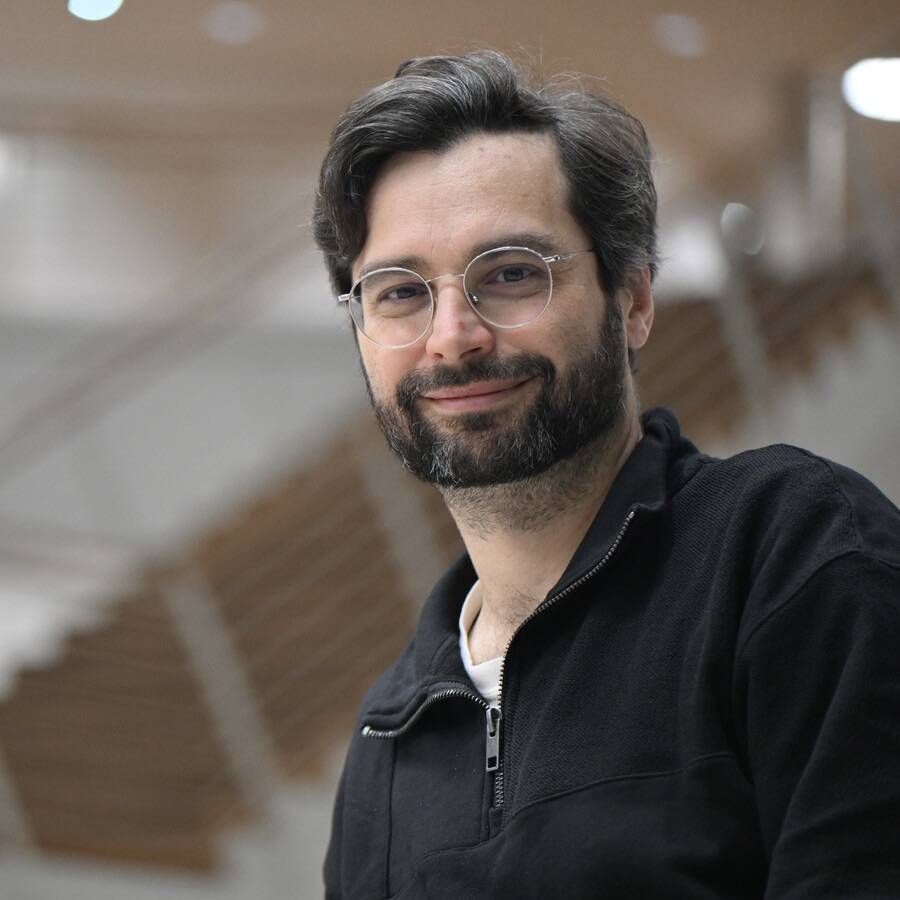}
\end{wrapfigure}
\noindent \textbf{Michael Ruggenthaler} obtained his Ph.D. in theoretical physics jointly from the Max-Planck Institute for Nuclear Physics and the University of Heidelberg, Germany, in 2009. As Schrödinger Fellow of the Austrian Science Fund he worked at the University of Jyväskylä, Finland, and then established an independent research group at the University of Innsbruck, Austria. In 2016 he moved to the Hamburg, where he took on the role of research group leader at the Max-Planck Institute for the Structure and Dynamics of Matter. He has been part of several prestigious national and international grants, e.g., the 2024 ERC Synergy Grant "UnMySt". His research group works on the theoretical and mathematical foundations of quantum many-body theories, in and out of equilibrium. His current research focus is on first-principles approaches to ab initio quantum electrodynamics, polaritonic chemistry and cavity materials engineering. \\

\begin{wrapfigure}{l}{32mm} 
\includegraphics[width=32mm,clip,keepaspectratio]{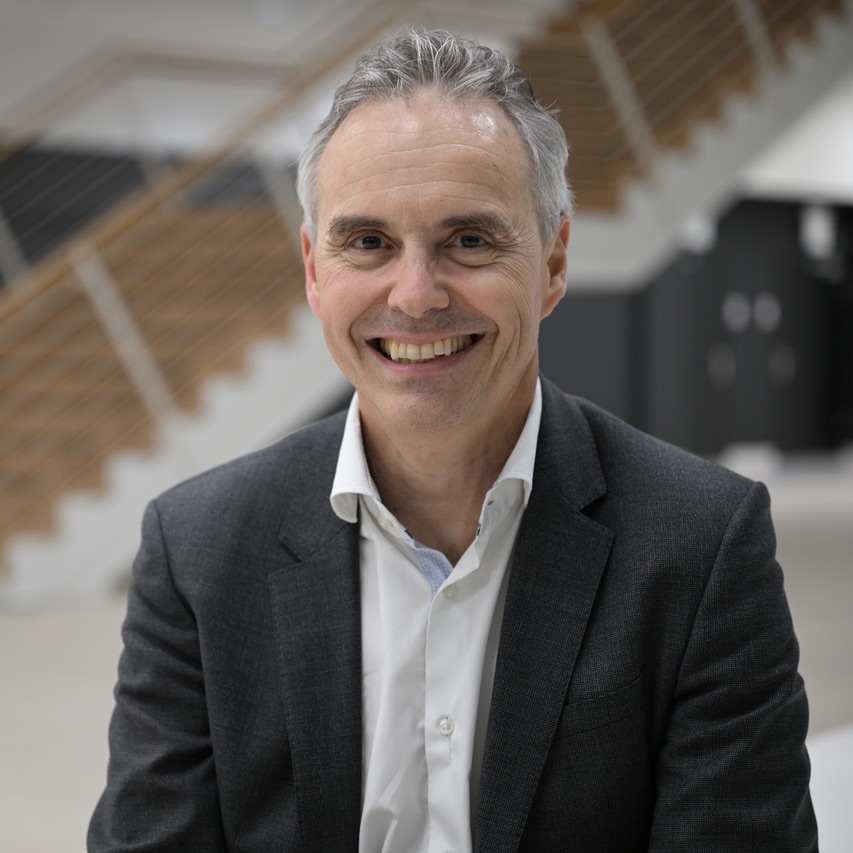}
\end{wrapfigure}
\noindent \textbf{Angel Rubio} is the Director of the Theory Department at the Max Planck Institute for Structure and Dynamics of Matter in Hamburg. His research focuses on modeling and understanding the electronic and structural properties of condensed matter. His group develops advanced theoretical frameworks and computational tools for the ab initio study and control of quantum many-body systems. A key achievement is the development of quantum electrodynamical density functional theory (QEDFT), an extension of TDDFT that predicts and controls non-equilibrium phases of quantum matter. This innovative framework enables the study of electronic and vibrational interactions with photons while preserving the materials' intrinsic properties. The group's work also spans polaritonic chemistry and the design of cavity and Floquet materials, pushing the boundaries of light-matter interactions and quantum electrodynamics in quantum materials. His work has been recognized by several awards, including the 2023 Spanish National Physics Prize “Blas Cabrera”, the  2018 Max Born medal and prize, 2016 Medal of the Spanish Royal Physical Society, the 2014 Premio Rey Jaime I for basic research, the 2006 DuPont Prize in nanotechnology, the 2005 Friedrich Wilhelm Bessel Research Award of the Humboldt Foundation, and  three European Research Council advanced grants. He is a fellow of the APS, EPS and AAAS, and member of  the Leopoldina Academy, BBAW, the European Academy of Sciences, the Academia Europaea, and a foreign associate member of the National Academy of Sciences (USA) and Chinese Academy of Sciences. \\

\end{document}